\documentclass[twocolumn]{aastex63}

\usepackage{amsfonts}
\usepackage{amsmath}
\usepackage{amssymb}
\usepackage{multirow}
\usepackage{dsfont}
\usepackage{color}
\usepackage{natbib}
\usepackage{hyperref}
\usepackage{xspace}
\usepackage{lineno}
\usepackage{graphicx}
\usepackage{enumitem}
\usepackage{xifthen}
\usepackage{hyperref}

\newcommand{\Teff}{T_{\rm eff}}
\newcommand{\logg}{\log g}
\newcommand{\vt}{\nu_t}
\newcommand{\MH}{\text{[M/H]}}

\newcommand{\xtrue}{x_{\textrm{True}}}
\newcommand{\Sigmatilde}{\widetilde\Sigma}
\newcommand{\wtilde}{\widetilde w}

\newcommand{\meanfe}{\left<\text{[Fe/H]}\right>}
\newcommand{\scatfe}{\sigma_{\text{Fe}}}

\newcommand{\code}[1]{\texttt{#1}\xspace}

\newcommand{\SSSSS}{${S}^5$\xspace}

\newcommand{\Ang}{\AA\xspace}
\newcommand{\Cisoratio}{^{12}\text{C}/^{13}\text{C}}
\newcommand{\kms}{\text{km}\,\text{s}^{-1}\xspace}
\newcommand{\isoel}[2]{$^{#1}$#2\xspace}

\newcommand{\Var}{\text{Var}}
\newcommand{\Cov}{\text{Cov}}
\usepackage{mathtools}

\DeclarePairedDelimiterX{\ExpArg}[1]{[}{]}{#1}

\received{}
\revised{}
\accepted{}
\submitjournal{AJ}

\shorttitle{\SSSSS High-resolution Spectroscopy}
\shortauthors{Ji et al.}
\graphicspath{{./}{figures/}}

\begin{document}

\title{The Southern Stellar Stream Spectroscopic Survey (\SSSSS):\\ Chemical Abundances of Seven Stellar Streams}

\correspondingauthor{Alexander P. Ji}
\email{aji@carnegiescience.edu}

% Author list file generated with: mkauthlist 1.2.3+16.gfd56256.dirty 
% mkauthlist -f -j apj highres-data-authors.csv authors.tex 

\author[0000-0002-4863-8842]{Alexander~P.~Ji}
\affiliation{Observatories of the Carnegie Institution for Science, 813 Santa Barbara St., Pasadena, CA 91101, USA}
\affiliation{Hubble Fellow}
\author[0000-0002-9110-6163]{Ting~S.~Li}
\affiliation{Observatories of the Carnegie Institution for Science, 813 Santa Barbara St., Pasadena, CA 91101, USA}
\affiliation{Department of Astrophysical Sciences, Princeton University, Princeton, NJ 08544, USA}
\affiliation{NHFP Einstein Fellow}
\author[0000-0001-6154-8983]{Terese~T.~Hansen}
\affiliation{George P. and Cynthia Woods Mitchell Institute for Fundamental Physics and Astronomy, and Department of Physics and Astronomy, Texas A\&M University, College Station, TX 77843, USA}
\affiliation{Department of Physics and Astronomy, Texas A\&M University, College Station, TX 77843, USA}
\author[0000-0003-0174-0564]{Andrew~R.~Casey}
\affiliation{School of Physics and Astronomy, Monash University, Wellington Rd, Clayton 3800, Victoria, Australia}
\author[0000-0003-2644-135X]{Sergey~E.~Koposov}
\affiliation{McWilliams Center for Cosmology, Carnegie Mellon University, 5000 Forbes Ave, Pittsburgh, PA 15213, USA}
\affiliation{Institute for Astronomy, University of Edinburgh, Royal Observatory, Blackford Hill, Edinburgh EH9 3HJ, UK}
\affiliation{Institute of Astronomy, University of Cambridge, Madingley Road, Cambridge CB3 0HA, UK}
\affiliation{Kavli Institute for Cosmology, University of Cambridge, Madingley Road, Cambridge CB3 0HA, UK}
\author[0000-0002-6021-8760]{Andrew~B.~Pace}
\affiliation{McWilliams Center for Cosmology, Carnegie Mellon University, 5000 Forbes Ave, Pittsburgh, PA 15213, USA}
\author[0000-0002-6529-8093]{Dougal~Mackey}
\affiliation{Research School of Astronomy and Astrophysics, Australian National University, Canberra, ACT 2611, Australia}
\author[0000-0003-3081-9319]{Geraint~F.~Lewis}
\affiliation{Sydney Institute for Astronomy, School of Physics, A28, The University of Sydney, NSW 2006, Australia}
\author[0000-0002-8165-2507]{Jeffrey~D.~Simpson}
\affiliation{School of Physics, UNSW, Sydney, NSW 2052, Australia}
\affiliation{Centre of Excellence for All-Sky Astrophysics in Three Dimensions (ASTRO 3D), Australia}
\author[0000-0001-7516-4016]{Joss~Bland-Hawthorn}
\affiliation{Sydney Institute for Astronomy, School of Physics, A28, The University of Sydney, NSW 2006, Australia}
\affiliation{Centre of Excellence for All-Sky Astrophysics in Three Dimensions (ASTRO 3D), Australia}
\author[0000-0001-8536-0547]{Lara~R.~Cullinane}
\affiliation{Research School of Astronomy and Astrophysics, Australian National University, Canberra, ACT 2611, Australia}
\author[0000-0001-7019-649X]{Gary.~S.~Da~Costa}
\affiliation{Research School of Astronomy and Astrophysics, Australian National University, Canberra, ACT 2611, Australia}
\author[0000-0001-6924-8862]{Kohei~Hattori}
\affiliation{McWilliams Center for Cosmology, Carnegie Mellon University, 5000 Forbes Ave, Pittsburgh, PA 15213, USA}
\author[0000-0002-3430-4163]{Sarah~L.~Martell}
\affiliation{School of Physics, UNSW, Sydney, NSW 2052, Australia}
\affiliation{Centre of Excellence for All-Sky Astrophysics in Three Dimensions (ASTRO 3D), Australia}
\author[0000-0003-0120-0808]{Kyler~Kuehn}
\affiliation{Lowell Observatory, 1400 W Mars Hill Rd, Flagstaff,  AZ 86001, USA}
\affiliation{Australian Astronomical Optics, Faculty of Science and Engineering, Macquarie University, Macquarie Park, NSW 2113, Australia}
\author[0000-0002-8448-5505]{Denis~Erkal}
\affiliation{Department of Physics, University of Surrey, Guildford GU2 7XH, UK}
\author[0000-0003-2497-091X]{Nora~Shipp}
\affiliation{Department of Astronomy \& Astrophysics, University of Chicago, 5640 S Ellis Avenue, Chicago, IL 60637, USA}
\affiliation{Kavli Institute for Cosmological Physics, University of Chicago, Chicago, IL 60637, USA}
\affiliation{Fermi National Accelerator Laboratory, P.O.\ Box 500, Batavia, IL 60510, USA}
\author[0000-0002-3105-3821]{Zhen~Wan}
\affiliation{Sydney Institute for Astronomy, School of Physics, A28, The University of Sydney, NSW 2006, Australia}
\author[0000-0003-1124-8477]{Daniel~B.~Zucker}
\affiliation{Department of Physics \& Astronomy, Macquarie University, Sydney, NSW 2109, Australia}
\affiliation{Macquarie University Research Centre for Astronomy, Astrophysics \& Astrophotonics, Sydney, NSW 2109, Australia}

\begin{abstract}

We present high-resolution Magellan/MIKE spectroscopy of 42 red giant stars in seven stellar streams confirmed by the Southern Stellar Stream Spectroscopic Survey (\SSSSS): 
ATLAS, Aliqa Uma, Chenab, Elqui, Indus, Jhelum, and Phoenix.
Abundances of 30 elements have been derived from over 10,000 individual line measurements or upper limits using photometric stellar parameters and a standard LTE analysis.
This is currently the most extensive set of element abundances for stars in stellar streams.
Three streams (ATLAS, Aliqa Uma, and Phoenix) are disrupted metal-poor globular clusters, although only weak evidence is seen for the light element anticorrelations commonly observed in globular clusters.
Four streams (Chenab, Elqui, Indus, and Jhelum) are disrupted dwarf galaxies, and their stars display abundance signatures that suggest progenitors with stellar masses ranging from $10^6-10^7 M_\odot$.
Extensive description is provided for the analysis methods, including the derivation of a new method for including the effect of stellar parameter correlations on each star's abundance and uncertainty.

This paper includes data gathered with the 6.5 meter Magellan Telescopes located at Las Campanas Observatory, Chile.
\end{abstract}


\section{Introduction} \label{sec:intro}

The Milky Way's stellar halo is a galactic graveyard that contains a record of past accretion events \citep[e.g.,][]{Freeman02, Johnston08, Helmi20}.
Dwarf galaxies and globular clusters fall into the Milky Way, become tidally unbound, and eventually mix into a smooth stellar halo.
Stellar streams are the intermediate stage, when an object is in the midst of tidal disruption, but its stars are still spatially and kinematically coherent.
Hundreds of streams from dozens of accreting objects are expected in the solar neighborhood \citep{Helmi99,Gomez13}, and indeed 
the number of known stellar streams has exploded in recent years \citep[e.g.,][]{Grillmair16b,Mateu18,Shipp18,Ibata19}, in large part thanks to large photometric surveys like the Sloan Digital Sky Survey (SDSS, \citealt{YorkSDSS, StoughtonSDSS}) and Dark Energy Survey (DES, \citealt{desdr1}); and more recently, all-sky proper motions from \textit{Gaia} \citep{GaiaSatellite,GaiaDR2}.

The detailed chemical abundances of stream stars are preserved even after the progenitor galaxy or cluster is disrupted.
Chemodynamic studies of stellar streams are thus a powerful way to investigate the nature of the progenitor systems and directly see the build up of the stellar halo through tidal disruption.
Abundances can be used to determine whether a stream's progenitor is a dwarf galaxy or a globular cluster \citep[e.g.,][]{Gratton04,Tolstoy09,Leaman12,Willman12,Casey14b,Fu18}.
They can also be used to confirm or reject an association between spatially separated stellar structures \citep[e.g.,][]{Freeman02,Kos18,Bergemann18,Marshall19}.
Furthermore, tidally disrupting globular clusters and dwarf galaxies may probe different parts of parameter space compared to their intact counterparts.
For example, metal-poor globular clusters might be more likely to be found as disrupted streams \citep[e.g.,][]{Kruijssen19}; while tidally disrupted dwarf galaxies may have had different accretion times or orbital histories compared to intact galaxies \citep[e.g.,][]{Rocha12}.

Although more than 60 streams have been discovered, only a few have actually been chemically characterized.
The Sagittarius Stream is one of the most prominent structures in the sky and thus has been the subject of many abundance studies \citep[e.g.,][]{Monaco07,Chou10,Keller10,Battaglia17,Carlin18,Hayes20}.
However, thus far, only seven other streams have been the subject of high-resolution spectroscopic abundance studies.
\citet{Casey14b} studied three stars in the Orphan stream, showing its progenitor was a dwarf galaxy;
\citet{Frebel13_300S} and \citet{Fu18} studied a total of seven stars in the 300S stream, also finding its progenitor was a dwarf galaxy;
\citet{Jahandar17} used APOGEE to study one likely stream member around the Palomar 1 globular cluster;
\citet{Marshall19} examined two stars in the stream around the actively disrupting ultra-faint dwarf galaxy Tucana~III, confirming similar abundances in the stream and the galaxy core;
\citet{Roederer19} studied two stars in the Sylgr stream, finding its progenitor was likely an extremely metal-poor globular cluster;
\citet{Simpson20} tagged five members of the Fimbulthul stream to the globular cluster $\omega$ Cen; and
\citet{Roederer10} examined 12 stars in the \citet{Helmi99} debris streams, finding these stars chemically resemble the bulk of the Milky Way's stellar halo.
With only 32 individual stars across seven streams, abundances in stellar streams are still rather sparse.
Eventually, streams become so spatially incoherent that they are considered to be part of the general stellar halo, although the halo can still be broken into discrete components like the Gaia-Enceladus-Sausage \citep[e.g.,][]{Belokurov18,Helmi18} and myriad other chemodynamic groups \citep[e.g.,][]{Kruijssen19b,Matsuno19,Myeong19,Mackereth20,Naidu20,Yuan20}.

The Southern Stellar Stream Spectroscopic Survey (\SSSSS) has been using 2-degree-Field fiber positioner and AAOmega spectrograph \citep{Lewis02,Sharp06} at the Anglo-Australian Telescope (AAT), along with proper motions from \textit{Gaia} \citep{GaiaSatellite,GaiaDR2}, to characterize the kinematics and metallicities of stars in stellar streams \citep{Li19S5, Shipp19}.
So far, \SSSSS has characterized twelve streams with the AAT, and in this work we focus on seven of the nine streams in the Dark Energy Survey footprint \citep[DES,][]{Shipp18, Li19S5}.
The ATLAS stream was initially discovered in the ATLAS survey \citep{Koposov14},
and the Phoenix stream was found in the Phoenix constellation with the first year of DES data \citep{Balbinot16}.
The other five streams (Aliqa Uma, Chenab, Elqui, Indus, and Jhelum) were discovered using the first three years of DES data and named after aquatic terms from different cultures \citep{Shipp18}.
All seven streams show clear tracks in position and velocity space that can be identified by eye \citep{Shipp19, Li19S5}.
\SSSSS has also serendipitously discovered a star with an extreme velocity \citep{Koposov20}.

This paper presents the results from high-resolution Magellan/MIKE \citep{Bernstein03} spectroscopic observations of 42 red giant stars selected from seven streams observed in the \SSSSS survey, including radial velocities and abundances for up to 35 species of 30 elements.
We have observed 5 stars in Aliqa Uma, 7 stars in ATLAS, 3 stars in Chenab, 4 stars in Elqui, 7 stars in Indus, 8 stars in Jhelum, and 8 stars in Phoenix.
Our results represent the most complete characterization of stellar stream abundances to date, doubling the total number of chemically characterized streams and the number of stars in those streams (excluding Sgr).
In this paper, we focus on a detailed description of our abundance analysis methodology. Science results will be presented in other papers \citep{Caseyprep, Hansenprep, Liprep, Paceprep}.
Section~\ref{sec:observations} presents the observation details and radial velocity measurements.
Sections~\ref{sec:stellarparams}~and~\ref{sec:analysis} present the stellar parameters and abundance analysis methods, with the resulting abundances presented in Section~\ref{sec:results} and detailed comments on each element in Section~\ref{sec:elements}.
Brief comments on the character of each individual stream are given in Section~\ref{sec:discussion} before concluding in Section~\ref{sec:conclusion}.
Appendix~\ref{app:params_comparison} compares the stellar parameters to other means of obtaining the parameters.
Appendix~\ref{app:estimator} gives a pedagogical description of calculating abundance uncertainties.
Appendix~\ref{app:equivalentwidths} shows internal validation of the equivalent width and abundances.
Appendix~\ref{app:params_errors} gives several figures showing abundance correlations with stellar parameters.

\section{Observations and Radial Velocities} \label{sec:observations}

The high-resolution targets were selected as the brightest ($r \lesssim 17.5$) member stars in these seven streams based on the kinematic and metallicity information from medium-resolution \SSSSS spectroscopy from the AAT \citep{Lewis02,Sharp06}. 
For the ATLAS, Aliqa Uma and Phoenix streams (globular cluster origins, thin and cold), member stars were selected with a simple cut in proper motion and radial velocity \citep{Liprep,Wan20}. 
For the other four dwarf galaxy origin streams, since the stream has much larger velocity dispersion and their phase space information is more blended with the Milky Way foreground, a selection based on membership probability is used \citep{Paceprep}. The membership probability of each star is calculated with a mixture model based on the spatial location of the star relative to the stream track, the proper motion, the radial velocity, and the metallicity. High membership probability ($P>0.7$) targets were selected for observations.
Note that due to the limited telescope time, not all bright members were observed, and stars with the highest membership probability tend to be mostly metal-poor stars, especially for dwarf galaxy streams where the metallicity spread is large. Therefore, the sample presented here might not be representative of the metallicity distribution for these dwarf galaxy streams. We defer this discussion to the medium-resolution data in other \SSSSS publications which contain a much larger sample of stream members with stellar metallicities.

These stars were observed with the Magellan/MIKE spectrograph \citep{Bernstein03} over four separate runs in 2018-2019, though most stars were observed in 2018 November and 2019 July (Table~\ref{tab:obs}).
The CCDs were binned 2x2, and slit widths of 0\farcs7 and 1\farcs0 were used depending on the seeing, resulting in typical resolutions of $R \sim $ 35k/28k and 28k/22k on the blue/red arms of MIKE, respectively.
Data from each run were reduced with CarPy \citep{Kelson03} and coadded separately.

Radial velocities for each star were measured by combining velocity measurements for individual echelle orders of both MIKE arms.
Only orders $51-88$ were considered, i.e., those with central wavelengths between 4000\,\AA\ and 6800\,\AA. The two bluest orders of the red arm were discarded due to low S/N.
Each order was normalized and the velocity was found using a weighted cross-correlation against a high-S/N spectrum of HD122563. This yielded a velocity and error for each order.
Orders with velocities more than five biweight scales away from the biweight average were iteratively sigma-clipped to remove outliers.
The final velocity is an inverse-variance weighted mean of the remaining order velocities, and we adopt the weighted standard deviation as the velocity error estimate.
Table~\ref{tab:obs} shows the final heliocentric velocity, velocity uncertainty, and the number of orders used to measure the velocity.

While the quoted velocity uncertainties represent the achievable precision, the uncertainties are likely larger due to systematic effects.
For instance, in some cases there were up to 1 $\kms$ offsets in the wavelength calibration between the blue and the red arms of the spectrograph. There were also sometimes trends in the velocities with wavelength, suggesting the atmospheric dispersion corrector did not completely remove the effect.
The maximum size of this range is 3 times the quoted $\sigma(v)$ for all stars, so we recommend any statistical investigation of velocities (e.g. for binarity) inflate the errors by that amount if not investigating the detailed systematic effects.

A few stars (Jhelum2\_15, Phoenix\_6, Phoenix\_10) were observed on multiple runs.
After measuring the velocities separately, there was no clear evidence for velocity variations. 
In all cases, most of the signal for the spectrum came from only one of the runs, and for clarity we report the observed date and MJD just for that run in Table~\ref{tab:obs}.
The velocity for these stars is a weighted average of the individual epochs.

\begin{figure}
    \centering
    \includegraphics[width=\linewidth]{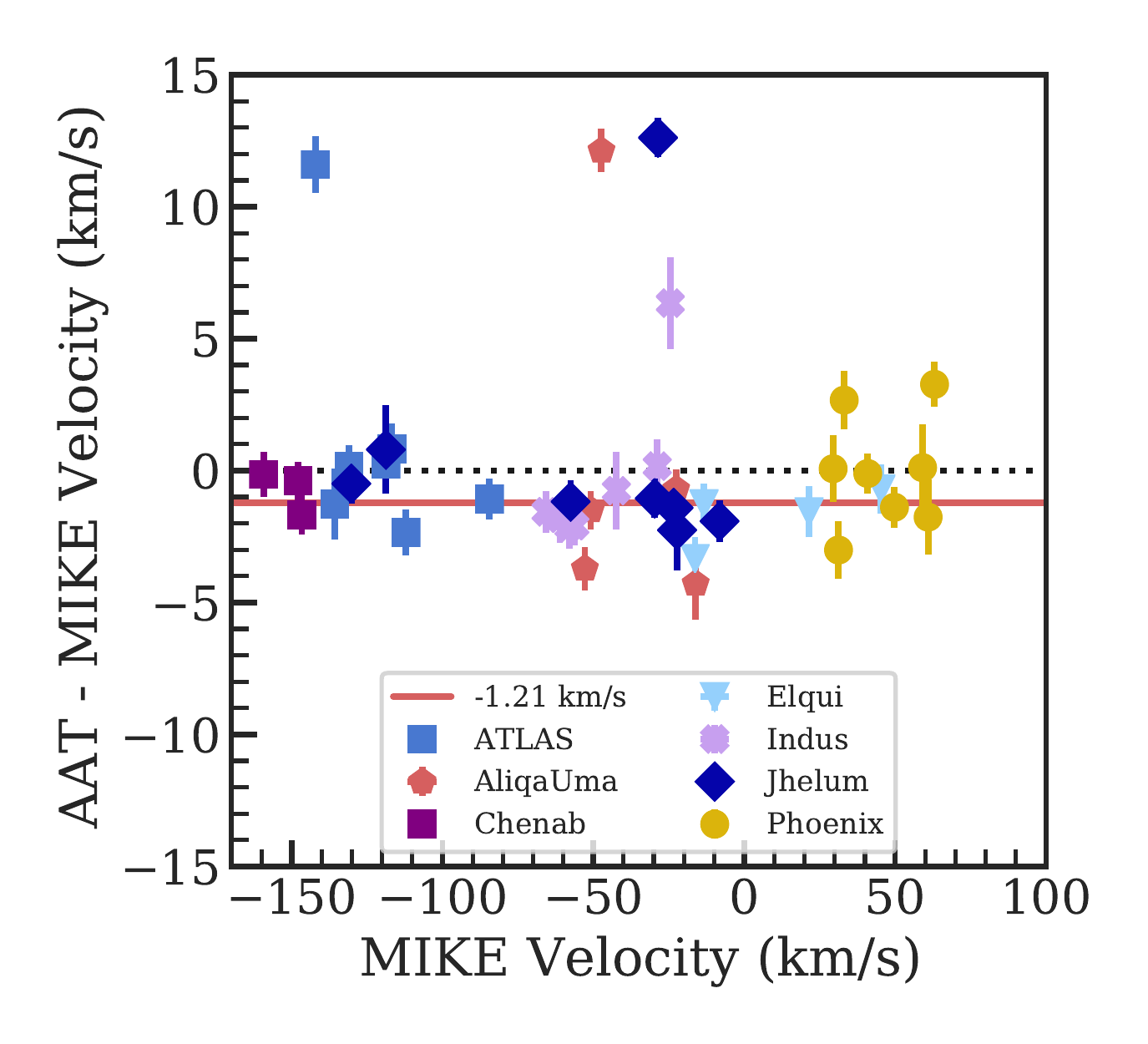}
    \caption{Difference between \SSSSS AAT velocities \citep{Li19S5} and MIKE velocities. After removing binaries, the remaining median offset is $-1.21 \kms$, indicated by the red line.}
    \label{fig:aat_rv}
\end{figure}

Figure~\ref{fig:aat_rv} shows the difference between our MIKE velocities and the \SSSSS AAT velocities \citep{Li19S5}.
The AAT spectra were visually inspected to ensure good quality velocity measurements, and the velocity precision is 0.7-1.7 km/s for all stars.
Three (eight) stars have velocity differences larger than 5$\sigma$ (3$\sigma$), suggesting these stars are likely (possible) binaries.
After removing the eight possible binaries, the median velocity offset is $-1.21\,\kms$, similar in magnitude to the $-1.11\,\kms$ global offset applied to the original \code{rvspecfit} velocities to match the absolute scale of APOGEE and \textit{Gaia}.
Changing between 5$\sigma$ and 3$\sigma$ binary candidates affects this offset by less than $0.05\,\kms$.
Since the absolute scale is uncertain, this offset is not applied in Table~\ref{tab:obs}, but any comparisons between the MIKE and AAT velocities should account for this.

\section{Stellar parameters}\label{sec:stellarparams}

Effective temperature $\Teff$ was determined photometrically using a dereddened $g-r$ color and color-temperature relations derived from the Dartmouth isochrones \citep{Dotter08}.
The photometry was from the Dark Energy Survey Data Release 1 (DR1) with color excess $E(B-V)$ from \citet{SFD98} and the extinction coefficients from DES DR1 \citep{desdr1}, namely, 
\begin{align}
    g_0 &= g - 3.186 E(B-V)_\textrm{SFD} \\
    r_0 &= r - 2.140 E(B-V)_\textrm{SFD}
\end{align}
The photometric uncertainties for our relatively bright stars are dominated by systematics, and we assume a 0.02 mag color uncertainty for all our stars that can be attributed to reddening error.
To convert the photometry to a temperature, the photometry was compared to 12 Gyr alpha-enhanced Dartmouth isochrones with $\mbox{[Fe/H]} = -2.5,\ -2.0,\ -1.5$. Using the isochrone with the closest predicted $g$ magnitude, $g-r$ was converted to $\Teff$.
The difference between the other isochrones was added to the $\Teff$ uncertainty, along with propagating the 0.02 mag color uncertainty.
Together, the typical $\Teff$ uncertainty is 50-60K.
At this level of uncertainty, using different old ages (10-14 Gyr) or alpha-normal isochrones makes negligible extra difference to the derived temperatures.

\begin{rotatetable*}
\begin{deluxetable*}{lccclcccrrrrrrrr}
\centerwidetable
\tablecolumns{16}
\tabletypesize{\scriptsize}
\tablecaption{\label{tab:obs}Observations}
        \tablehead{Name & source\_id & RA & Dec & Obs Date & MJD & $g$ & $r$ & $t_{\text{exp}}$ & Slit & SNR & SNR & $v_{\text{hel}}$ & $\sigma(v)$ & $N_{\text{ord}}$ & $v_{\text{AAT}}$ \\ 
 & & (h:m:s) & (d:m:s) & & & (mag) & (mag) & (min) & (arcsec) & 4500\AA & 6500\AA & (km/s) & (km/s) & & (km/s)} 
\startdata
ATLAS\_0        & 2345957664457105408 & 00:58:40.08 & $-$23:51:49.7 & 2019-07-27 & 58691.29 & 16.17 & 15.47 &  20 & 0.7 & 18 & 35 &$-131.0$&  0.3 & 35 & -130.8 \\ 
ATLAS\_1        & 2349268564550587904 & 00:48:54.95 & $-$22:44:58.0 & 2018-09-30 & 58391.07 & 16.90 & 16.32 & 120 & 0.7 & 18 & 33 &$-135.7$&  0.5 & 33 & -137.0 \\ 
ATLAS\_12       & 5022844307121290752 & 01:40:08.67 & $-$29:52:14.6 & 2018-09-30 & 58391.29 & 15.71 & 14.88 &  80 & 0.7 & 13 & 29 &$ -84.4$&  0.4 & 19 &  -85.5 \\ 
ATLAS\_22       & 5039838702437479936 & 01:16:27.10 & $-$26:07:01.0 & 2018-10-01 & 58392.35 & 16.22 & 15.50 &  20 & 0.7 & 19 & 37 &$-112.1$&  0.5 & 35 & -114.5 \\ 
ATLAS\_25       & 5040671754294144512 & 01:12:21.84 & $-$25:44:52.2 & 2019-07-28 & 58692.26 & 16.56 & 15.92 &  56 & 0.7 & 22 & 45 &$-116.7$&  0.2 & 35 & -115.9 \\ 
ATLAS\_26       & 5040976937490509184 & 01:11:13.03 & $-$24:44:48.3 & 2019-07-27 & 58691.31 & 16.43 & 15.83 &  35 & 0.7 & 23 & 43 &$-118.7$&  0.2 & 34 & -118.5 \\ 
ATLAS\_27       & 2346224467824940544 & 00:52:59.32 & $-$22:54:15.5 & 2018-10-01 & 58392.07 & 16.73 & 16.12 &  35 & 0.7 & 17 & 31 &$-142.2$&  0.4 & 34 & -130.6 \\ 
AliqaUma\_0     & 4953695608534281088 & 02:35:26.13 & $-$37:22:30.2 & 2019-10-19 & 58775.29 & 17.37 & 16.80 & 120 & 1.0 & 17 & 29 &$ -16.1$&  0.7 & 17 &  -20.4 \\ 
AliqaUma\_5     & 4966915105554905344 & 02:26:26.20 & $-$35:22:26.1 & 2018-10-01 & 58392.29 & 16.19 & 15.34 &  23 & 0.7 & 23 & 46 &$ -22.6$&  0.3 & 35 &  -23.3 \\ 
AliqaUma\_7     & 4969961611757057536 & 02:16:18.92 & $-$34:06:22.9 & 2019-07-28 & 58692.34 & 17.03 & 16.45 &  90 & 0.7 & 27 & 49 &$ -47.3$&  0.3 & 34 &  -35.2 \\ 
AliqaUma\_9     & 4971176778264340352 & 02:09:08.30 & $-$32:46:06.1 & 2018-10-01 & 58392.30 & 16.10 & 15.30 &  25 & 0.7 & 24 & 47 &$ -50.9$&  0.2 & 33 &  -52.4 \\ 
AliqaUma\_10    & 4971328167270778496 & 02:09:58.92 & $-$32:05:40.0 & 2018-10-01 & 58392.37 & 16.58 & 15.85 &  28 & 0.7 & 18 & 35 &$ -52.8$&  0.3 & 35 &  -56.5 \\ 
Chenab\_10      & 6558441247408890240 & 21:48:16.11 & $-$52:11:43.8 & 2019-07-26 & 58690.15 & 16.34 & 15.42 &  39 & 0.7 & 24 & 54 &$-147.9$&  0.3 & 35 & -148.2 \\ 
Chenab\_12      & 6558460660661091456 & 21:46:14.54 & $-$52:01:06.6 & 2019-07-28 & 58692.08 & 16.36 & 15.26 &  72 & 0.7 & 20 & 53 &$-146.6$&  0.3 & 35 & -148.3 \\ 
Chenab\_16      & 6559165825572005120 & 21:54:15.45 & $-$49:38:05.0 & 2019-07-27 & 58691.11 & 17.15 & 16.43 &  80 & 0.7 & 20 & 42 &$-159.3$&  0.3 & 35 & -159.4 \\ 
Elqui\_0        & 4935696500108507776 & 01:23:24.63 & $-$43:33:20.0 & 2019-07-26 & 58690.34 & 17.21 & 16.22 &  81 & 0.7 & 19 & 49 &$  45.3$&  0.5 & 35 &  +44.6 \\ 
Elqui\_1        & 4983776837921214336 & 01:19:05.70 & $-$42:07:20.1 & 2018-10-01 & 58392.10 & 16.57 & 15.53 &  49 & 0.7 & 22 & 54 &$ -13.3$&  0.3 & 34 &  -14.6 \\ 
Elqui\_3        & 4984107138085841664 & 01:21:23.11 & $-$42:02:09.2 & 2018-10-01 & 58392.32 & 17.41 & 16.42 &  40 & 0.7 & 15 & 35 &$  21.5$&  0.6 & 10 &  +19.9 \\ 
Elqui\_4        & 4984799005777773952 & 01:12:21.84 & $-$41:33:23.9 & 2018-10-01 & 58392.12 & 17.39 & 16.60 &  60 & 0.7 & 19 & 39 &$ -16.3$&  0.2 & 33 &  -19.6 \\ 
Indus\_0        & 6390575508661401216 & 23:24:01.75 & $-$64:02:20.8 & 2019-07-27 & 58691.16 & 16.20 & 15.60 &  50 & 0.7 & 25 & 47 &$ -28.8$&  0.2 & 34 &  -28.7 \\ 
Indus\_6        & 6394607108562733440 & 22:56:57.69 & $-$62:41:37.5 & 2019-07-25 & 58689.28 & 17.17 & 16.64 &  90 & 0.7 & 21 & 40 &$ -24.5$&  0.3 & 35 &  -18.2 \\ 
Indus\_8        & 6407002315459841152 & 22:43:18.18 & $-$60:54:22.5 & 2019-07-26 & 58690.21 & 17.37 & 16.83 & 120 & 0.7 & 22 & 41 &$ -42.4$&  0.3 & 13 &  -43.1 \\ 
Indus\_12       & 6411531547451908480 & 22:11:27.25 & $-$58:04:44.7 & 2019-07-26 & 58690.25 & 15.45 & 14.70 &  25 & 0.7 & 24 & 49 &$ -56.2$&  0.3 & 35 &  -58.3 \\ 
Indus\_13       & 6412626111276193920 & 22:05:30.97 & $-$56:30:53.4 & 2019-07-25 & 58689.23 & 17.05 & 16.46 &  86 & 0.7 & 23 & 43 &$ -58.0$&  0.3 & 35 &  -60.2 \\ 
Indus\_14       & 6412885389863009152 & 22:00:10.50 & $-$56:04:10.3 & 2019-07-26 & 58690.18 & 16.64 & 16.01 &  40 & 0.7 & 22 & 42 &$ -65.7$&  0.3 & 35 &  -67.2 \\ 
Indus\_15       & 6461006409605852416 & 21:54:09.09 & $-$55:18:35.0 & 2019-07-28 & 58692.15 & 16.94 & 16.30 &  55 & 0.7 & 21 & 40 &$ -61.1$&  0.3 & 35 &  -62.9 \\ 
Jhelum\_0       & 6502308120794799616 & 23:12:15.12 & $-$51:09:36.6 & 2019-07-26 & 58690.30 & 16.38 & 15.81 &  51 & 0.7 & 24 & 43 &$  -8.2$&  0.3 & 34 &  -10.1 \\ 
Jhelum2\_2      & 6501458404465460992 & 23:18:34.74 & $-$52:02:10.2 & 2019-07-28 & 58692.24 & 16.15 & 15.53 &  36 & 0.7 & 24 & 45 &$ -28.6$&  0.2 & 34 &  -16.0 \\ 
Jhelum1\_5      & 6511949016704646272 & 22:16:19.56 & $-$50:00:21.1 & 2019-07-26 & 58690.28 & 16.06 & 15.45 &  35 & 0.7 & 23 & 43 &$ -23.4$&  0.3 & 34 &  -24.8 \\ 
Jhelum1\_8      & 6513867905012445696 & 22:41:59.04 & $-$50:13:01.8 & 2019-07-27 & 58691.20 & 16.70 & 16.17 &  60 & 0.7 & 23 & 41 &$ -22.3$&  0.6 & 33 &  -24.5 \\ 
Jhelum2\_10     & 6514001358235953280 & 22:48:24.41 & $-$50:19:51.6 & 2019-07-25 & 58689.18 & 16.50 & 15.93 &  60 & 0.7 & 25 & 46 &$ -29.7$&  0.3 & 35 &  -30.7 \\ 
Jhelum2\_11     & 6516771371624716288 & 22:36:30.00 & $-$50:24:42.3 & 2019-07-27 & 58691.25 & 16.69 & 16.16 &  60 & 0.7 & 23 & 41 &$ -57.5$&  0.3 & 34 &  -58.7 \\ 
Jhelum2\_14     & 6562728071447798784 & 21:41:13.45 & $-$47:29:02.0 & 2019-07-25 & 58689.10 & 16.43 & 15.88 &  60 & 0.7 & 23 & 44 &$-118.8$&  0.3 & 34 & -118.0 \\ 
Jhelum2\_15     & 6563842426481787264 & 21:33:27.15 & $-$46:06:32.6 & 2019-06-24 & 58658.21 & 15.86 & 15.24 &  64 & 0.7 & 23 & 43 &$-130.5$&  0.3 & 28 & -130.7 \\ 
Phoenix\_1      & 4914426859986001920 & 01:23:48.36 & $-$53:57:27.4 & 2018-10-01 & 58392.18 & 16.98 & 16.39 &  50 & 0.7 & 23 & 39 &$  63.1$&  0.3 & 33 &  +66.3 \\ 
Phoenix\_2      & 4914446067079706624 & 01:24:36.27 & $-$53:40:01.2 & 2019-10-19 & 58775.20 & 17.65 & 17.12 & 120 & 0.7 & 13 & 21 &$  60.9$&  0.6 & 26 &  +59.2 \\ 
Phoenix\_3      & 4914527911976567424 & 01:25:55.15 & $-$53:17:35.1 & 2019-07-25 & 58689.34 & 17.57 & 17.05 & 120 & 0.7 & 22 & 40 &$  59.1$&  0.4 & 32 &  +59.2 \\ 
Phoenix\_6      & 4917862490225433984 & 01:39:20.84 & $-$49:09:11.7 & 2018-09-30 & 58391.16 & 15.96 & 15.30 &  90 & 1.0 & 16 & 29 &$  49.7$&  0.3 & 28 &  +48.4 \\ 
Phoenix\_7      & 4954034292475361280 & 01:42:44.22 & $-$47:29:05.2 & 2018-10-01 & 58392.25 & 16.28 & 15.65 &  30 & 0.7 & 26 & 46 &$  40.9$&  0.2 & 13 &  +40.8 \\ 
Phoenix\_8      & 4954245123830234240 & 01:41:53.37 & $-$47:06:51.6 & 2019-07-27 & 58691.34 & 17.71 & 17.20 & 120 & 0.7 & 21 & 38 &$  33.1$&  0.4 & 31 &  +35.8 \\ 
Phoenix\_9      & 4955727815260641408 & 01:48:16.06 & $-$44:20:53.8 & 2018-10-01 & 58392.21 & 16.99 & 16.43 &  40 & 0.7 & 20 & 35 &$  31.2$&  0.3 & 31 &  +28.2 \\ 
Phoenix\_10     & 4956084950380306816 & 01:51:02.50 & $-$43:02:41.0 & 2018-09-30 & 58391.23 & 16.64 & 16.12 & 134 & 1.0 & 24 & 43 &$  29.2$&  0.6 & 26 &  +29.5 \\ 
\enddata
\end{deluxetable*}
\end{rotatetable*}

Surface gravity $\logg$ was determined photometrically from the DES $g$ magnitude using the equation \citep{Venn17}
\begin{equation}\label{eq:logg}
\begin{split}
    \log g = 4.44 &+ \log M_\star + 4 \log (T_{\rm eff}/5780\text{K}) \\
                  &+ 0.4 (g_0 - \mu + BC(g) - 4.75)
\end{split}
\end{equation}
The \citet{Casagrande14} bolometric corrections ($BC(g)$) were used for SDSS magnitudes, which are not significantly different from DES magnitudes for this purpose. All stars were assumed to have mass $M_\star = 0.75 \pm 0.1 M_\odot$, as typical for an old red giant. The distance moduli $\mu$ were assumed to be constant for each stream, using the values from \citet{Shipp18}. Since some streams exhibit significant distance gradients up to 0.3 mag \citep{Liprep}, we assume a $1\sigma$ distance modulus uncertainty of 0.3 mag.
The final $\logg$ uncertainty is derived by propagating individual uncertainties in Equation~\ref{eq:logg} and is dominated by the distance modulus uncertainty.
The typical $\logg$ uncertainty is 0.16 dex.

After fixing $\Teff$ and $\logg$ and measuring equivalent widths, the microturbulence $\vt$ was determined for each star by balancing the abundance of Fe II lines vs their reduced equivalent width.
We used Fe II instead of Fe I because all our stars have at least 8 Fe II lines spanning a wide range of line strengths (typically $-5.4 < \log \text{EQW}/\lambda < -4.6$, while Fe I lines spanned $-5.4 < \log \text{EQW}/\lambda < -4.5$), and using photometric temperatures has a significant impact on the microturbulence derived from Fe I lines.
This is because an LTE analysis using photometric temperatures will not satisfy excitation equilibrium, and there are correlations between excitation potential and reduced equivalent width.
Using Fe I instead of Fe II typically results in ${\approx}0.3$ km/s higher microturbulence.
The $\vt$ uncertainty is estimated by varying $\vt$ until the slope changed by one standard error on the slope.
The typical $\vt$ uncertainty is $0.21\,\kms$, though in two stars was as high as ${\sim}0.6\kms$. Those stars have lower S/N ratios, resulting relatively few (${\sim}10$) noisier Fe II lines that do not span as wide a range of reduced equivalent widths.

The model metallicity was set to match the simple average of Fe II lines, and $\mbox{[$\alpha$/Fe]} = +0.4$ unless [Mg/Fe] was significantly lower. We used $\mbox{[$\alpha$/Fe]} = +0.0$ for Elqui\_3 and Elqui\_4, $\mbox{[$\alpha$/Fe]} = +0.2$ for Elqui\_0 and AliqaUma\_0, and $\mbox{[$\alpha$/Fe]} = +0.1$ for Jhelum2\_14.
A model metallicity uncertainty of 0.2 dex was adopted for all stars.
The $\mbox{[$\alpha$/Fe]}$ and $\mbox{[M/H]}$ values used do not affect the abundances nearly as much as the temperature, surface gravity, and microturbulence.

\begin{figure}
    \centering
    \includegraphics[width=\linewidth]{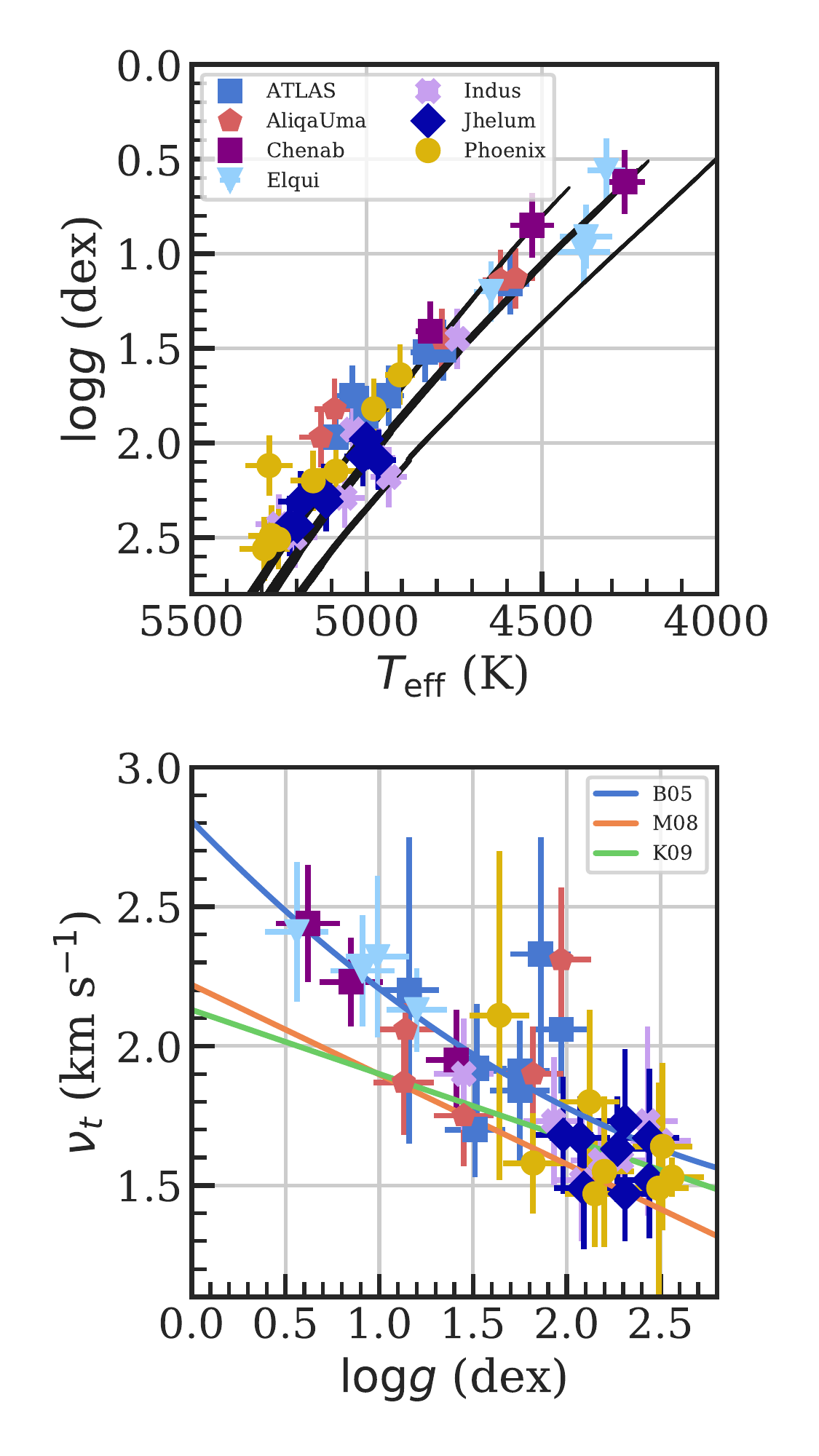}
    \caption{Stellar parameters for all analyzed stars compared to scaling relations.
    Top shows $\Teff$ vs $\logg$ compared to Dartmouth isochrones of three different metallicities ($\mbox{[Fe/H]} = -2.5,\ -2.0,\ -1.5$). Isochrones of different ages and $\alpha$-enhancements have also been plotted, but they are essentially identical for these red giants.
    Bottom shows $\logg$ vs $\vt$ compared to three empirical $\logg$ to $\vt$ fits \citep{Barklem05,Marino08,Kirby09}.}
    \label{fig:sp}
\end{figure}

The resulting stellar parameters are given in Table~\ref{tab:sp} and plotted in Figure~\ref{fig:sp}.
The top panel shows $\Teff$ vs. $\logg$ for our stars, which are well-matched to the Dartmouth isochrones.
The bottom panel shows $\vt$ vs. $\logg$ for our stars, which lie near empirical fits to other high-resolution samples \citep{Barklem05,Marino08,Kirby09}.
In general, the results are well-matched to the \citet{Barklem05} fit, as expected since this fit was derived using the largest number of cool and metal-poor giants.

\begin{deluxetable}{lcccccc}
\tablecolumns{7}
\tabletypesize{\scriptsize}
\tablecaption{\label{tab:sp}Stellar Parameters}
\tablehead{Star & $\Teff$ (K) & $\logg$ (dex) & $\nu_t$ ($\kms$) & [M/H]}
\startdata
AliqaUma\_0  & $5131 \pm 62$ & $1.97 \pm 0.16$ & $2.31 \pm 0.26$ & $-2.40$ \\
AliqaUma\_10 & $4785 \pm 39$ & $1.45 \pm 0.16$ & $1.75 \pm 0.18$ & $-2.28$ \\
AliqaUma\_5  & $4575 \pm 55$ & $1.13 \pm 0.16$ & $1.87 \pm 0.19$ & $-2.34$ \\
AliqaUma\_7  & $5092 \pm 58$ & $1.82 \pm 0.16$ & $1.90 \pm 0.17$ & $-2.37$ \\
AliqaUma\_9  & $4618 \pm 52$ & $1.14 \pm 0.16$ & $2.06 \pm 0.17$ & $-2.46$ \\
ATLAS\_0     & $4833 \pm 41$ & $1.52 \pm 0.16$ & $1.92 \pm 0.23$ & $-2.47$ \\
ATLAS\_1     & $5088 \pm 57$ & $1.97 \pm 0.16$ & $2.06 \pm 0.23$ & $-2.43$ \\
ATLAS\_12    & $4590 \pm 54$ & $1.16 \pm 0.16$ & $2.20 \pm 0.55$ & $-2.16$ \\
ATLAS\_22    & $4781 \pm 44$ & $1.51 \pm 0.16$ & $1.70 \pm 0.17$ & $-2.18$ \\
ATLAS\_25    & $4937 \pm 43$ & $1.75 \pm 0.16$ & $1.84 \pm 0.25$ & $-2.36$ \\
ATLAS\_26    & $5042 \pm 47$ & $1.75 \pm 0.16$ & $1.91 \pm 0.16$ & $-2.26$ \\
ATLAS\_27    & $5002 \pm 44$ & $1.86 \pm 0.16$ & $2.33 \pm 0.42$ & $-2.55$ \\
Chenab\_10   & $4528 \pm 63$ & $0.85 \pm 0.17$ & $2.23 \pm 0.16$ & $-1.94$ \\
Chenab\_12   & $4263 \pm 57$ & $0.62 \pm 0.17$ & $2.44 \pm 0.21$ & $-1.80$ \\
Chenab\_16   & $4819 \pm 41$ & $1.41 \pm 0.16$ & $1.95 \pm 0.18$ & $-2.15$ \\
Elqui\_0     & $4374 \pm 75$ & $0.91 \pm 0.17$ & $2.27 \pm 0.20$ & $-2.02$ \\
Elqui\_1     & $4316 \pm 54$ & $0.56 \pm 0.17$ & $2.41 \pm 0.25$ & $-2.91$ \\
Elqui\_3     & $4380 \pm 74$ & $0.99 \pm 0.17$ & $2.32 \pm 0.29$ & $-1.81$ \\
Elqui\_4     & $4645 \pm 50$ & $1.20 \pm 0.16$ & $2.13 \pm 0.15$ & $-2.03$ \\
Indus\_0     & $5040 \pm 47$ & $1.93 \pm 0.16$ & $1.73 \pm 0.23$ & $-2.41$ \\
Indus\_12    & $4741 \pm 46$ & $1.45 \pm 0.16$ & $1.90 \pm 0.20$ & $-2.14$ \\
Indus\_13    & $5063 \pm 58$ & $2.29 \pm 0.16$ & $1.59 \pm 0.16$ & $-1.91$ \\
Indus\_14    & $4969 \pm 51$ & $2.08 \pm 0.16$ & $1.52 \pm 0.22$ & $-1.98$ \\
Indus\_15    & $4937 \pm 52$ & $2.18 \pm 0.16$ & $1.59 \pm 0.14$ & $-1.71$ \\
Indus\_6     & $5251 \pm 66$ & $2.43 \pm 0.16$ & $1.73 \pm 0.34$ & $-2.45$ \\
Indus\_8     & $5206 \pm 65$ & $2.50 \pm 0.16$ & $1.66 \pm 0.21$ & $-2.02$ \\
Jhelum\_0    & $5122 \pm 58$ & $2.27 \pm 0.16$ & $1.63 \pm 0.19$ & $-2.02$ \\
Jhelum1\_5   & $5011 \pm 53$ & $2.07 \pm 0.16$ & $1.67 \pm 0.15$ & $-2.12$ \\
Jhelum1\_8   & $5199 \pm 66$ & $2.44 \pm 0.16$ & $1.52 \pm 0.21$ & $-2.42$ \\
Jhelum2\_10  & $5116 \pm 58$ & $2.31 \pm 0.16$ & $1.47 \pm 0.17$ & $-2.01$ \\
Jhelum2\_11  & $5220 \pm 65$ & $2.44 \pm 0.16$ & $1.67 \pm 0.25$ & $-2.17$ \\
Jhelum2\_14  & $5188 \pm 66$ & $2.31 \pm 0.16$ & $1.73 \pm 0.26$ & $-2.48$ \\
Jhelum2\_15  & $5001 \pm 52$ & $1.98 \pm 0.16$ & $1.68 \pm 0.21$ & $-2.14$ \\
Jhelum2\_2   & $4967 \pm 51$ & $2.09 \pm 0.16$ & $1.49 \pm 0.22$ & $-1.62$ \\
Phoenix\_1   & $5088 \pm 57$ & $2.15 \pm 0.16$ & $1.47 \pm 0.19$ & $-2.52$ \\
Phoenix\_10  & $5279 \pm 68$ & $2.12 \pm 0.16$ & $1.80 \pm 0.33$ & $-2.93$ \\
Phoenix\_2   & $5252 \pm 66$ & $2.51 \pm 0.16$ & $1.64 \pm 0.30$ & $-2.67$ \\
Phoenix\_3   & $5272 \pm 67$ & $2.49 \pm 0.16$ & $1.49 \pm 0.38$ & $-2.76$ \\
Phoenix\_6   & $4905 \pm 43$ & $1.64 \pm 0.16$ & $2.11 \pm 0.59$ & $-2.68$ \\
Phoenix\_7   & $4980 \pm 45$ & $1.82 \pm 0.16$ & $1.58 \pm 0.18$ & $-2.62$ \\
Phoenix\_8   & $5292 \pm 71$ & $2.56 \pm 0.17$ & $1.53 \pm 0.07$ & $-2.79$ \\
Phoenix\_9   & $5153 \pm 64$ & $2.20 \pm 0.16$ & $1.55 \pm 0.27$ & $-2.70$ \\
\enddata
\tablecomments{All [M/H] errors are taken to be 0.2 dex.}
\vspace{-3em}
\end{deluxetable}

The stellar parameters are compared to a standard 1D-LTE spectroscopic analysis and the AAT spectra analyzed by \code{rvspecfit} in Appendix~\ref{app:params_comparison}, finding good agreement after accounting for expected systematic uncertainties.
It is clear there are no foreground dwarf stars in our sample, validating the use of photometric stellar parameters.

\section{Abundance Analysis} \label{sec:analysis}

A standard abundance analysis was performed with the 2017 version of the 1D LTE radiative transfer code \code{MOOG} that includes scattering \citep{Sneden73,Sobeck11}\footnote{\url{https://github.com/alexji/moog17scat}} and the \code{ATLAS} model atmospheres \citep{Castelli04}.
The analysis code \code{SMHR}\footnote{\url{https://github.com/andycasey/smhr}} (first described in \citealt{Casey14}) was used to measure equivalent widths, interpolate model atmospheres, run \code{MOOG}, and fit syntheses.
We have implemented a new error analysis formalism in \code{SMHR} that is described in Appendix~\ref{app:estimator}.

\subsection{Atomic data}

The base line lists are adapted from \code{linemake}\footnote{\url{https://github.com/vmplacco/linemake}}.
These start with the Kurucz line lists \citep{Kurucz95}\footnote{\url{http://kurucz.harvard.edu/linelists.html}}, then replace individual lines with those from laboratory measurements (summaries in \citealt{Sneden09} for neutron-capture elements; \citealt{Sneden16} for iron peak elements).
The most recent update is to Fe II lines \citep{DenHartog19Fe2}.
We also used NIST to update many light elements (sodium, magnesium, aluminum, silicon, potassium; \citealt{NIST}); VALD to update calcium lines \citep{VALD}; and \citet{Caffau08oxygen} for the oxygen lines.
For molecular lines, the default Kurucz CH lists were replaced with those from \citet{Masseron14}, and the CN lists from \citet{Sneden14}.
Any hyperfine splitting is also taken from \code{linemake}.
The full atomic data and references are given in Table \ref{tab:atomicdata}.

\begin{table}
\caption{\label{tab:atomicdata}Atomic Data}
\begin{tabular}{lcccc}
\hline \hline
Elem. & Wave & ExPot & loggf & Ref \\ \hline
  C-H & 4310.00 &\nodata&\nodata&  1 \\
  C-H & 4323.00 &\nodata&\nodata&  1 \\
  C-N & 3876.00 &\nodata&\nodata&  2 \\
  O I & 6300.30 & 0.00 &$  -9.82$&  3 \\
  O I & 6363.78 & 0.02 &$ -10.30$&  3 \\
 Na I & 5682.63 & 2.10 &$  -0.71$&  4 \\
 Na I & 5688.20 & 2.10 &$  -0.41$&  4 \\
\end{tabular}
\tablerefs{(1) \citet{Masseron14};
(2) \citet{Sneden14};
(3) \citet{Caffau08oxygen}; 
(4) \citet{NIST};
(5) \citet{VALD}; 
(6) \citet{Lawler89}, using hfs from \citet{Kurucz95};
(7) \citet{Lawler13Ti};
(8) \citet{Wood13Ti};
(9) \citet{Lawler14V};
(10) \citet{Wood14V};
(11) \citet{Sobeck07Cr};
(12) \citet{Lawler17Cr};
(13) \citet{DenHartog11Mn};
(14) \citet{Belmonte17Fe};
(15) \citet{DenHartog14Fe};
(16) \citet{Obrian91Fe};
(17) \citet{Ruffoni14Fe};
(18) \citet{Melendez09Fe2};
(19) \citet{DenHartog19Fe2};
(20) \citet{Lawler15Co};
(21) \citet{Wood14Ni};
(22) \citet{Roederer12Zn};
(23) \citet{Biemont11Y};
(24) \citet{Hannaford82Y};
(25) \citet{Ljung06Zr};
(26) \citet{McWilliam98};
(27) \citet{Lawler01La};
(28) \citet{Lawler09Ce};
(29) \citet{DenHartog03Nd};
(30) \citet{Lawler06Sm};
(31) \citet{Lawler01Eu};
(32) \citet{DenHartog06Gd};
(33) \citet{Sneden09}}

This table is available in its entirety in a machine-readable form in the online journal. A portion is shown here for guidance regarding its form and content.
\end{table}

For future reference, we recommend using the Sc II $\log gf$ and hyperfine structure from \citet{Lawler19Sc}, rather than the older \citet{Lawler89} values.
This choice does not affect our results because only UV lines and the 5700\,{\AA} multiplet have significant differences in \citet{Lawler19Sc}, and we did not use any of those lines. The oscillator strengths for the lines we used differ by no more than 0.03 dex in the updated data, within the measurement uncertainty.

R-process isotopes were assumed for Ba and Eu \citep{Sneden08}, and a $\Cisoratio = 9$.
These choices and their impact are discussed in Section~\ref{sec:elements}.

\subsection{Equivalent widths}

Equivalent widths were measured semi-automatically using \code{SMHR}. Each absorption line was fit with a model that includes a (usually Gaussian, sometimes Voigt; see Section~\ref{sec:elements}) absorption profile multiplied by a linear continuum model. After these parameters are optimized, the algorithm identifies groups ($>3$) of neighboring pixels that are significantly discrepant ($>3\sigma$) from the fitted model, and tries to improve the fit by including an absorption profile centered on the group with the profile width matched to the absorption line of interest. This procedure occurs iteratively and minimizes the effects of nearby absorption lines biasing the local continuum determination. 
After this, all measurements were manually inspected to verify each line, primarily to add extra masks as necessary or reject lines with reduction artifacts.
The final equivalent width uncertainties include continuum placement uncertainty.

To verify the equivalent widths from \code{SMHR}, we also independently measured equivalent widths using \code{IRAF}\footnote{IRAF is distributed by the National Optical Astronomy Observatory, which is operated by the Association of Universities for Research in Astronomy (AURA) under a cooperative agreement with the National Science Foundation.} in 2/3 of our target stars.
The differences are consistent with spectrum noise and described in Appendix~\ref{app:equivalentwidths}.

\subsection{Syntheses}

Abundances of synthesized lines were automatically fit using \code{SMHR}. The fitting algorithm does a $\chi^2$ minimization jointly optimizing the abundance of one element, the local continuum (which is usually a linear model), a Gaussian smoothing kernel, and a radial velocity offset that is bounded to be small.
To reduce the number of \code{MOOG} calls, local grids of spectra are synthesized and linearly interpolated within this grid during optimization. Each fit was visually examined, and poor-fitting spectral regions were masked and re-fit.
The final abundance uncertainties include the uncertainty in the local continuum fit, smoothing, and radial velocity.
To verify our results, we also independently synthesized lines for stars spanning the signal-to-noise and stellar parameter range. The differences are mostly consistent with noise and described in Appendix~\ref{app:equivalentwidths}.
For a few elements (Al, Sc, Mn, Ba), this verification suggests the synthesis statistical uncertainties are not sufficient to describe the spectrum noise. An extra systematic uncertainty is added in quadrature for these elements (described in detail in Section~\ref{sec:elements}).

\subsection{Upper limits}\label{sec:limits}

Upper limits were derived with spectral synthesis following the procedure in \citet{Ji20}.
For each feature, a synthetic spectrum was fit to match the continuum, radial velocity, and smoothing of the observed spectrum. Then holding the continuum and smoothing fixed, the abundance was increased until
$\Delta\chi^2 = 25$. This is formally a $5\sigma$ upper limit, though it does not include uncertainties for the continuum placement.
While this works well for individual isolated lines, the provided upper limits for molecular features CH and CN are likely over-confident because they do not account for continuum placement.

\subsection{Combining lines and error analysis}\label{sec:summarizelines}

We have applied a new method to combine individual line measurements and uncertainties in a way that fully and self-consistently propagates statistical and stellar parameter uncertainties for individual line measurements.
A full derivation and justification is described in Appendix~\ref{app:estimator}, but the procedure is described here.

For a given star, let each species X have $N$ lines indexed by $i = 1, \ldots, N$.
Each line has a measured abundance $x_i$ (in units of $\log \epsilon(\text{X})$), statistical uncertainty $e_i$, and stellar parameter differences $\delta_{i,k}$ where $k$ is one of the stellar parameters $\Teff$, $\logg$, $\vt$, or $\MH$.
Additionally, each species X has a systematic uncertainty $s_X \geq 0$, such that the total uncertainty on an individual line is $\sigma_i^2 = e_i^2 + s_X^2$.
Rather than directly combining the lines (e.g. with a straight or inverse variance weighted average), we now include the fact that the lines $x_i$ are correlated due to stellar parameters.

The stellar parameters $\theta = \left(\Teff, \logg, \vt, \MH\right)$ are drawn from a multivariate distribution with covariance matrix $\Sigma_\theta$.
We construct this noting that $\Sigma_{\theta, kl} = \sigma_k\sigma_l\rho_{kl}$ where $\sigma_k$ and $\sigma_l$ are individual stellar parameter uncertainties (from Table~\ref{tab:sp}), and $\rho_{kl}$ is the correlation matrix between these parameters \citep[e.g.,][]{McWilliam13}:
\begin{equation}\label{eq:rhomat}
    \rho =
    \begin{pmatrix}
        1 & \rho_{\Teff,\logg} & \rho_{\Teff,\vt} & \rho_{\Teff,\MH} \\
        \rho_{\Teff,\logg} & 1 & \rho_{\logg,\vt} & \rho_{\logg,\MH} \\
        \rho_{\Teff,\vt} & \rho_{\logg,\vt} & 1 & \rho_{\vt,\MH} \\
        \rho_{\Teff,\MH} & \rho_{\logg,\MH} & \rho_{\vt,\MH} & 1
    \end{pmatrix}
\end{equation}
Since our data are a reasonably large sample of metal-poor red giants, the stellar parameter correlations $\rho$ were estimated by taking the Pearson correlation of our stars' parameters using \code{scipy.stats.pearsonr}, reported in Table~\ref{tab:spcorr}.
The strong correlation between $\Teff$ and $\logg$ matches other isochrone-based determinations \citep{McWilliam13}.

With these values, the $N \times N$ covariance matrix is constructed with
\begin{equation}\label{eq:sigmatilde}
    \Sigmatilde = \text{diag}(\sigma_i^2) + \delta \rho \delta^T
\end{equation}
where $\delta$ is the $N \times 4$ matrix of $\delta_{i,k}$, and $\delta^T$ is the transposed matrix.
The matrix is then inverted to calculate an effective weight for each line:
\begin{equation}\label{eq:wtilde}
    \wtilde_i = \sum_j \Sigma_{ij}^{-1}
\end{equation}
Note that the individual $\wtilde_i$ can be negative, but the sum $\sum_i \wtilde_i$ is always positive.
Also, $\wtilde_i$ must be recomputed if using a subset of lines.
Then the best estimate $\hat x$ of the average abundance of $X$, accounting for all stellar parameter correlations and statistical uncertainties, is:
\begin{equation}\label{eq:xhat}
    \hat x = \frac{\sum_i \wtilde_i x_i}{\sum_i \wtilde_i}
\end{equation}
while the variance on $\hat x$ is given by
\begin{equation}\label{eq:varxhat}
    \Var(\hat x) = \frac{1}{\sum_i \wtilde_i}
\end{equation}
and the error on $X$ is $\sqrt{\Var(\hat x)}$.

Table~\ref{tab:lines} contains all of the individual line measurements.
For each line $i$, it has the line abundance $\log\epsilon_i = x_i$; all the statistical ($e_i$), systematic ($s_X$), and stellar parameter ($\delta_{i,k}$) errors needed to compute $\Sigmatilde$ and $\wtilde$; and the actual value of $\wtilde_i$ for each line.
In the example table, two Fe I lines that have opposite signs for $\wtilde$ are shown.
This means that stellar parameters have a differential effect on the lines relative to the mean abundance. In this case, one Fe line is much stronger than the other, so errors in microturbulence have a substantial differential effect that causes the different signs.
The table also has an example of three Mg I lines with very different weights. The 4703\Ang counts much more because it has a significantly lower statistical uncertainty and moderately less dependence on stellar parameters. The 5172\Ang line has almost no weight, because it is near saturation and a small equivalent error corresponds to a large abundance error.
This illustrates one major benefit of including line-by-line uncertainties, i.e. that known dependencies on stellar parameters and signal-to-noise are automatically taken into account. The final abundances are thus much less dependent on the specific set of lines chosen for abundance measurements.

The final combined abundances are tabulated in Table~\ref{tab:abunds};
$\log\epsilon$ is the result of Equation~\ref{eq:xhat}.
The standard spectroscopic notation $\mbox{[X/H]} = \log\epsilon(\text{X}) - \log\epsilon_\odot(\text{X})$ is normalized using solar abundances from \citet{Asplund09}.
Uncertainties in the solar normalization were not propagated, so the [X/H] uncertainties are the same as the $\log\epsilon$ uncertainties.
$\sigma_{\textrm{[X/H]}}$ is the result of Equation~\ref{eq:varxhat}.

The [X/Fe] values have two complications: a choice must be made between Fe I and Fe II, and correlated uncertainties in X and Fe must be propagated.
By default in this paper, we have decided to use Fe I for neutral species and Fe II for ionized species (e.g., [Mg I/Fe I] or [Ti II/Fe II]).
This is because neutral and ionized species usually have similar dependencies on stellar parameters, maximizing the precision on the final [X/Fe] ratio \citep[e.g.,][]{Roederer14c}.
For the correlated uncertainties, first note that [X/Fe] $=$ [X/H] - [Fe/H].
Thus $\Var(\text{[X/Fe]}) = \Var(\text{X}) + \Var(\text{Fe}) - 2 \Cov(\text{X},\text{Fe})$.
For any two different species X and Y, the covariance in $\log\epsilon(\text{X})$ and $\log\epsilon(\text{Y})$ is given by 
\begin{equation}\label{eq:covxy}
    \Cov(\hat x, \hat y) = \Delta_X \rho \Delta_Y
\end{equation}
where $\Delta_X$ is a vector of the $\Delta_{X,k}$ for $k = T, g, v, M$ given in Table~\ref{tab:abunds} and $\rho$ is from Equation~\ref{eq:rhomat}. 
$\Delta_X$ is the weighted response of species X to the stellar parameter errors in Table~\ref{tab:sp}, defined in detail in Appendix~\ref{app:estimator}.
The error $\sigma_{\text{[X/Fe]}}$ in Table~\ref{tab:abunds} is then calculated using Equations~\ref{eq:varxhat} and \ref{eq:covxy}.
Note that Equation~\ref{eq:covxy} is not correct if X=Y, use Equation~\ref{eq:varxhat} instead.

There are sometimes mild differences between [M/H] and [Fe/H] because the stellar parameter determination did not include the effect of weighted lines. However, the resulting differences in the model metallicity are much less than $<0.2$ dex, which is included in the error propagation. Model metallicity uncertainties also make negligible difference to the results compared to other sources of uncertainty.
In general [Fe~I/H] and [Fe~II/H] agree, with a typical difference of $-0.08 \pm 0.11$ dex, where [Fe I/H] is lower as expected from NLTE effects \citep[e.g.,][]{Ezzeddine17}.
However, four stars have particularly large differences: Elqui\_0, Elqui\_3, Elqui\_4, and ATLAS\_12 have $\mbox{[Fe I/H]} - \mbox{[Fe II/H]} < -0.20$ (see Section~\ref{sec:fe}).

\begin{table}
\caption{\label{tab:spcorr}Stellar Parameter Correlations}
\centering
\begin{tabular}{lc}
\hline \hline
Variables & Value \\ \hline
$\rho_{\Teff,\logg}$ & $+0.96$ \\
$\rho_{\Teff,\vt}$   & $-0.82$ \\
$\rho_{\Teff,\MH}$   & $-0.37$ \\
$\rho_{\logg,\vt}$   & $-0.87$ \\
$\rho_{\logg,\MH}$   & $-0.21$ \\
$\rho_{\vt,\MH}$     & $+0.01$ \\
\end{tabular}
\end{table}

\section{Abundance Results} \label{sec:results}

Table~\ref{tab:lines} has every individual line measurement for our stars, including upper limits.
Each row contains the star name, the wavelength $\lambda$ of the relevant feature in \Ang, the \code{MOOG} species (ID), the excitation potential and $\log gf$, the equivalent width and uncertainty when available (EW, $\sigma$(EW)), the full width half max (FWHM in \Ang), an upper limit flag (ul), the measured abundance $\log \epsilon_i$, a total abundance uncertainty $\sigma_i$, a statistical uncertainty $e_i$ that propagates spectrum noise, a systematic uncertainty $s_X$ that accounts for line-to-line scatter in excess of the abundance uncertainties (see Appendix~\ref{app:estimator}), the stellar parameter abundance differences $\delta_{i,k}$, and an effective weight $\wtilde_i$.

Table~\ref{tab:abunds} has the final abundances for our stars. 
Each row contains the star name; the element measured (El.); the number of lines used ($N$); an upper limit flag (ul); the abundance ($\log \epsilon$); the [X/H] value relative to the \citet{Asplund09} solar abundances; the uncertainty on $\log\epsilon(X)$ and [X/H] that includes both statistical and stellar parameter uncertainties ($\sigma_{\text{[X/H]}}$);
the [X/Fe] value and uncertainty (where Fe is Fe I if X is neutral and  Fe II if X is ionized); and the abundance differences due to a $1\sigma$ change in stellar parameters $\Delta_k$.
Several important elements and their abundance uncertainties are summarized for all stars in Table~\ref{tab:abundsummary}.

Figure~\ref{fig:xfegrid} shows most of the element abundances measured in this paper.
This figure uses [Fe I/H] on the x-axis, and [X/Fe] ratios where Fe can be either Fe I or Fe II.
We use the species Ti II, V I, Cr I, and Sr II for those elements, and C-H and C-N for the C and N abundances.
Cu, Ce, Nd, Sm, and Gd have not been plotted.
The error ellipses are the proper covariances between [X/Fe] and [Fe I/H], where any correlation is introduced solely through stellar parameters.

Individual correlations with stellar parameters are shown in Appendix~\ref{app:params_errors}.
Salient features of these figures will be discussed in Section~\ref{sec:elements}.
In brief summary, the elements C, N, Al, Sc, V, Mn, Co, Cu, Sr, Y, Zr, Ba, La, Eu, Dy were measured with spectral synthesis, while the other elements O, Na, Mg, Si, K, Ca, Ti, Cr, Fe, Ni, Zn, Ce, Nd, Sm, and Gd were measured with equivalent widths.
Species having known significant non-local thermodynamic equilibrium (NLTE) effects potentially in excess of 0.2 dex include Na I, Al I, K I, Ti I, Cr I, Mn I, and Fe I. The NLTE effects have not been included in this analysis.

\section{Comments on Specific Elements} \label{sec:elements}

This section contains comments useful for interpreting the abundances of these elements, such as how the abundances were derived, and relevant caveats such as sensitivity to stellar parameters or NLTE effects.

\subsection{Carbon, Nitrogen, Oxygen}
C is measured from spectral synthesis of the CH molecular features at 4313\,\Ang and 4323\,\Ang, where each of these regions is treated independently.
AliqaUma\_0 has too low S/N to measure a C abundance, so upper limits were placed.
[C/Fe] clearly decreases as $\logg$ decreases, which is expected for red giants as they ascend the giant branch \citep[e.g.,][]{Placco14}.

Oxygen affects the C abundance through CO molecular equilibrium, but we have only measured it in two stars.
Thus $\mbox{[O/Fe]} = +0.4$ was assumed throughout.
Reducing to $\mbox{[O/Fe]} = 0.0$ decreases the [C/Fe] abundance by less than 0.05\,dex for all our stars, which we regard as negligible.
Increasing to $\mbox{[O/Fe]} = 1.0$ increases [C/Fe] by less than 0.1 dex for most stars.
We thus add an extra uncertainty of 0.1 dex in quadrature to the statistical [C/Fe] error ($e_i$ in Table~\ref{tab:lines}).
This is mostly sufficient, but three of the coolest and most metal-rich stars ($\Teff \lesssim 4300$ K, $\mbox{[Fe/H]} \gtrsim -1.9$) have much larger [C/Fe] differences when changing [O/Fe]:
Chenab\_12, Elqui\_0, and Elqui\_3 have [C/Fe] increase by 0.32, 0.18, and 0.29 dex respectively when increasing [O/Fe] to $+1$.
For consistency, the systematic error was kept at 0.1 dex for these three stars.

For isotopes, the ratio $\Cisoratio = 9$ is assumed throughout. This value is chosen because all analyzed stars are RGB stars and have been through the first dredge-up that produces an equilibrium value of $\Cisoratio$ close to 9.
Visually comparing synthetic spectra with different isotope ratios around 4224\Ang and 4323\Ang shows this is a good assumption.
In many cases a typical higher value of $\Cisoratio = 99$ might provide a moderately better fit, and the stars Chenab\_16 and Elqui\_1 might have a $\Cisoratio$ as low as 4. However, the data generally do not have enough S/N to place a meaningful constraint on the isotope ratio.

\begin{deluxetable*}{lrrrrrrrrrrrrrrrrrr}
\tablecolumns{18}
\tabletypesize{\footnotesize}
\tablecaption{\label{tab:lines}Line Measurements}
\tablehead{Star & $\lambda$ & ID & $\chi$ & $\log gf$ & EW & $\sigma$(EW) & FWHM & ul & $\log \epsilon_i$ & $\sigma_i$ & $e_i$ & $s_X$ & $\delta_{i,\Teff}$ & $\delta_{i,\logg}$ & $\delta_{i,\nu_t}$ & $\delta_{i,\text{[M/H]}}$ &$\widetilde{w}_i$}
\startdata
ATLAS\_1    & 4702.99 &  12.0 &  4.35 & -0.44 &  85.7 &  3.9 &  0.17 &   &  5.56 &  0.06 &  0.06 &  0.00 & +0.04 & -0.02 & -0.04 & +0.00 & 121.19\\
ATLAS\_1    & 5172.68 &  12.0 &  2.71 & -0.39 & 194.9 & 38.4 &  0.30 &   &  5.37 &  0.30 &  0.30 &  0.00 & +0.07 & -0.06 & -0.08 & -0.00 &  1.08\\
%ATLAS\_1    & 5183.60 &  12.0 &  2.72 & -0.17 & 308.0 & 62.8 &  0.33 &   &  5.82 &  0.21 &  0.21 &  0.00 & +0.08 & -0.08 & -0.03 & +0.01 & 17.38\\
%ATLAS\_1    & 5194.94 &  26.0 &  1.56 & -2.02 &  59.8 &  7.1 &  0.16 &   &  4.80 &  0.27 &  0.12 &  0.24 & +0.07 & -0.01 & -0.04 & -0.00 &  4.67\\
%ATLAS\_1    & 5202.34 &  26.0 &  2.17 & -1.87 &  35.4 &  7.6 &  0.16 &   &  4.94 &  0.28 &  0.13 &  0.24 & +0.06 & -0.01 & -0.02 & -0.00 &  6.75\\
%ATLAS\_1    & 5206.02 &  24.0 &  0.94 &  0.02 &  83.3 &  7.9 &  0.20 &   &  2.94 &  0.23 &  0.16 &  0.17 & +0.07 & -0.01 & -0.08 & -0.00 &  4.83\\
%ATLAS\_1    & 5210.38 &  22.0 &  0.05 & -0.82 &  38.9 &  6.1 &  0.15 &   &  2.85 &  0.10 &  0.10 &  0.00 & +0.08 & -0.01 & -0.02 & +0.00 & 22.22\\
%ATLAS\_1    & 5215.18 &  26.0 &  3.26 & -0.86 &  57.1 &  9.5 &  0.24 &   &  5.50 &  0.29 &  0.16 &  0.24 & +0.05 & -0.00 & -0.04 & +0.00 &  5.88\\
%ATLAS\_1    & 5216.27 &  26.0 &  1.61 & -2.08 &  67.4 &  5.8 &  0.15 &   &  5.04 &  0.26 &  0.10 &  0.24 & +0.07 & -0.01 & -0.06 & -0.00 &  3.59\\
ATLAS\_1    & 5227.19 &  26.0 &  1.56 & -1.23 & 139.7 &  9.4 &  0.25 &   &  5.62 &  0.30 &  0.18 &  0.24 & +0.09 & -0.03 & -0.19 & -0.02 & -4.86\\
ATLAS\_1    & 5250.65 &  26.0 &  2.20 & -2.18 &  50.8 &  6.1 &  0.13 &   &  5.53 &  0.26 &  0.10 &  0.24 & +0.06 & -0.00 & -0.03 & +0.00 &  6.39\\
ATLAS\_1    & 5528.40 &  12.0 &  4.35 & -0.50 &  92.1 &  6.8 &  0.24 &   &  5.70 &  0.10 &  0.10 &  0.00 & +0.04 & -0.01 & -0.05 & +0.00 & 26.70\\
ATLAS\_1    & 4310.00 & 106.0 &\nodata&\nodata&  syn  &  syn &  0.14 &   &  6.43 &  0.10 &  0.10 &  0.00 & +0.12 & -0.06 & +0.00 & +0.05 & 64.05\\
\enddata
\tablecomments{A portion of this table is shown for form. The full version is available online.}
\end{deluxetable*}

\begin{deluxetable*}{llccrrrrrrrrrr}
\tablecolumns{14}
\tabletypesize{\footnotesize}
\tablecaption{\label{tab:abunds}Stellar Abundances}
\tablehead{Star & El. & $N$ & ul & $\log \epsilon$ & [X/H] & $\sigma_{\text{[X/H]}}$ & [X/Fe] & $\sigma_{\text{[X/Fe]}}$ & $\Delta_T$ & $\Delta_g$ & $\Delta_v$ & $\Delta_M$ & $s_X$}
\startdata
ATLAS\_1        & C-H   &   2 &     &$+6.43$&$-2.00$&  0.09 &$+0.41$&  0.10 &  0.11 & -0.06 &  0.01 &  0.05 &  0.00 \\
ATLAS\_1        & C-N   &   1 & $<$ &$+6.21$&$-1.62$&\nodata&$+0.78$&\nodata&\nodata&\nodata&\nodata&\nodata&\nodata\\
ATLAS\_1        & O I   &   1 & $<$ &$+8.18$&$-0.51$&\nodata&$+1.89$&\nodata&\nodata&\nodata&\nodata&\nodata&\nodata\\
ATLAS\_1        & Na I  &   2 &     &$+4.45$&$-1.79$&  0.13 &$+0.61$&  0.12 &  0.08 & -0.06 & -0.09 & -0.01 &  0.00 \\
ATLAS\_1        & Mg I  &   6 &     &$+5.60$&$-2.00$&  0.07 &$+0.40$&  0.08 &  0.04 & -0.02 & -0.04 &  0.00 &  0.00 \\
ATLAS\_1        & Al I  &   2 &     &$+3.10$&$-3.35$&  0.50 &$-0.95$&  0.50 &  0.11 & -0.04 & -0.06 &  0.02 &  0.59 \\
ATLAS\_1        & Si I  &   2 &     &$+5.76$&$-1.75$&  0.14 &$+0.66$&  0.15 &  0.02 & -0.04 & -0.06 &  0.01 &  0.00 \\
ATLAS\_1        & K I   &   2 &     &$+3.40$&$-1.63$&  0.10 &$+0.77$&  0.10 &  0.05 & -0.01 & -0.03 & -0.00 &  0.00 \\
ATLAS\_1        & Ca I  &  16 &     &$+4.29$&$-2.05$&  0.08 &$+0.35$&  0.09 &  0.04 & -0.00 & -0.03 & -0.00 &  0.17 \\
ATLAS\_1        & Sc II &   7 &     &$+0.72$&$-2.43$&  0.10 &$+0.05$&  0.10 &  0.01 &  0.04 & -0.01 &  0.01 &  0.10 \\
ATLAS\_1        & Ti I  &  11 &     &$+2.92$&$-2.03$&  0.09 &$+0.38$&  0.09 &  0.06 & -0.01 & -0.01 &  0.00 &  0.00 \\
ATLAS\_1        & Ti II &  26 &     &$+2.91$&$-2.04$&  0.09 &$+0.44$&  0.10 &  0.02 &  0.05 &  0.01 &  0.02 &  0.21 \\
ATLAS\_1        & V I   &   1 &     &$+1.75$&$-2.18$&  0.12 &$+0.22$&  0.13 &  0.02 &  0.02 &  0.01 & -0.02 &  0.00 \\
ATLAS\_1        & V II  &   1 &     &$+1.75$&$-2.18$&  0.21 &$+0.30$&  0.20 & -0.01 &  0.16 &  0.04 &  0.01 &  0.00 \\
ATLAS\_1        & Cr I  &   5 &     &$+3.21$&$-2.42$&  0.11 &$-0.02$&  0.11 &  0.06 & -0.01 & -0.02 & -0.00 &  0.17 \\
ATLAS\_1        & Cr II &   1 &     &$+3.44$&$-2.20$&  0.10 &$+0.28$&  0.10 & -0.01 &  0.05 & -0.02 &  0.01 &  0.00 \\
ATLAS\_1        & Mn I  &   1 & $<$ &$+3.53$&$-1.90$&\nodata&$+0.51$&\nodata&\nodata&\nodata&\nodata&\nodata&\nodata\\
ATLAS\_1        & Fe I  &  91 &     &$+5.10$&$-2.40$&  0.06 &$+0.00$&  0.00 &  0.05 & -0.00 &  0.02 &  0.01 &  0.24 \\
ATLAS\_1        & Fe II &  10 &     &$+5.02$&$-2.48$&  0.09 &$+0.00$&  0.00 &  0.00 &  0.05 & -0.00 &  0.02 &  0.07 \\
ATLAS\_1        & Co I  &   4 &     &$+3.00$&$-1.99$&  0.16 &$+0.41$&  0.16 &  0.06 &  0.01 & -0.02 & -0.01 &  0.15 \\
ATLAS\_1        & Ni I  &   8 &     &$+4.07$&$-2.15$&  0.12 &$+0.26$&  0.12 &  0.05 & -0.00 & -0.01 &  0.00 &  0.28 \\
ATLAS\_1        & Cu I  &   1 & $<$ &$+2.70$&$-1.49$&\nodata&$+0.92$&\nodata&\nodata&\nodata&\nodata&\nodata&\nodata\\
ATLAS\_1        & Zn I  &   1 & $<$ &$+2.80$&$-1.76$&\nodata&$+0.65$&\nodata&\nodata&\nodata&\nodata&\nodata&\nodata\\
ATLAS\_1        & Sr II &   2 &     &$+0.20$&$-2.67$&  0.26 &$-0.19$&  0.25 & -0.02 &  0.06 & -0.10 & -0.02 &  0.17 \\
ATLAS\_1        & Y II  &   2 &     &$-0.26$&$-2.47$&  0.12 &$+0.01$&  0.11 &  0.02 &  0.05 & -0.01 &  0.02 &  0.00 \\
ATLAS\_1        & Zr II &   1 &     &$+0.59$&$-1.99$&  0.22 &$+0.49$&  0.22 & -0.01 &  0.06 & -0.02 & -0.01 &  0.00 \\
ATLAS\_1        & Ba II &   5 &     &$-0.51$&$-2.69$&  0.14 &$-0.21$&  0.12 &  0.03 &  0.04 & -0.04 &  0.01 &  0.11 \\
ATLAS\_1        & La II &   1 & $<$ &$+0.21$&$-0.89$&\nodata&$+1.59$&\nodata&\nodata&\nodata&\nodata&\nodata&\nodata\\
ATLAS\_1        & Eu II &   2 & $<$ &$-1.12$&$-1.64$&\nodata&$+0.84$&\nodata&\nodata&\nodata&\nodata&\nodata&\nodata\\
\enddata
\tablecomments{One star from this table is shown for form. The full version is available online.}
\end{deluxetable*}

\begin{figure*}
    \centering
    \includegraphics[width=\linewidth]{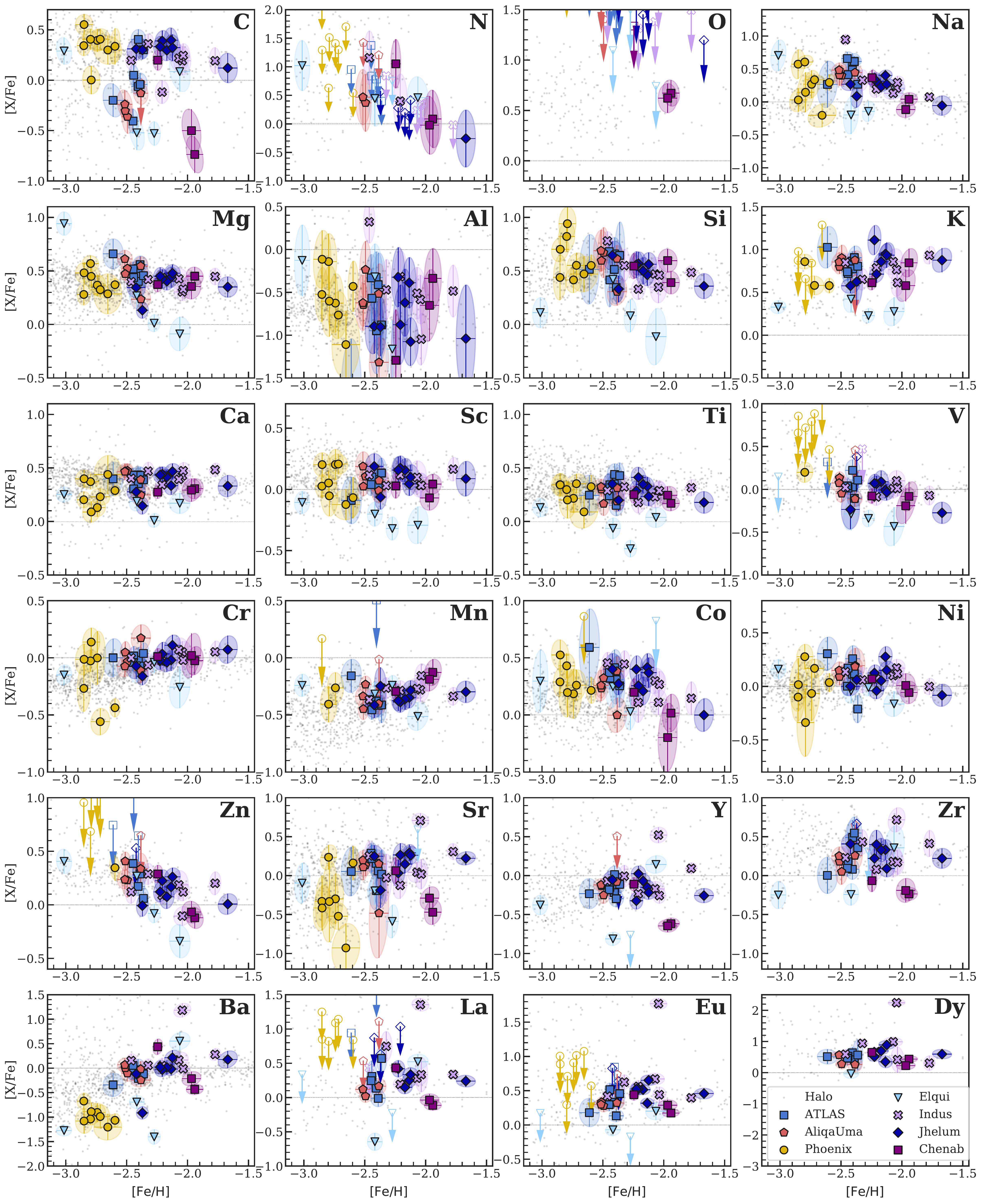}
    \caption{[X/Fe] vs [Fe~I/H] for most elements measured in this paper.
    Cu, Ce, Nd, Sm, and Gd have not been included here.
    Solid colored points indicate measurements, where the error ellipse represents the correlated [X/Fe] vs [Fe~I/H] errors after propagating stellar parameter uncertainties.
    Open symbols with downward pointing arrows indicate upper limits.
    Grey points in background are halo stars from JINAbase \citep{jinabase}.
    }
    \label{fig:xfegrid}
\end{figure*}

\begin{rotatetable*}
\begin{deluxetable*}{l|rr|rr|rr|rr|rr|rr|rr|rr|rr|rr|rr}
\centerwidetable
\tablecolumns{23}
\tabletypesize{\scriptsize}
\tablecaption{\label{tab:abundsummary}Abundance Summary}
\tablehead{Star & \multicolumn{2}{c}{[Fe I/H]} & \multicolumn{2}{c}{[Fe II/H]} & \multicolumn{2}{c}{[C/Fe]} & \multicolumn{2}{c}{[Na/Fe]} & \multicolumn{2}{c}{[Mg/Fe]} & \multicolumn{2}{c}{[Ca/Fe]} & \multicolumn{2}{c}{[Ti II/Fe]} & \multicolumn{2}{c}{[Ni/Fe]} & \multicolumn{2}{c}{[Sr/Fe]} & \multicolumn{2}{c}{[Ba/Fe]} & \multicolumn{2}{c}{[Eu/Fe]}}
\startdata
\hline
AliqaUma\_0    &$-2.38$& $0.08$&$-2.39$& $0.08$&$-0.03$&  lim  &\nodata&\nodata&$+0.24$& $0.10$&$+0.25$& $0.10$&$+0.22$& $0.13$&$+0.19$& $0.16$&$-0.48$& $0.58$&$-0.24$& $0.14$&$+0.74$&  lim  \\
AliqaUma\_10   &$-2.38$& $0.06$&$-2.31$& $0.08$&$-0.13$& $0.11$&$+0.45$& $0.10$&$+0.55$& $0.09$&$+0.39$& $0.08$&$+0.27$& $0.09$&$+0.18$& $0.08$&$+0.15$& $0.14$&$-0.02$& $0.16$&$+0.32$& $0.09$\\
AliqaUma\_5    &$-2.49$& $0.05$&$-2.33$& $0.11$&$-0.36$& $0.17$&$+0.40$& $0.16$&$+0.52$& $0.08$&$+0.47$& $0.07$&$+0.23$& $0.09$&$+0.12$& $0.07$&$+0.13$& $0.14$&$-0.10$& $0.17$&$+0.28$& $0.09$\\
AliqaUma\_7    &$-2.51$& $0.04$&$-2.40$& $0.06$&$-0.30$& $0.10$&$+0.41$& $0.16$&$+0.47$& $0.06$&$+0.47$& $0.06$&$+0.46$& $0.07$&$+0.09$& $0.08$&$+0.11$& $0.17$&$+0.00$& $0.12$&$+0.33$& $0.08$\\
AliqaUma\_9    &$-2.52$& $0.06$&$-2.47$& $0.07$&$-0.24$& $0.14$&$+0.48$& $0.11$&$+0.61$& $0.08$&$+0.47$& $0.08$&$+0.18$& $0.09$&$+0.15$& $0.08$&$+0.20$& $0.13$&$+0.07$& $0.14$&$+0.30$& $0.07$\\
\hline
ATLAS\_0       &$-2.41$& $0.06$&$-2.39$& $0.09$&$-0.06$& $0.10$&$+0.49$& $0.11$&$+0.52$& $0.08$&$+0.44$& $0.08$&$+0.28$& $0.08$&$+0.11$& $0.07$&$+0.16$& $0.18$&$-0.04$& $0.22$&$+0.33$& $0.09$\\
ATLAS\_1       &$-2.40$& $0.06$&$-2.48$& $0.09$&$+0.41$& $0.10$&$+0.61$& $0.12$&$+0.40$& $0.08$&$+0.35$& $0.09$&$+0.38$& $0.09$&$+0.26$& $0.12$&$-0.19$& $0.25$&$-0.21$& $0.12$&$+0.84$&  lim  \\
ATLAS\_12      &$-2.61$& $0.08$&$-2.32$& $0.19$&$-0.20$& $0.17$&$+0.27$& $0.35$&$+0.66$& $0.14$&$+0.40$& $0.16$&$+0.33$& $0.18$&$+0.30$& $0.16$&$+0.05$& $0.32$&$-0.34$& $0.24$&$+0.18$& $0.21$\\
ATLAS\_22      &$-2.39$& $0.06$&$-2.20$& $0.09$&$-0.04$& $0.11$&$+0.61$& $0.08$&$+0.56$& $0.08$&$+0.43$& $0.08$&$+0.20$& $0.09$&$+0.12$& $0.10$&$+0.06$& $0.11$&$-0.07$& $0.15$&$+0.13$& $0.09$\\
ATLAS\_25      &$-2.44$& $0.04$&$-2.43$& $0.09$&$+0.05$& $0.10$&$+0.41$& $0.15$&$+0.44$& $0.08$&$+0.30$& $0.07$&$+0.20$& $0.07$&$+0.17$& $0.06$&$+0.22$& $0.18$&$-0.05$& $0.14$&$+0.52$& $0.10$\\
ATLAS\_26      &$-2.45$& $0.05$&$-2.29$& $0.07$&$-0.41$& $0.12$&$+0.66$& $0.10$&$+0.52$& $0.07$&$+0.43$& $0.07$&$+0.37$& $0.06$&$+0.00$& $0.09$&$+0.29$& $0.13$&$+0.06$& $0.14$&$+0.30$& $0.09$\\
ATLAS\_27      &$-2.36$& $0.04$&$-2.47$& $0.12$&$+0.30$& $0.09$&$+0.27$& $0.22$&$+0.46$& $0.11$&$+0.39$& $0.08$&$+0.24$& $0.10$&$-0.21$& $0.13$&$+0.02$& $0.31$&$-0.12$& $0.16$&$+0.46$& $0.20$\\
\hline
Chenab\_10     &$-1.94$& $0.07$&$-1.98$& $0.08$&$-0.74$& $0.18$&$+0.04$& $0.10$&$+0.45$& $0.09$&$+0.31$& $0.09$&$+0.14$& $0.10$&$-0.06$& $0.10$&$-0.47$& $0.16$&$-0.43$& $0.11$&$+0.17$& $0.06$\\
Chenab\_12     &$-1.97$& $0.08$&$-1.87$& $0.10$&$-0.50$& $0.21$&$-0.12$& $0.12$&$+0.35$& $0.12$&$+0.29$& $0.11$&$+0.35$& $0.14$&$+0.01$& $0.11$&$-0.29$& $0.14$&$-0.21$& $0.14$&$+0.29$& $0.08$\\
Chenab\_16     &$-2.25$& $0.04$&$-2.17$& $0.08$&$+0.20$& $0.10$&$+0.37$& $0.12$&$+0.37$& $0.07$&$+0.27$& $0.06$&$+0.21$& $0.07$&$+0.07$& $0.06$&$+0.06$& $0.17$&$+0.44$& $0.16$&$+0.44$& $0.08$\\
\hline
Elqui\_0       &$-2.42$& $0.06$&$-2.08$& $0.09$&$-0.52$& $0.17$&$-0.20$& $0.29$&$+0.27$& $0.12$&$+0.19$& $0.09$&$-0.14$& $0.13$&$-0.04$& $0.12$&$-0.20$& $0.17$&$-0.69$& $0.14$&$-0.07$& $0.09$\\
Elqui\_1       &$-3.01$& $0.06$&$-2.91$& $0.09$&$+0.29$& $0.13$&$+0.70$& $0.24$&$+0.94$& $0.11$&$+0.25$& $0.07$&$+0.06$& $0.10$&$+0.16$& $0.09$&$-0.10$& $0.26$&$-1.27$& $0.11$&$+0.17$&  lim  \\
Elqui\_3       &$-2.06$& $0.09$&$-1.81$& $0.09$&$+0.09$& $0.20$&$+0.20$& $0.15$&$-0.09$& $0.16$&$+0.17$& $0.10$&$-0.08$& $0.15$&$-0.16$& $0.12$&$+0.59$&  lim  &$+0.55$& $0.18$&$+0.20$& $0.13$\\
Elqui\_4       &$-2.27$& $0.05$&$-2.05$& $0.08$&$-0.53$& $0.12$&$-0.14$& $0.15$&$+0.01$& $0.08$&$+0.01$& $0.07$&$-0.11$& $0.10$&$-0.01$& $0.08$&$-0.59$& $0.21$&$-1.40$& $0.15$&$-0.17$&  lim  \\
\hline
Indus\_0       &$-2.46$& $0.05$&$-2.44$& $0.09$&$+0.20$& $0.09$&$+0.94$& $0.08$&$+0.40$& $0.07$&$+0.40$& $0.07$&$+0.24$& $0.08$&$+0.14$& $0.08$&$+0.25$& $0.19$&$+0.15$& $0.14$&$+0.42$& $0.09$\\
Indus\_12      &$-2.21$& $0.06$&$-2.13$& $0.10$&$-0.12$& $0.12$&$+0.20$& $0.13$&$+0.44$& $0.07$&$+0.34$& $0.07$&$+0.23$& $0.08$&$+0.07$& $0.08$&$-0.13$& $0.17$&$+0.05$& $0.16$&$+0.56$& $0.09$\\
Indus\_13      &$-2.04$& $0.07$&$-1.92$& $0.09$&$+0.19$& $0.14$&$+0.21$& $0.09$&$+0.30$& $0.08$&$+0.43$& $0.08$&$+0.33$& $0.09$&$+0.17$& $0.08$&$+0.71$& $0.09$&$+1.18$& $0.14$&$+1.77$& $0.08$\\
Indus\_14      &$-2.07$& $0.08$&$-1.98$& $0.10$&$+0.22$& $0.14$&$+0.13$& $0.09$&$+0.43$& $0.10$&$+0.39$& $0.10$&$+0.22$& $0.10$&$+0.10$& $0.10$&$+0.04$& $0.13$&$+0.17$& $0.16$&$+0.67$& $0.09$\\
Indus\_15      &$-1.77$& $0.05$&$-1.69$& $0.07$&$+0.19$& $0.13$&$+0.07$& $0.05$&$+0.45$& $0.08$&$+0.48$& $0.06$&$+0.29$& $0.09$&$-0.00$& $0.08$&$+0.31$& $0.10$&$+0.28$& $0.12$&$+0.40$& $0.06$\\
Indus\_6       &$-2.32$& $0.05$&$-2.35$& $0.10$&$+0.36$& $0.10$&$+0.33$& $0.16$&$+0.42$& $0.09$&$+0.47$& $0.08$&$+0.33$& $0.10$&$+0.06$& $0.12$&$-0.03$& $0.31$&$+0.06$& $0.13$&$+0.62$& $0.11$\\
Indus\_8       &$-2.04$& $0.05$&$-2.02$& $0.10$&$+0.24$& $0.12$&$+0.29$& $0.10$&$+0.33$& $0.08$&$+0.47$& $0.08$&$+0.24$& $0.10$&$+0.12$& $0.09$&$+0.02$& $0.19$&$-0.01$& $0.13$&$+0.44$& $0.11$\\
\hline
Jhelum\_0      &$-2.13$& $0.06$&$-2.04$& $0.08$&$+0.40$& $0.12$&$+0.40$& $0.08$&$+0.44$& $0.10$&$+0.33$& $0.08$&$+0.28$& $0.10$&$+0.28$& $0.10$&$+0.29$& $0.11$&$-0.02$& $0.13$&$+0.32$& $0.08$\\
Jhelum\_1\_5   &$-2.17$& $0.06$&$-2.13$& $0.07$&$+0.30$& $0.10$&$+0.23$& $0.08$&$+0.40$& $0.08$&$+0.42$& $0.07$&$+0.34$& $0.09$&$+0.05$& $0.10$&$+0.15$& $0.12$&$+0.02$& $0.14$&$+0.51$& $0.07$\\
Jhelum\_1\_8   &$-2.42$& $0.07$&$-2.42$& $0.11$&$+0.31$& $0.12$&$+0.27$& $0.10$&$+0.34$& $0.09$&$+0.28$& $0.08$&$+0.37$& $0.10$&$+0.00$& $0.09$&$+0.25$& $0.14$&$-0.12$& $0.13$&$+0.82$&  lim  \\
Jhelum\_2\_10  &$-2.12$& $0.07$&$-2.03$& $0.08$&$+0.32$& $0.13$&$+0.27$& $0.08$&$+0.48$& $0.08$&$+0.47$& $0.07$&$+0.25$& $0.10$&$+0.15$& $0.09$&$+0.26$& $0.11$&$+0.21$& $0.15$&$+0.65$& $0.08$\\
Jhelum\_2\_11  &$-2.21$& $0.06$&$-2.19$& $0.10$&$+0.39$& $0.10$&$+0.31$& $0.11$&$+0.43$& $0.09$&$+0.44$& $0.07$&$+0.50$& $0.09$&$-0.04$& $0.08$&$+0.27$& $0.15$&$-0.03$& $0.14$&$+0.53$& $0.09$\\
Jhelum\_2\_14  &$-2.37$& $0.05$&$-2.39$& $0.09$&$+0.30$& $0.09$&$+0.09$& $0.17$&$+0.13$& $0.08$&$+0.15$& $0.08$&$+0.17$& $0.12$&$+0.06$& $0.10$&$-0.19$& $0.23$&$-0.91$& $0.13$&$+0.64$&  lim  \\
Jhelum\_2\_15  &$-2.23$& $0.05$&$-2.13$& $0.09$&$+0.33$& $0.10$&$+0.35$& $0.10$&$+0.45$& $0.08$&$+0.43$& $0.07$&$+0.25$& $0.09$&$+0.12$& $0.08$&$+0.03$& $0.17$&$+0.02$& $0.14$&$+0.52$& $0.09$\\
Jhelum\_2\_2   &$-1.67$& $0.08$&$-1.62$& $0.10$&$+0.12$& $0.15$&$-0.06$& $0.15$&$+0.35$& $0.10$&$+0.33$& $0.10$&$+0.17$& $0.10$&$-0.08$& $0.10$&$+0.22$& $0.09$&$+0.18$& $0.18$&$+0.46$& $0.08$\\
\hline
Phoenix\_1     &$-2.60$& $0.04$&$-2.52$& $0.16$&$+0.34$& $0.18$&$+0.29$& $0.12$&$+0.37$& $0.06$&$+0.29$& $0.05$&$+0.27$& $0.07$&$+0.04$& $0.09$&$+0.16$& $0.22$&$-1.06$& $0.14$&$+0.57$&  lim  \\
Phoenix\_10    &$-2.85$& $0.06$&$-2.93$& $0.09$&$+0.34$& $0.10$&$+0.57$& $0.22$&$+0.28$& $0.10$&$+0.20$& $0.08$&$+0.38$& $0.10$&$-0.10$& $0.23$&$-0.33$& $0.34$&$-0.67$& $0.17$&$+1.01$&  lim  \\
Phoenix\_2     &$-2.65$& $0.12$&$-2.62$& $0.10$&$+0.30$& $0.14$&$-0.20$& $0.18$&$+0.29$& $0.19$&$+0.44$& $0.18$&\nodata&\nodata&$+1.59$&  lim  &$-0.93$& $0.31$&$-1.20$& $0.27$&$+1.07$&  lim  \\
Phoenix\_3     &$-2.74$& $0.07$&$-2.70$& $0.12$&$+0.40$& $0.11$&$+0.27$& $0.21$&$+0.37$& $0.10$&$+0.13$& $0.09$&$+0.13$& $0.13$&$-0.07$& $0.26$&$-0.30$& $0.32$&$-0.91$& $0.19$&$+0.91$&  lim  \\
Phoenix\_6     &$-2.79$& $0.07$&$-2.68$& $0.15$&$+0.00$& $0.14$&$+0.15$& $0.30$&$+0.45$& $0.10$&$+0.09$& $0.10$&$+0.24$& $0.16$&$-0.34$& $0.32$&$-0.33$& $0.52$&$-0.89$& $0.19$&$+0.71$&  lim  \\
Phoenix\_7     &$-2.80$& $0.07$&$-2.62$& $0.08$&$+0.41$& $0.10$&$+0.61$& $0.10$&$+0.57$& $0.07$&$+0.37$& $0.07$&$+0.44$& $0.08$&$+0.28$& $0.12$&$+0.24$& $0.16$&$-1.04$& $0.12$&$+0.30$&  lim  \\
Phoenix\_8     &$-2.72$& $0.08$&$-2.79$& $0.09$&$+0.41$& $0.10$&$+0.34$& $0.07$&$+0.33$& $0.07$&$+0.23$& $0.08$&$+0.31$& $0.07$&$+0.17$& $0.09$&$-0.52$& $0.17$&$-0.99$& $0.24$&$+1.02$&  lim  \\
Phoenix\_9     &$-2.85$& $0.07$&$-2.71$& $0.08$&$+0.55$& $0.10$&$+0.03$& $0.18$&$+0.48$& $0.10$&$+0.40$& $0.08$&$+0.32$& $0.12$&$+0.02$& $0.25$&$-0.42$& $0.29$&$-1.08$& $0.25$&$+0.89$&  lim  \\
\enddata
\end{deluxetable*}
\end{rotatetable*}

When possible, N is measured by synthesizing the CN bands at 3865-3885\Ang. This is done after measuring C from the CH bands.
These bands are often detected in the cooler stars ($\Teff < 4800$\,K).
Where not detected, an upper limit is synthesized, reported in Table~\ref{tab:lines}. However as mentioned in Section~\ref{sec:analysis}, upper limits for molecular features are likely underestimated because they do not include continuum placement uncertainty.
CN has some dependence on the C abundance, and due to this and the overall low S/N in the CN band region, we have applied a minimum 0.3 dex floor to the CN abundance uncertainty.

Two cool and relatively metal-rich stars in Chenab have O measured from equivalent widths of the forbidden lines at 6300\Ang and 6363\Ang.
The two line abundances agree, but they are near telluric regions and affected by several systematic blends \citep{Asplund04} so should be regarded with caution.
For the other stars, O upper limits are found using the 6300\Ang line.

\subsection{Magnesium, Silicon, Calcium, Titanium}
Mg is measured with equivalent widths of up to 9 lines, with four lines detected in all stars (4702\Ang, 5172\Ang, 5183\Ang, and 5528\Ang).
The Mg b lines are often saturated and require fitting Voigt profiles to get an accurate equivalent width. After using Voigt profiles, their abundances agree with the other lines.
The 4702\Ang line tends to have the largest weight and thus the most influence on the final abundance.
Note there is a moderate anticorrelation between [Mg/Fe] and $\logg$.

Si is the least reliable $\alpha$-element measured.
Across our sample, the 3905\Ang and 4102\Ang Si lines are always detected. Neither line provides a very reliable abundance, since the 3905\Ang line is both saturated and blended while the 4102\Ang line is in the wing of a Balmer line.
However the resulting Si abundances tend to be reasonably close, though the 3905 is biased lower.
In stars with $\mbox{[Fe/H]} \gtrsim -1.9$, Si can be detected with lines from 5690--6000\Ang.
The 3905\Ang line is synthesized due to a carbon blend, with equivalent widths used for the others.

Ca I is measured using equivalent widths of 25 lines. 
We specifically updated the Ca $\log gf$ values in \code{linemake} using VALD, because the original $\log gf$ values in \code{linemake} resulted in large Ca abundance scatter in standard stars.
The number of measured Ca lines per star varies from 4 to 23, but restricting to the most common Ca lines (used in at least 30 of our 42 stars) makes a negligible $-0.02 \pm 0.03$ dex difference. We consider Ca to be the most reliably measured $\alpha$-element.

Ti is usually considered as an $\alpha$-element, although nucleosynthetically it may be closer to Fe-peak elements like Sc and V \citep{Cowan20}.
Both Ti I and Ti II lines are measured using equivalent widths. The Ti II abundances are $0.09 \pm 0.13$ dex higher than the Ti I abundances. In metal-poor giants, Ti II abundances are more trustworthy than Ti I. There are more and stronger lines, and Ti I may be significantly affected by NLTE.
52 unique Ti I lines were measured, of which only six are present in more than 30 stars of our sample. If we were to derive Ti I abundances using only these six lines, the abundances would change by $-0.03 \pm 0.11$ dex, where the 0.11 dex scatter suggests that line selection can significantly affect a star's Ti I abundance (though not on average).
For Ti II, 17 out of 65 lines are measured in more than 30 stars of our sample. Using just these lines results in abundances that change only by $0.04 \pm 0.06$ dex, further indicating that the Ti II abundances are more reliable.

\subsection{Sodium, Aluminum, Potassium, Scandium}
For cool and metal-poor giants, Na is almost always measured only from equivalent widths of two Na D resonance lines.
The exception is the star AliqaUma\_0, which has strong sky line residuals preventing a measurement or useful upper limit.
The Na D lines often have slight negative NLTE corrections of $-0.1$ to $-0.4$ dex for cool and metal-poor stars that have not been applied here \citep{Lind11}.
The weaker Na lines at 5682\Ang and 5688\Ang are also detected in the cooler and more metal-rich stars ($\mbox{[Fe/H]} > -2$), where they agree with the Na D lines within 0.1 dex.

The only detectable Al lines in our spectra are the 3944\Ang and 3961\Ang lines, which are measured using spectral synthesis. It is unfortunately difficult to derive a reliable abundance from either line. Both lines are in the blue where the S/N is lower, near strong hydrogen lines that affect continuum placement, and heavily affected by NLTE corrections of ${\sim}+0.7$ dex \citep{Nordlander17}. Furthermore, the 3944\Ang line is heavily blended.
We have added an extra 0.3 dex minimum systematic uncertainty to each Al line to account for the significant continuum modeling issues.
Still, we encourage strong caution in using any of our Al abundances, as the abundance uncertainties are large and may still be underestimated.

K is measured from equivalent widths of the resonant K lines at 7665\Ang and 7699\Ang. The 7665\Ang line is often blended with telluric absorption, in which case that line is not used. In one star (ATLAS\_22), the 7665\Ang line is clean but the 7699\Ang has a clear telluric blend.
When the 7699\Ang line is not detected, an upper limit is synthesized.
There are moderate negative NLTE corrections for K that range from $-0.0$ to $-0.4$ dex \citep{Reggiani19}.

Five bluer Sc II lines from 4246\Ang to 4415\Ang are detected in essentially all our stars, while three redder lines are detected in most stars.
These lines have hyperfine structure, and the bluest lines are often quite blended with carbon, so all Sc lines are synthesized.
An extra 0.1 dex minimum systematic uncertainty per Sc line is added, because the hyperfine structure causes these line abundances to be sensitive to the synthesis smoothing kernel.
The two bluest lines tend to have much lower weight than the other lines.

\subsection{Vanadium, Chromium, Manganese}
Two V I lines and two V II lines are measured using spectral syntheses, due to hyperfine structure and strong or minor blends for all the V lines under consideration.
The V I 4379\Ang line is the best line, though it has a minor blend with $^{12}$CH.
The V I 4384\Ang line is often detected but heavily blended with a Sc and Fe line.
The V II 4005\Ang line is adjacent to and slightly blended with some strong Fe lines.
The V II 3952\Ang line is not usually measured because the S/N is lower and it is hard to determine the continuum in this region, but we report it when possible.
Note that our error estimation does not propagate abundance uncertainties in the blending features, so the errors are likely underestimated for V.
When both V I and V II are measured in a star, the V II abundances are $0.30 \pm 0.23$ dex higher than the V I abundances. This is larger than the individual V I or V II error, but it is similar to the [V II/V I] ratio found in \citet{Roederer14c}.
Because the V I lines are stronger in our stars, we use this is as the default V abundance in this paper's figures.

Equivalent widths of 17 Cr I lines in 41 stars and 6 Cr II lines in 33 stars are measured.
The Cr II abundances are larger than Cr I by $0.18 \pm 0.24$ dex.
Cr I is affected by NLTE \citep{Bergemann10}, so the Cr II abundances should have fewer systematic errors although the lines are noisier and detected less often.
However, because Cr II is not detected in all of our stars, we default to the Cr I abundance in this paper's figures.

Up to 6 different Mn I lines are synthesized, at least one of which is detected in 35 of our stars. The resonant Mn triplet at 4030\Ang is seen in all our stars, but we never use these lines. Mn is significantly affected by NLTE \citep{Bergemann19}, and it is likely the Mn triplet has a significantly different LTE-to-NLTE zero-point than the other lines. Even ignoring the Mn triplet, it is likely that the Mn abundances have a $+0.4$ to $+0.6$ dex correction.
Like Sc, an extra 0.1 dex minimum systematic uncertainty is added per Mn line because the hyperfine structure causes these lines to be sensitive to the synthesis smoothing kernel.

\subsection{Iron} \label{sec:fe}
Equivalent widths of plenty of Fe I and Fe II lines are measured in all our stars, considering 175 Fe I lines and 30 Fe II lines.
Typically 100 Fe I lines are measured in each star, although as few as 29 and as many as 130.
The median number of Fe II lines is 18, with at least 8 Fe II lines measured in all stars.
The Fe II lines have been used to determine the microturbulence and model atmosphere metallicities of our stars.

We did not explicitly balance ionization states, and the Fe I and Fe II abundances thus usually differ by $0.08$ dex with $0.11$ dex scatter.
Four stars have [Fe I/H] over 0.2 dex lower than [Fe II/H]: ATLAS\_12 (0.29 dex), Elqui\_0 (0.34 dex), Elqui\_3 (0.25 dex), and Elqui\_4 (0.22 dex).
Such a difference is expected for the Elqui stars, as they are the coolest stars and thus significant NLTE corrections apply \citep{Bergemann12,Mashonkina16,Ezzeddine17}.
The ATLAS\_12 star had only 7 Fe II lines, resulting in an unusually large microturbulence error that lowers the Fe I abundances but also substantially increases the [Fe I/H] error bar.

\subsection{Cobalt, Nickel, Copper, Zinc}

Four lines of Co at 4020\Ang, 4110\Ang, 4118\Ang, and 4121\Ang are considered.
These are synthesized to account for hyperfine structure.
The Co lines often disagree substantially with each other, suggesting a possibly unaccounted for systematic in their abundances or in the line lists. The source of this discrepancy is not clear, but the quoted abundance errors reflect this disagreement by adding per-line systematic uncertainties to match the line-to-line scatter (Section~\ref{sec:summarizelines}, Appendix~\ref{app:estimator}).

The Ni I abundance is measured from equivalent widths.
Up to 24 lines are measured in any individual star, though only 2--4 lines are detected in most stars.
The strongest 5476\Ang line is always detected or used to set an upper limit, with the next strongest lines at 4714\Ang, 6643\Ang, and 6767\Ang.

One Cu I line at 5105\Ang is detected in three of our most Fe-rich stars and measured using equivalent widths.
A Cu upper limit is synthesized for the other stars.

Two Zn I lines are measured at 4810\Ang and 4722\Ang using equivalent widths. When both are present they agree well, and sometimes only the 4810\Ang line is present. We synthesize an upper limit with the 4810\Ang line when neither is detected.

\subsection{Strontium, Yttrium, Zirconium}

The Sr II lines at 4077\Ang and 4215\Ang are detected in all but one of our stars. The exception is Elqui\_3, a cool and metal-rich star with enough molecular absorption that these Sr lines cannot be measured reliably.
However in this star and two other stars, the Sr I line at 4607\Ang is detected. When both are measured, the Sr I line has a lower abundance by $0.15-0.30$ dex than Sr II.
The Sr II lines are measured using spectral synthesis, while the Sr I line is from an equivalent width.
The Sr II lines are generally saturated, so they are strongly affected by microturbulence.
As a result, Y and Zr are better tracers of a similar nucleosynthetic process when they are detected, although Sr provides good dynamic range.

A synthesis measurement or upper limit for Y II is found by examining three Y II lines in all our stars (4398\Ang, 4883\Ang, and 4900\Ang). If these are clearly detected, up to five other Y lines are measured.
No Y lines are detected in the metal-poor Phoenix stream, and we do not place upper limits as the Sr abundance is too low to expect a useful Y measurement or limit.

Only a single Zr II line at 4208\Ang is measured, either synthesizing or placing an upper limit. Similar to Y, Zr is not considered in the Phoenix stream as the limit is not meaningful.

\subsection{Barium, Lanthanum}
Ba II has five strong lines. The 4554\Ang line is detected in every one of our stars, and the 4934\Ang is detected in all but a few Phoenix stars. The other three redder lines are weaker but generally detected in all but the Phoenix stars.
The presence of hyperfine structure and isotopic splitting means that all Ba lines must be synthesized.

The isotope ratio (or specifically the even-to-odd isotope ratio $f_{\rm odd}$) can significantly impact the abundance derived from the two strongest Ba lines.
In general, the detailed results require full 3D and NLTE modeling, as well as much higher S/N and resolution than achieved here \citep{Gallagher15}.
Thus for simplicity, $r$-process isotope ratios were assumed for all our stars \citep{Sneden08}.
If the Solar Ba isotope ratios were used instead, the Ba abundance from the 4554\Ang line would increase by up to 0.25 dex \citep{Mashonkina19}.
Note that when the weaker Ba lines are detected, the abundance difference using just those lines is only $0.06 \pm 0.09$ dex higher compared to using all five lines.
To account for the possible effect of isotope ratios, we have decided to add an extra uncertainty of 0.20 dex in quadrature to the error of the two strongest Ba lines.
Because the abundance is somewhat dependent on the smoothing kernel, we have added an additional 0.1 dex systematic uncertainty to all Ba lines.

The production of La II is highly correlated with Ba, and when detected it is better than Ba because it is less saturated and not affected by isotopic ratios \citep{Simmerer04}.
La has hyperfine splitting so is measured with spectral synthesis.
La is detected in about half our stars, and up to six La lines are considered, with the strongest one at 4086\Ang. Since Ba is detected in all of our stars, a La limit is placed using the 4086\Ang line in all of the stars, though it is often a very weak limit.

\subsection{Europium, Dysprosium}

Eu and Dy are elements that primarily trace the $r$-process. In the solar system, over 98\% of Eu and 88\% of Dy comes from the $r$-process \citep[e.g.,][]{Sneden08}.

Up to five lines of Eu II are synthesized at 4129\Ang, 4205\Ang, 4435\Ang, 4523\Ang, and 6645\Ang. Usually, only the two bluest lines are detected and sufficiently strong to be used. Hyperfine structure and isotope splitting are included assuming the \citet{Sneden08} isotope ratios.

Dy II is one of the most abundant $r$-process elements \citep[e.g.,][]{Sneden08} and two particularly strong lines are considered, one near the Sr 4077\Ang line and one in the red wing of the 4102\Ang Balmer line.
Both of these lines are synthesized. We do not put upper limits on the Dy abundance, since when it is not detected the Eu abundance is a more useful constraint on the $r$-process abundance of a star.

\subsection{Other neutron-capture elements}

Indus\_13 is an $r$-process enhanced star, and the continuum is substantially affected by the $r$-process elements. Ce, Nd, Sm, and Gd were thus also measured for this star. Many of these elements make a substantial impact to the overall continuum, which is the main reason these elements were measured.
\citet{Hansenprep} will present a more detailed analysis of this star.

Note that when considering all stars in all our streams, many neutron-capture elements (Y, Zr, La, Eu, Dy) appear to have significant trends with the stellar parameters (see Appendix~\ref{app:params_errors}).
This is not a systematic effect, but rather reflects the fact that each stream has intrinsically different neutron-capture element abundances, and due to their differing distances span a different range of stellar parameters.
It just so happens that in this sample, stars in the furthest streams (i.e., coolest, lowest gravity, highest microturbulence stars) have lower overall neutron-capture element abundances than stars in closer streams.

\section{Discussion}\label{sec:discussion}

We first consider the metallicity distributions of the seven streams from high-resolution spectroscopy, providing some evidence for separating them into three thin globular cluster streams (ATLAS, Aliqa Uma, and Phoenix) and four thick dwarf galaxy streams (Chenab, Elqui, Indus, and Jhelum).
We then briefly discuss each stream's abundances individually in the context of literature abundances of globular clusters and dwarf spheroidal galaxies.

\subsection{Stream progenitors from metallicity spread}

\citet{Shipp18} classified the progenitors of the seven streams considered here as either globular clusters or dwarf galaxies.
The classification was based on a mass-to-light ratio estimate, where the dynamical mass was inferred from the stream width and the luminous mass was inferred using isochrone models of the observed color-magnitude diagrams.
These classifications can be refined by examining the metallicity dispersions. Globular clusters display spreads of Fe peak elements at a level of ${\sim}0.03$ dex \citep[e.g.,][]{Gratton04,Yong13b}, which will be undetectable at our precision. Dwarf galaxies display significant [Fe/H] spreads in excess of 0.2 dex \citep[e.g.,][]{Tolstoy09,Leaman12,Willman12,Simon19}.

Here, we investigate the mean metallicity $\meanfe$ and metallicity dispersion $\scatfe$ of these streams using the metallicities from high-resolution spectroscopy.
Compared to the metallicities from the AAT medium-resolution spectroscopy \citep{Li19S5}, the high-resolution abundances are moderately more precise and likely more accurate.
However, the sample sizes are smaller, with 3--8 stars per stream.
For the thick streams (Chenab, Elqui, Indus, Jhelum), our target selection could have missed metal-rich member stars that are harder to separate from the Milky Way foreground (see Section~\ref{sec:observations}, Section~\ref{sec:discdsph}).
A detailed consideration of these effects will be discussed in subsequent work \citep{Paceprep}.

The metallicity distribution of each stream was modeled as having an unknown mean abundance $\meanfe$ and intrinsic scatter $\scatfe$.
The Fe~II abundance is used for [Fe/H], which is slightly less precise than Fe~I due to having fewer lines but negligibly affected by sytematic NLTE effects \citep[e.g.,][]{Ezzeddine17}.
Each star's observed metallicity was assumed to be drawn from this Gaussian distribution, plus Gaussian observational noise from $\sigma_{\text{[X/H]}}$ from Table~\ref{tab:abunds}.
We used an improper uniform prior for $\meanfe$
and a uniform prior on $\log{\scatfe} \sim \mathcal{U}\left(-3, 0\right)$.
The Hamiltonian Monte Carlo sampler implemented in \code{Stan} \citep{stan} was used to draw posterior samples for the mean and scatter for each stream.

\begin{figure}
    \centering
    \includegraphics[width=\linewidth]{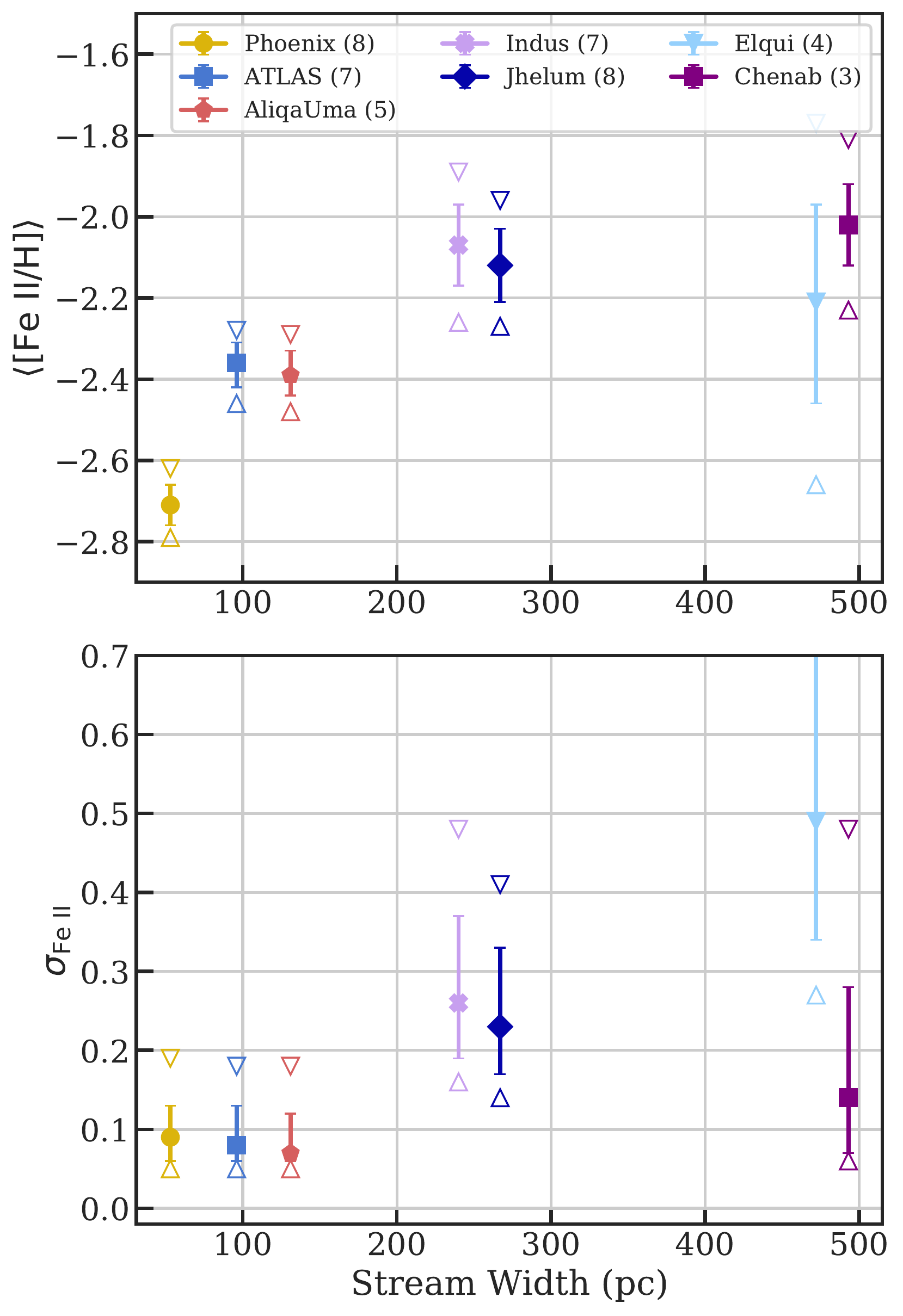}
    \caption{Mean metallicity (top panel) and metallicity dispersion (bottom panel) for high-resolution abundances in seven streams, plotted against the stream width \citep{Shipp18}. The solid colored point indicates the median of the posterior samples, the error bars indicate the middle 68\% scatter, and the open triangles indicate the 5th and 95th percentiles.
    The legend shows how many stars were analyzed in this paper for each stream.
    The three thinner streams have unresolved metallicity spreads, confirming their progenitors to be globular clusters.
    }
    \label{fig:mdfs}
\end{figure}

The results are shown in Figure~\ref{fig:mdfs}. The y-axis shows percentiles of the posterior distributions for $\meanfe$ and $\scatfe$. The 5th/95th, 16th/84th, and 50th percentiles are shown as open triangles, error bars, and a solid point, respectively.
The x-axis plots the physical stream width derived in \citet{Shipp18}. The legend shows how many stars were observed with MIKE in each stream.
The top panel shows the mean metallicities for the streams, which are all between $-3 < \mbox{[Fe/H]} < -2$. The Phoenix stream's progenitor would have been the lowest metallicity globular cluster known \citep{Wan20}.

The bottom panel of Figure~\ref{fig:mdfs} shows that the three thin streams have unresolved metallicity dispersions, with a 95\% upper limit of about 0.2 dex.
In contrast, the thicker streams mostly have clearly resolved metallicity dispersion. The exception is Chenab, which has only three stars, but it is still clearly a dwarf galaxy stream due to its connection with the Orphan stream (Section~\ref{sec:discdsph}).
Note that Aliqa Uma was tentatively classified as a possible dwarf galaxy stream based on its mass-to-light ratio \citep{Shipp18}, but it is clearly a globular cluster stream and in fact an extension of ATLAS \citep{Liprep}.
The metallicity dispersions here confirm that thin streams tend to be globular clusters, while thick streams tend to be dwarf galaxies.

Note that the exact value of the metallicity dispersion upper limit in our three globular cluster streams has some dependence on the prior, particularly the lower limit on $\log\scatfe$.
Increasing the prior's lower limit to $10^{-2}$ dex would cause the 95\% upper limits for ATLAS, Aliqa Uma, and Phoenix to increase by about 0.1 dex.
Decreasing the lower limit to $10^{-4}$ dex would decrease the upper limits by about 0.05 dex.
The smallest detected metallicity dispersions in star clusters are about 0.02 dex \citep{Yong13b,Krumholz19}, so the minimum prior value must be less than 0.02. We have thus chosen a minimum of 0.001 to allow the result to reach a near-zero dispersion without artificially concentrating the prior near zero dispersion.

\subsection{Globular Cluster Streams} \label{sec:discgc}

Three streams (ATLAS, Aliqa Uma, Phoenix) have thin morphologies and small velocity and metallicity dispersions that suggest they are disrupted globular clusters \citep[GCs, ][]{Shipp18,Li19S5}.
GCs show light element variations (C through Si) that vary in specific patterns due to the CNO, Ne-Na, and Mg-Al proton capture cycles.
In general, the abundances of \isoel{13}{C}, \isoel{14}{N}, \isoel{23}{Na}, \isoel{27}{Al}, and \isoel{28}{Si} increase, while the abundances of \isoel{12}{C}, \isoel{16}{O}, \isoel{20}{Ne}, and \isoel{24}{Mg} decrease \citep[e.g.,][]{Gratton12,Gratton19}.
In NGC 2419 and NGC 2808, some unknown process also induces an Mg-K anticorrelation (\citealt{Cohen12}, see discussion in \citealt{Kemp18}).

\begin{figure*}
    \centering
    \includegraphics[width=\linewidth]{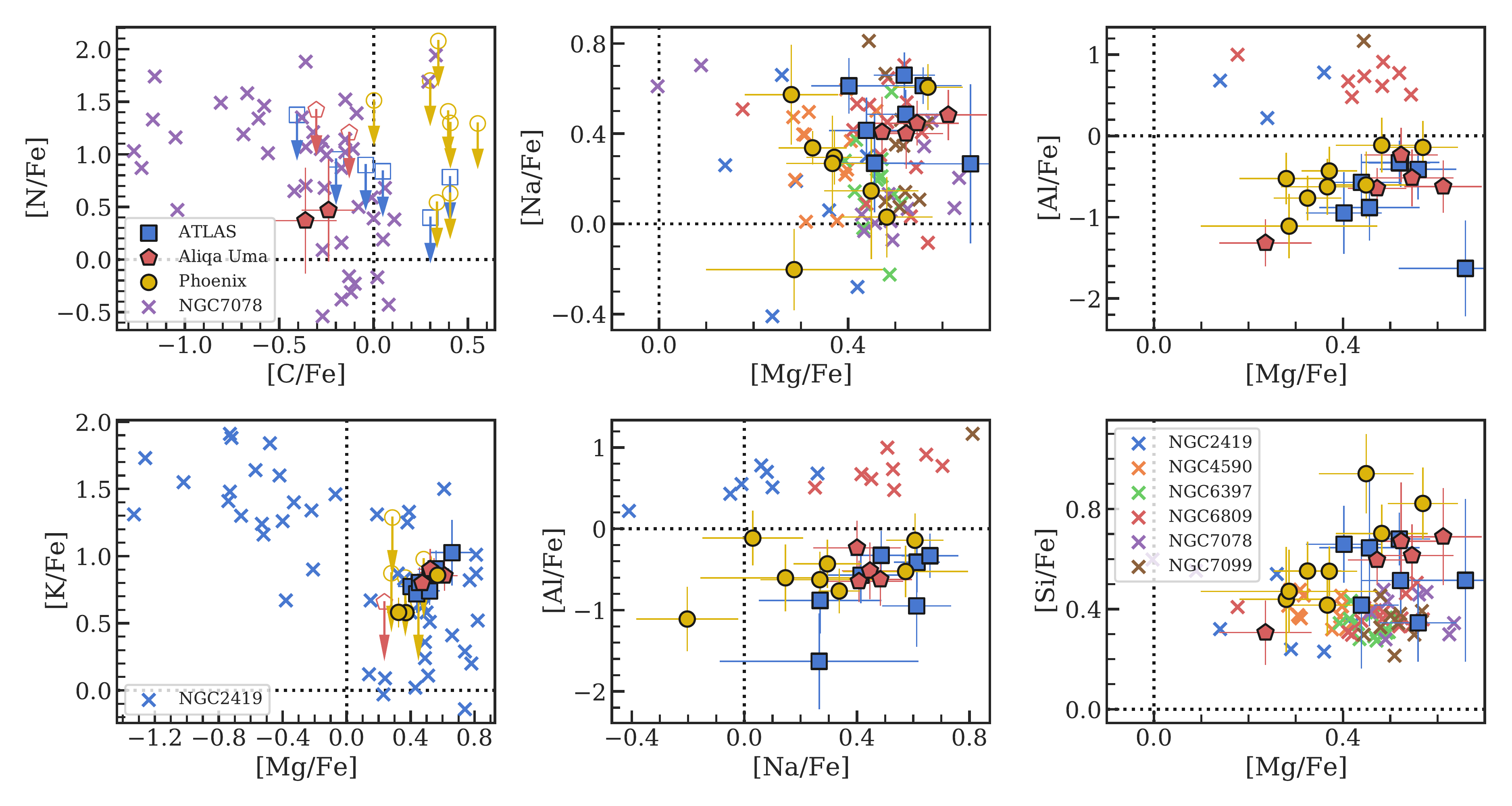}
    \caption{Globular cluster element trends compared to our measurements for globular cluster streams ATLAS, Aliqa Uma, and Phoenix. Cross symbols indicate literature abundances for globular clusters with $\mbox{[Fe/H]} < -1.9$
    (\citealt{Roediger14} for top-left C-N panel; \citealt{Mucciarelli12} for bottom-left Mg-K panel;  \citealt{Carretta09b} for the rest), while open symbols with a downward pointing arrow indicate upper limits in our streams (as in Figure~\ref{fig:xfegrid}).
    Note the large zero-point offset in [Al/Fe] is due to NLTE effects (see Section~\ref{sec:elements}).
    \label{fig:gcabunds}}
\end{figure*}

Figure~\ref{fig:gcabunds} shows the relevant measurable elements for our globular cluster streams.
Of these elements, C, Na, and Mg are the most reliably measured elements in our GC streams.
In a few stars, N can be measured from the CN bands; the rest have upper limits that should be treated with caution (Section~\ref{sec:limits}).
Si and K are only measured from $1-2$ lines, but these should be reliable.
However, Al is measured from the 3944\Ang and 3961\Ang lines, with a large NLTE correction that should be considered in any interpretation.
For comparison, GC abundances from \citet{Carretta09b} are plotted as open circles; C and N abundances for NGC 7078 from \citet{Roediger14} as open squares; and K and Mg abundances in NGC 2419 from \citet{Mucciarelli12} as open squares. We have only included the most metal-poor GCs with $\mbox{[Fe/H]} < -1.9$, matching our stream metallicities.

No clear evidence is seen for the expected GC abundance trends for any elements in our stellar streams. The most significant trend is the Na-Mg anticorrelation, which may be present in ATLAS and Phoenix, but is consistent with noise.
This is not especially surprising given the abundance uncertainties, relatively small number of stars, and the fact that metal-poor globular clusters tend to have the least extreme abundance differences \citep{Carretta09a}.
In particular, due to the logarithmic nature of abundance measurements, we are only likely to detect the abundance increases for the odd-Z elements N, Na, and Al.
This is because the proton capture cycles convert abundant elements (O, Ne, Mg) to underabundant elements (N, Na, Al) while conserving the total heavy element nuclei.
In other words, since the cycle inputs O, Ne, and Mg are intrinsically ${\gtrsim}10{\times}$ more abundant than the cycle products N, Na, and Al; logarithmic increases in N, Na and Al will be seen before significant logarithmic decreases in O, Ne, or Mg.
More detailed quantification is reserved for future work \citep{Caseyprep}.

\subsubsection{ATLAS and Aliqa Uma}
These two streams are spatially and kinematically consistent with being a single stream whose progenitor is a globular cluster \citep{Li19S5,Liprep}.
ATLAS and Aliqa Uma form a continuous track in radial velocity and proper motion on the sky, but, as discussed in \citet{Liprep}, a massive perturber created a spatial kink that caused them to be initially classified as two separate streams in \citet{Shipp18}.
The stellar abundances support this conclusion: both streams have essentially identical abundance character in all elements, with no detected metallicity spread and nearly-identical abundance ratios \citep{Liprep}.
There is weak evidence for larger scatter in the light elements Na and Mg, and they are anti-correlated in the direction that would be expected for a globular cluster.
Like most globular clusters, all [X/Fe] ratios of the heavier elements are consistent with those seen in the stellar halo \citep[e.g.,][]{Pritzl05}.
Combining all the stars in both streams gives a metallicity dispersion 95\% confidence upper limit of 0.12 dex.

\subsubsection{Phoenix}
The progenitor of the thin Phoenix stream is likely a globular cluster. Its low inferred metallicity of $\mbox{[Fe/H]} = -2.7$ is below the globular cluster floor of $-2.4$, demonstrating that globular clusters below the metallicity floor previously existed, but they have probably mostly been tidally disrupted during Galactic evolution \citep{Wan20,Kruijssen19}.
The mean abundance ratios are mostly consistent with the stellar halo, with the exception being [Ba/Fe], which is significantly lower and suggests Phoenix's progenitor was born in a lower mass galaxy than most globular clusters.
In addition, one star is clearly a lithium-rich giant. The abundances of this stream are discussed in detail by \citet{Caseyprep}.

\subsection{Dwarf Galaxy Streams}\label{sec:discdsph}

Four of our streams have thick morphologies, as well as significant metallicity dispersions and larger velocity dispersions that imply they are disrupting dwarf galaxies \citep{Shipp18,Li19S5}.

\begin{figure*}
    \centering
    \includegraphics[width=\linewidth]{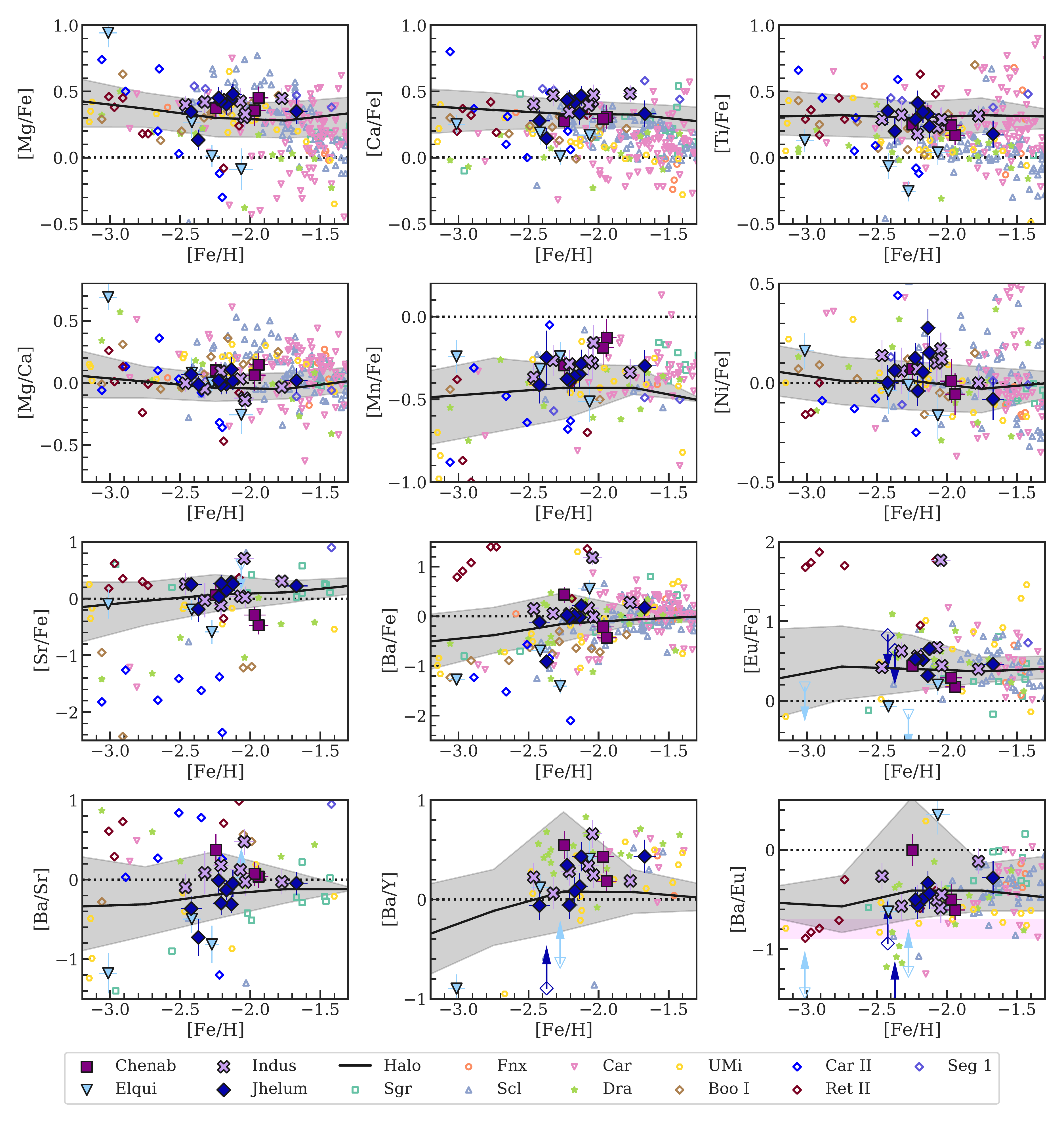}
    \caption{Abundance ratios of thick dSph streams (large filled points, upper limits with arrows) compared to dSph galaxies (small open points) and the halo median and 68\% scatter (black line and shaded region). Shaded pink region in [Ba/Eu] figure shows a pure $r$-process ratio.}
    \label{fig:dsphabunds}
\end{figure*}

Figure~\ref{fig:dsphabunds} compares several relevant element abundances to literature dwarf spheroidal (dSph) abundances spanning the full range of satellite galaxy luminosities:
Sagittarius \citep{Majewski17, JHansen18};
Fornax \citep{Letarte10, Shetrone03, Tafelmeyer10}; 
Sculptor \citep{Hill19,Jablonka15,Shetrone03,Simon15Scl,Geisler05,Skuladottir15,Kirby12,Frebel10a};
Carina \citep{Norris17b,Shetrone03,Venn12,Lemasle12},
Draco \citep{Cohen09,Tsujimoto17,Shetrone01,Tsujimoto15a,Fulbright04},
Ursa Minor \citep{Cohen10,Shetrone01,Ural15,Kirby12,Aoki07b},
Bo\"{o}tes I \citep{Frebel16,Ishigaki14,Gilmore13,Norris10a,Norris10b},
Carina II \citep{Ji20},
Reticulum II \citep{Ji16c,Roederer16b},
and Segue 1 \citep{Frebel14}.
For clarity, no upper limits are plotted for the literature sample.

Many of the Sgr stars come from APOGEE DR16 (H. J\"onsson et al. in prep; \citealt{Majewski17,Nidever15,Wilson19,GarciaPerez16,Shetrone15,Zakowski17}).
These are selected using the quality cuts \code{STARFLAG = ASPCAPFLAG = 0}, \code{VERR} $< 0.2 \kms$, \code{SNR} $>$ 70, $\Teff > 3700$K, and $\logg < 3.5$ \citep{Hayes20}.
Only stars within 1.5 half light radii of the Sgr center, or 514.05 arcsec of $(\alpha,\delta) = (283.747, -30.4606)$ \citep{Majewski03}, are considered.
After inspection, Milky Way foreground stars are removed with velocity and proper motion cuts of $100 < \mbox{\code{VHELIO\_AVG}} < 180 \kms$, $-3.2 < \mbox{\code{GAIA\_PMRA}} < -2.25$ mas/yr, and $-1.9 < \mbox{\code{GAIA\_PMDEC}} < -0.9$ mas/yr.
The final APOGEE selection has 400 stars.

Figure~\ref{fig:dsphabunds}
also shows abundances in the Milky Way halo and disk collected in JINAbase \citep{jinabase}, using only data from \citet{Fulbright00,Barklem05,Aoki09,Cohen13,Roederer14c}. For clarity, the halo stars are grouped in bins of 0.5 dex, plotting the median (black line) and 68\% scatter (shaded grey region) of each bin.

The top row of Figure~\ref{fig:dsphabunds} shows [Mg, Ca, Ti/Fe] vs [Fe/H], which track the general star formation efficiency of a dwarf galaxy \citep[e.g.,][]{Tinsley80,Matteucci90,Tolstoy09,Kirby11}.
The only stream showing significant declines in [$\alpha$/Fe] is Elqui, while the other streams are generally consistent with the halo median.
There is also significant evolution in [Mg/Ca] with [Fe/H] in Elqui, while the other streams generally match the flat halo trend.

[Mn, Ni/Fe] are also shown, which can track changes in Type Ia supernova enrichment \citep{McWilliam18,Kirby19,DeLosReyes20}.
These elements do not display any large trends with [Fe/H], although there is a hint that Chenab's metal-rich stars have higher [Mn/Fe].

The bottom two rows show the neutron-capture elements.
[Sr, Ba, Eu/Fe] are shown as the most easily measured tracers of elements from the first, second, and rare-earth neutron-capture peaks.
In terms of neutron-capture element abundances, the stream stars are very similar to the luminous dSph galaxies but they differ from the ultra-faint dSphs.
The dSphs differ from the halo primarily in Ba, which is substantially lower than the halo at $\mbox{[Fe/H]} \lesssim -2.2$.
The bottom row shows [Ba/Sr] and [Ba/Y]. The high-Fe dSph stars have elevated [Ba/Y] ratios compared to the halo, which is often interpreted as evidence for a metal-poor s-process taking place in dwarf galaxies \citep[e.g.,][]{Shetrone03,Venn04}.
[Ba/Eu] indicates the relative ratio of s- and r-process, where the shaded pink region is a pure r-process [Ba/Eu] and higher values indicate some amount of s-process contamination \citep[e.g.,][]{Sneden08}.

\subsubsection{Chenab}
The Chenab stream is a Southern hemisphere extension of the Orphan stream \citep{Koposov19}.
Using RRL star counts, the progenitor is estimated to have a luminosity $M_V = -10.8 \pm 1.3$, placing its mass as similar to Sculptor and between that of Sextans and Leo I \citep{Munoz18,Koposov19}.
This matches the expectations found through high-resolution spectroscopic study of three Orphan stream stars by \citet{Casey14b}, and our three new stars confirm previous conclusions, especially in having high [Ba/Y] ratios characteristic of intact dwarf galaxies.

Unlike \citet{Casey14b} we do not find a decreasing [$\alpha$/Fe] trend with metallicity, but our stars span a smaller [Fe/H] range and may still be on the [$\alpha$/Fe] plateau, implying an [$\alpha$/Fe] knee somewhere between $-2.0 < \mbox{[Fe/H]} < -1.5$, consistent with Sculptor \citep{Hill19}.
There is some evidence in our data for an upturn in [Mn/Fe] for the two more Fe-rich stars, a trend that continues in the stars from \citet{Casey14b}. This could indicate a transition from sub-Chandrasekhar to Chandrasekhar mass Type~Ia supernovae \citep{DeLosReyes20}, although there is not a corresponding rise in [Ni/Fe] \citep{Kirby19}.

\subsubsection{Elqui}
The Elqui stream's dwarf galaxy progenitor is likely the lowest mass galaxy progenitor of the streams studied here.
Morphologically, this was already suggested using the progenitor masses derived in \citet{Shipp18}.
The four Elqui stars range from $-3 < \mbox{[Fe/H]} < -2$, and the most metal-rich stars in Elqui have $\mbox{[$\alpha$/Fe]} \sim 0$, distinctly lower than the other streams and the stellar halo at this metallicity, but similar to that of low mass galaxies like Draco.
The neutron-capture elements in Elqui display solar $\mbox{[Sr/Fe]} \sim 0$, much higher than Sr in most lower-mass ultra-faint dwarf galaxies. The exception is Reticulum~II, which has very different $\mbox{[Ba/Fe]}$ from Elqui.
Together, these trends suggest Elqui's progenitor galaxy's stellar mass was at the low end of classical dSph galaxies, around $10^6 M_\odot$ or $M_V \sim -9$.

Elqui\_3 has a clear s-process signature with moderately enhanced Ba and C and [Ba/Eu] $> 0$.
It is not clear if this is due to binary mass transfer or ISM enrichment: no velocity variations are found in this star, and the enhancements are not as extreme as the CEMP-s stars that are clearly results of mass transfer \citep[e.g.,][]{Hansen16}. 

Elqui\_1 is the most Fe-poor star in our sample at $\mbox{[Fe/H]} \sim -3$.
This star is likely C-enhanced, as it has $\Teff \sim 4300$K but $\mbox{[C/Fe]} \sim 0.3$.
The \citet{Placco14} correction\footnote{\url{http://vplacco.pythonanywhere.com/}} for this star gives $\mbox{[C/Fe]} \sim +1.0$. This star also has a very high $\mbox{[Mg/Fe]} \sim 1.0$ but low $\mbox{[Si/Fe}] \sim 0.1$ and $\mbox{[Ca/Fe]} \sim 0.2$, possibly suggesting it is a carbon-enhanced star primarily enriched by a very massive star.
Indeed, the [Fe/H], [Mg/C], [N/Na], and [Sc/Mn] abundances all suggest this star has a high chance of being enriched by only one Population~III supernova, according to the models in \citet{Hartwig18}.
Furthermore, Elqui displays a much more rapid decline in [Mg/Fe] vs [Fe/H] compared to [Ca/Fe] vs [Fe/H], reminiscent of a few other dwarf galaxies like Sgr and Carina~II \citep{McWilliam13,Hasselquist17,Ji20}.

\subsubsection{Indus and Jhelum}

We consider Indus and Jhelum together because it has been suggested that they are two wraps of the same stream \citep{Shipp18,Bonaca19}.
Jhelum also may have two separate spatial and/or kinematic populations \citep{Bonaca19, Shipp19}.
Differences in elemental abundances could help verify whether the kinematic and spatial populations are in fact different systems, but by eye the stars in these two streams have very similar abundances to each other and to the background stellar halo.
A more detailed analysis will be presented in \citet{Paceprep}.

There is a mild discrepancy between the median metallicity of our Indus and Jhelum stars and the [$\alpha$/Fe] ratios observed in those stars.
Most of the observed stars in these two streams have $\mbox{[Fe/H]} \sim -2$.
Intact dwarf galaxies with $\left<\mbox{[Fe/H]}\right> \sim -2$ have luminosities $-10 \lesssim M_V \lesssim -8$ (Carina, Ursa Minor, Sextans, Draco, Canes Venatici I, from the compilation in \citealt{Munoz18,Simon19}).
However, all the stars in these two streams are $\alpha$-enhanced, with $\mbox{[Mg,Ca,Ti/Fe]} \sim +0.3$ to $+0.4$.
Only relatively luminous galaxies, $M_V \gtrsim -10$, have enhanced [$\alpha$/Fe] at $\mbox{[Fe/H]} \sim -2$ \citep[e.g.,][]{Kirby11}.
The most likely explanation for this discrepancy is that the stars observed here are somewhat biased towards lower metallicity compared to all possible Indus and Jhelum members.
\citet{Paceprep} and \citet{Hansenprep} will discuss this in more detail.

One star in Indus (Indus\_13) has extremely high levels of $r$-process enhancement, with $\mbox{[Eu/Fe]} \sim +1.8$ and $\mbox{[Fe/H]} \sim -2.0$.
This is one of the most Fe-rich $r$-process-enhanced stars known, though similar stars have been found in Ursa Minor and the stellar halo \citep{Aoki07b,Sakari18}.
Additionally, one star in Indus (Indus\_0) has high N, Na, and Al consistent with globular cluster abundance anomalies.
Stars in dSphs showing these anomalies are rare, though the anomalies are known to occur in the globular clusters associated with the Fornax dSph \citep[e.g.,][]{Larsen14,Hendricks16}.
\citet{Hansenprep} will discuss these stars and their implications for the formation of Indus's progenitor.

\section{Summary}\label{sec:conclusion}

We have presented results from high-resolution spectroscopy of 42 red giant stars in seven stellar streams, including abundances of up to 30 elements.
Three streams are from disrupted globular clusters with $\mbox{[Fe/H]} < -2$ (ATLAS, Aliqa Uma, and Phoenix).
Four streams (Chenab, Elqui, Indus, and Jhelum) are disrupted dwarf galaxies with chemical evolution histories suggesting progenitor masses between Draco and Sculptor ($M_\star \sim 10^{6-7} M_\odot$).

The primary aim of this work was to present the detailed abundance analysis methodology. The main results are shown in Figure~\ref{fig:xfegrid}.
The stellar parameters were derived using photometric temperatures and surface gravities, while microturbulence was inferred from Fe II lines (Table~\ref{tab:sp}).
A 1D LTE abundance analysis was performed using \code{MOOG} and \code{ATLAS} model atmospheres, propagating individual line uncertainties (Table~\ref{tab:lines}) and accounting for correlated stellar parameters (Tables~\ref{tab:spcorr} and \ref{tab:abunds}, see Appendix~\ref{app:estimator}).
We recommend that those using the abundances in this paper read through Section~\ref{sec:elements} to understand how the abundances were derived, and consider the figures in Appendix~\ref{app:params_errors} to see if those correlations affect interpretations.

Figure~\ref{fig:mdfs} shows the relation between stream widths and metallicity dispersions, showing a separation between the thin globular cluster streams with unresolved metallicity dispersions and the thicker dwarf galaxy streams with resolved metallicity dispersions.
Figures~\ref{fig:gcabunds} and \ref{fig:dsphabunds} show our results compared to literature values for intact globular clusters and dwarf galaxies.
This paper has made minimal scientific interpretations, and future work will discuss those comparisons in detail.

\acknowledgments
We thank the referee for important and thorough comments regarding this long paper.
APJ and TSL are supported by NASA through Hubble Fellowship grant HST-HF2-51393.001 and HST-HF2-51439.001 respectively, awarded by the Space Telescope Science Institute, which is operated by the Association of Universities for Research in Astronomy, Inc., for NASA, under contract NAS5-26555.
APJ is partially supported by the Thacher Research Award in Astronomy.
ARC is supported in part by the Australian Research Council through a Discovery Early Career Researcher Award (DE190100656). Parts of this research were supported by the Australian Research Council Centre of Excellence for All Sky Astrophysics in 3 Dimensions (ASTRO 3D), through project number CE170100013.
ABP acknowledges support from NSF grant AST-1813881.
SK is partially supported by NSF grants AST-1813881, AST-1909584 and Heising-Simons foundation grant 2018-1030.
SLM and JDS acknowledge support from the Australian Research Council through Discovery Project grant DP180101791.

% SIMBAD
This research has made use of the SIMBAD database, operated at CDS, Strasbourg, France \citep{Simbad}, and NASA's Astrophysics Data System Bibliographic Services.
% VALD
This work has made use of the VALD database, operated at Uppsala University, the Institute of Astronomy RAS in Moscow, and the University of Vienna.
% Gaia
This work presents results from the European Space Agency (ESA) space
mission Gaia. Gaia data are being processed by the Gaia Data
Processing and Analysis Consortium (DPAC). Funding for the DPAC is
provided by national institutions, in particular the institutions
participating in the Gaia MultiLateral Agreement (MLA). The Gaia
mission website is https://www.cosmos.esa.int/gaia. The Gaia archive
website is https://archives.esac.esa.int/gaia.
%DES DR1
This project used public archival data from the Dark Energy Survey
(DES). Funding for the DES Projects has been provided by the
U.S. Department of Energy, the U.S. National Science Foundation, the
Ministry of Science and Education of Spain, the Science and Technology
Facilities Council of the United Kingdom, the Higher Education Funding
Council for England, the National Center for Supercomputing
Applications at the University of Illinois at Urbana-Champaign, the
Kavli Institute of Cosmological Physics at the University of Chicago,
the Center for Cosmology and Astro-Particle Physics at the Ohio State
University, the Mitchell Institute for Fundamental Physics and
Astronomy at Texas A\&M University, Financiadora de Estudos e
Projetos, Funda{\c c}{\~a}o Carlos Chagas Filho de Amparo {\`a}
Pesquisa do Estado do Rio de Janeiro, Conselho Nacional de
Desenvolvimento Cient{\'i}fico e Tecnol{\'o}gico and the
Minist{\'e}rio da Ci{\^e}ncia, Tecnologia e Inova{\c c}{\~a}o, the
Deutsche Forschungsgemeinschaft, and the Collaborating Institutions in
the Dark Energy Survey.  The Collaborating Institutions are Argonne
National Laboratory, the University of California at Santa Cruz, the
University of Cambridge, Centro de Investigaciones Energ{\'e}ticas,
Medioambientales y Tecnol{\'o}gicas-Madrid, the University of Chicago,
University College London, the DES-Brazil Consortium, the University
of Edinburgh, the Eidgen{\"o}ssische Technische Hochschule (ETH)
Z{\"u}rich, Fermi National Accelerator Laboratory, the University of
Illinois at Urbana-Champaign, the Institut de Ci{\`e}ncies de l'Espai
(IEEC/CSIC), the Institut de F{\'i}sica d'Altes Energies, Lawrence
Berkeley National Laboratory, the Ludwig-Maximilians Universit{\"a}t
M{\"u}nchen and the associated Excellence Cluster Universe, the
University of Michigan, the National Optical Astronomy Observatory,
the University of Nottingham, The Ohio State University, the OzDES
Membership Consortium, the University of Pennsylvania, the University
of Portsmouth, SLAC National Accelerator Laboratory, Stanford
University, the University of Sussex, and Texas A\&M University.
Based in part on observations at Cerro Tololo Inter-American
Observatory, National Optical Astronomy Observatory, which is operated
by the Association of Universities for Research in Astronomy (AURA)
under a cooperative agreement with the National Science Foundation.

{\it Facilities:} 
{Anglo-Australian Telescope (AAOmega+2dF); Magellan/Clay (MIKE)}

{\it Software:} 
{\code{numpy} \citep{numpy}, 
\code{scipy} \citep{scipy}, 
\code{matplotlib} \citep{matplotlib}, 
\code{pandas} \citep{pandas},
\code{seaborn} \citep{seaborn},
\code{astropy} \citep{astropy,astropy:2018},
\code{emcee} \citep{emcee},
\code{CarPy} \citep{Kelson03}, 
\code{MOOG} \citep{Sneden73,Sobeck11},
\code{SMHR} \citep{Casey14},
\code{rvspecfit} \citep{rvspecfit},
\code{PyStan} \citep{stan,pystan}
%\code{sklearn} \citep{Pedregosa2011}}
}

\bibliographystyle{aasjournal}

\begin{thebibliography}{}
\expandafter\ifx\csname natexlab\endcsname\relax\def\natexlab#1{#1}\fi
\providecommand{\url}[1]{\href{#1}{#1}}
\providecommand{\dodoi}[1]{doi:~\href{http://doi.org/#1}{\nolinkurl{#1}}}
\providecommand{\doeprint}[1]{\href{http://ascl.net/#1}{\nolinkurl{http://ascl.net/#1}}}
\providecommand{\doarXiv}[1]{\href{https://arxiv.org/abs/#1}{\nolinkurl{https://arxiv.org/abs/#1}}}

\bibitem[{{Abohalima} \& {Frebel}(2018)}]{jinabase}
{Abohalima}, A., \& {Frebel}, A. 2018, \apjs, 238, 36,
  \dodoi{10.3847/1538-4365/aadfe9}

\bibitem[{{Aoki} {et~al.}(2007){Aoki}, {Honda}, {Sadakane}, \&
  {Arimoto}}]{Aoki07b}
{Aoki}, W., {Honda}, S., {Sadakane}, K., \& {Arimoto}, N. 2007, \pasj, 59, L15,
  \dodoi{10.1093/pasj/59.3.L15}

\bibitem[{{Aoki} {et~al.}(2009){Aoki}, {Arimoto}, {Sadakane}, {Tolstoy},
  {Battaglia}, {Jablonka}, {Shetrone}, {Letarte}, {Irwin}, {Hill}, {Francois},
  {Venn}, {Primas}, {Helmi}, {Kaufer}, {Tafelmeyer}, {Szeifert}, \&
  {Babusiaux}}]{Aoki09}
{Aoki}, W., {Arimoto}, N., {Sadakane}, K., {et~al.} 2009, \aap, 502, 569,
  \dodoi{10.1051/0004-6361/200911959}

\bibitem[{{Asplund} {et~al.}(2004){Asplund}, {Grevesse}, {Sauval}, {Allende
  Prieto}, \& {Kiselman}}]{Asplund04}
{Asplund}, M., {Grevesse}, N., {Sauval}, A.~J., {Allende Prieto}, C., \&
  {Kiselman}, D. 2004, \aap, 417, 751, \dodoi{10.1051/0004-6361:20034328}

\bibitem[{{Asplund} {et~al.}(2009){Asplund}, {Grevesse}, {Sauval}, \&
  {Scott}}]{Asplund09}
{Asplund}, M., {Grevesse}, N., {Sauval}, A.~J., \& {Scott}, P. 2009, \araa, 47,
  481, \dodoi{10.1146/annurev.astro.46.060407.145222}

\bibitem[{{Astropy Collaboration} {et~al.}(2013){Astropy Collaboration},
  {Robitaille}, {Tollerud}, {Greenfield}, {Droettboom}, {Bray}, {Aldcroft},
  {Davis}, {Ginsburg}, {Price-Whelan}, {Kerzendorf}, {Conley}, {Crighton},
  {Barbary}, {Muna}, {Ferguson}, {Grollier}, {Parikh}, {Nair}, {Unther},
  {Deil}, {Woillez}, {Conseil}, {Kramer}, {Turner}, {Singer}, {Fox}, {Weaver},
  {Zabalza}, {Edwards}, {Azalee Bostroem}, {Burke}, {Casey}, {Crawford},
  {Dencheva}, {Ely}, {Jenness}, {Labrie}, {Lim}, {Pierfederici}, {Pontzen},
  {Ptak}, {Refsdal}, {Servillat}, \& {Streicher}}]{astropy}
{Astropy Collaboration}, {Robitaille}, T.~P., {Tollerud}, E.~J., {et~al.} 2013,
  \aap, 558, A33, \dodoi{10.1051/0004-6361/201322068}

\bibitem[{{Balbinot} {et~al.}(2016){Balbinot}, {Yanny}, {Li}, {Santiago},
  {Marshall}, {Finley}, {Pieres}, {Abbott}, {Abdalla}, {Allam},
  {Benoit-L{\'e}vy}, {Bernstein}, {Bertin}, {Brooks}, {Burke}, {Carnero
  Rosell}, {Carrasco Kind}, {Carretero}, {Cunha}, {da Costa}, {DePoy}, {Desai},
  {Diehl}, {Doel}, {Estrada}, {Flaugher}, {Frieman}, {Gerdes}, {Gruen},
  {Gruendl}, {Honscheid}, {James}, {Kuehn}, {Kuropatkin}, {Lahav}, {March},
  {Martini}, {Miquel}, {Nichol}, {Ogando}, {Romer}, {Sanchez}, {Schubnell},
  {Sevilla-Noarbe}, {Smith}, {Soares-Santos}, {Sobreira}, {Suchyta}, {Tarle},
  {Thomas}, {Tucker}, {Walker}, \& {DES Collaboration}}]{Balbinot16}
{Balbinot}, E., {Yanny}, B., {Li}, T.~S., {et~al.} 2016, \apj, 820, 58,
  \dodoi{10.3847/0004-637X/820/1/58}

\bibitem[{{Barklem} {et~al.}(2005){Barklem}, {Christlieb}, {Beers}, {Hill},
  {Bessell}, {Holmberg}, {Marsteller}, {Rossi}, {Zickgraf}, \&
  {Reimers}}]{Barklem05}
{Barklem}, P.~S., {Christlieb}, N., {Beers}, T.~C., {et~al.} 2005, \aap, 439,
  129, \dodoi{10.1051/0004-6361:20052967}

\bibitem[{{Battaglia} {et~al.}(2017){Battaglia}, {North}, {Jablonka},
  {Shetrone}, {Minniti}, {D{\'\i}az}, {Starkenburg}, \& {Savoy}}]{Battaglia17}
{Battaglia}, G., {North}, P., {Jablonka}, P., {et~al.} 2017, \aap, 608, A145,
  \dodoi{10.1051/0004-6361/201731879}

\bibitem[{{Belmonte} {et~al.}(2017){Belmonte}, {Pickering}, {Ruffoni}, {Den
  Hartog}, {Lawler}, {Guzman}, \& {Heiter}}]{Belmonte17Fe}
{Belmonte}, M.~T., {Pickering}, J.~C., {Ruffoni}, M.~P., {et~al.} 2017, \apj,
  848, 125, \dodoi{10.3847/1538-4357/aa8cd3}

\bibitem[{{Belokurov} {et~al.}(2018){Belokurov}, {Erkal}, {Evans}, {Koposov},
  \& {Deason}}]{Belokurov18}
{Belokurov}, V., {Erkal}, D., {Evans}, N.~W., {Koposov}, S.~E., \& {Deason},
  A.~J. 2018, \mnras, 478, 611, \dodoi{10.1093/mnras/sty982}

\bibitem[{{Bergemann} \& {Cescutti}(2010)}]{Bergemann10}
{Bergemann}, M., \& {Cescutti}, G. 2010, \aap, 522, A9,
  \dodoi{10.1051/0004-6361/201014250}

\bibitem[{{Bergemann} {et~al.}(2012){Bergemann}, {Lind}, {Collet}, {Magic}, \&
  {Asplund}}]{Bergemann12}
{Bergemann}, M., {Lind}, K., {Collet}, R., {Magic}, Z., \& {Asplund}, M. 2012,
  \mnras, 427, 27, \dodoi{10.1111/j.1365-2966.2012.21687.x}

\bibitem[{{Bergemann} {et~al.}(2018){Bergemann}, {Sesar}, {Cohen}, {Serenelli},
  {Sheffield}, {Li}, {Casagrande}, {Johnston}, {Laporte}, {Price-Whelan},
  {Sch{\"o}nrich}, \& {Gould}}]{Bergemann18}
{Bergemann}, M., {Sesar}, B., {Cohen}, J.~G., {et~al.} 2018, \nat, 555, 334,
  \dodoi{10.1038/nature25490}

\bibitem[{{Bergemann} {et~al.}(2019){Bergemann}, {Gallagher}, {Eitner},
  {Bautista}, {Collet}, {Yakovleva}, {Mayriedl}, {Plez}, {Carlsson},
  {Leenaarts}, {Belyaev}, \& {Hansen}}]{Bergemann19}
{Bergemann}, M., {Gallagher}, A.~J., {Eitner}, P., {et~al.} 2019, arXiv
  e-prints, arXiv:1905.05200.
\newblock \doarXiv{1905.05200}

\bibitem[{{Bernstein} {et~al.}(2003){Bernstein}, {Shectman}, {Gunnels},
  {Mochnacki}, \& {Athey}}]{Bernstein03}
{Bernstein}, R., {Shectman}, S.~A., {Gunnels}, S.~M., {Mochnacki}, S., \&
  {Athey}, A.~E. 2003, Proc. SPIE, 4841, 1694, \dodoi{10.1117/12.461502}

\bibitem[{{Bi{\'e}mont} {et~al.}(2011){Bi{\'e}mont}, {Blagoev}, {Engstr{\"o}m},
  {Hartman}, {Lundberg}, {Malcheva}, {Nilsson}, {Whitehead}, {Palmeri}, \&
  {Quinet}}]{Biemont11Y}
{Bi{\'e}mont}, {\'E}., {Blagoev}, K., {Engstr{\"o}m}, L., {et~al.} 2011,
  \mnras, 414, 3350, \dodoi{10.1111/j.1365-2966.2011.18637.x}

\bibitem[{{Bonaca} {et~al.}(2019){Bonaca}, {Conroy}, {Price-Whelan}, \&
  {Hogg}}]{Bonaca19}
{Bonaca}, A., {Conroy}, C., {Price-Whelan}, A.~M., \& {Hogg}, D.~W. 2019,
  \apjl, 881, L37, \dodoi{10.3847/2041-8213/ab36ba}

\bibitem[{{Caffau} {et~al.}(2008){Caffau}, {Ludwig}, {Steffen}, {Ayres},
  {Bonifacio}, {Cayrel}, {Freytag}, \& {Plez}}]{Caffau08oxygen}
{Caffau}, E., {Ludwig}, H.~G., {Steffen}, M., {et~al.} 2008, \aap, 488, 1031,
  \dodoi{10.1051/0004-6361:200809885}

\bibitem[{{Carlin} {et~al.}(2018){Carlin}, {Sheffield}, {Cunha}, \&
  {Smith}}]{Carlin18}
{Carlin}, J.~L., {Sheffield}, A.~A., {Cunha}, K., \& {Smith}, V.~V. 2018,
  \apjl, 859, L10, \dodoi{10.3847/2041-8213/aac3d8}

\bibitem[{Carpenter {et~al.}(2017)Carpenter, Gelman, Hoffman, Lee, Goodrich,
  Betancourt, Brubaker, Guo, Li, \& Riddell}]{stan}
Carpenter, B., Gelman, A., Hoffman, M.~D., {et~al.} 2017, Journal of
  statistical software, 76

\bibitem[{{Carretta} {et~al.}(2009{\natexlab{a}}){Carretta}, {Bragaglia},
  {Gratton}, \& {Lucatello}}]{Carretta09b}
{Carretta}, E., {Bragaglia}, A., {Gratton}, R., \& {Lucatello}, S.
  2009{\natexlab{a}}, \aap, 505, 139, \dodoi{10.1051/0004-6361/200912097}

\bibitem[{{Carretta} {et~al.}(2009{\natexlab{b}}){Carretta}, {Bragaglia},
  {Gratton}, {Lucatello}, {Catanzaro}, {Leone}, {Bellazzini}, {Claudi},
  {D'Orazi}, {Momany}, {Ortolani}, {Pancino}, {Piotto}, {Recio-Blanco}, \&
  {Sabbi}}]{Carretta09a}
{Carretta}, E., {Bragaglia}, A., {Gratton}, R.~G., {et~al.} 2009{\natexlab{b}},
  \aap, 505, 117, \dodoi{10.1051/0004-6361/200912096}

\bibitem[{{Casagrande} \& {VandenBerg}(2014)}]{Casagrande14}
{Casagrande}, L., \& {VandenBerg}, D.~A. 2014, \mnras, 444, 392,
  \dodoi{10.1093/mnras/stu1476}

\bibitem[{{Casey}(2014)}]{Casey14}
{Casey}, A.~R. 2014, ArXiv e-prints.
\newblock \doarXiv{1405.5968}

\bibitem[{{Casey} {et~al.}(in prep){Casey}, {Ji}, \& {S5
  Collaboration}}]{Caseyprep}
{Casey}, A.~R., {Ji}, A.~P., \& {S5 Collaboration}. in prep

\bibitem[{{Casey} {et~al.}(2014){Casey}, {Keller}, {Da Costa}, {Frebel}, \&
  {Maunder}}]{Casey14b}
{Casey}, A.~R., {Keller}, S.~C., {Da Costa}, G., {Frebel}, A., \& {Maunder}, E.
  2014, \apj, 784, 19, \dodoi{10.1088/0004-637X/784/1/19}

\bibitem[{{Castelli} \& {Kurucz}(2004)}]{Castelli04}
{Castelli}, F., \& {Kurucz}, R.~L. 2004, ArXiv Astrophysics e-prints

\bibitem[{{Chou} {et~al.}(2010){Chou}, {Cunha}, {Majewski}, {Smith},
  {Patterson}, {Mart{\'\i}nez-Delgado}, \& {Geisler}}]{Chou10}
{Chou}, M.-Y., {Cunha}, K., {Majewski}, S.~R., {et~al.} 2010, \apj, 708, 1290,
  \dodoi{10.1088/0004-637X/708/2/1290}

\bibitem[{{Cohen} {et~al.}(2013){Cohen}, {Christlieb}, {Thompson}, {McWilliam},
  {Shectman}, {Reimers}, {Wisotzki}, \& {Kirby}}]{Cohen13}
{Cohen}, J.~G., {Christlieb}, N., {Thompson}, I., {et~al.} 2013, \apj, 778, 56,
  \dodoi{10.1088/0004-637X/778/1/56}

\bibitem[{{Cohen} \& {Huang}(2009)}]{Cohen09}
{Cohen}, J.~G., \& {Huang}, W. 2009, \apj, 701, 1053,
  \dodoi{10.1088/0004-637X/701/2/1053}

\bibitem[{{Cohen} \& {Huang}(2010)}]{Cohen10}
---. 2010, \apj, 719, 931, \dodoi{10.1088/0004-637X/719/1/931}

\bibitem[{{Cohen} \& {Kirby}(2012)}]{Cohen12}
{Cohen}, J.~G., \& {Kirby}, E.~N. 2012, \apj, 760, 86,
  \dodoi{10.1088/0004-637X/760/1/86}

\bibitem[{{Cowan} {et~al.}(2020){Cowan}, {Sneden}, {Roederer}, {Lawler},
  {Hartog}, {Sobeck}, \& {Boesgaard}}]{Cowan20}
{Cowan}, J.~J., {Sneden}, C., {Roederer}, I.~U., {et~al.} 2020, \apj, 890, 119,
  \dodoi{10.3847/1538-4357/ab6aa9}

\bibitem[{{de los Reyes} {et~al.}(2020){de los Reyes}, {Kirby}, {Seitenzahl},
  \& {Shen}}]{DeLosReyes20}
{de los Reyes}, M. A.~C., {Kirby}, E.~N., {Seitenzahl}, I.~R., \& {Shen}, K.~J.
  2020, \apj, 891, 85, \dodoi{10.3847/1538-4357/ab736f}

\bibitem[{{Den Hartog} {et~al.}(2003){Den Hartog}, {Lawler}, {Sneden}, \&
  {Cowan}}]{DenHartog03Nd}
{Den Hartog}, E.~A., {Lawler}, J.~E., {Sneden}, C., \& {Cowan}, J.~J. 2003,
  \apjs, 148, 543, \dodoi{10.1086/376940}

\bibitem[{{Den Hartog} {et~al.}(2006){Den Hartog}, {Lawler}, {Sneden}, \&
  {Cowan}}]{DenHartog06Gd}
---. 2006, \apjs, 167, 292, \dodoi{10.1086/508262}

\bibitem[{{Den Hartog} {et~al.}(2019){Den Hartog}, {Lawler}, {Sneden}, {Cowan},
  \& {Brukhovesky}}]{DenHartog19Fe2}
{Den Hartog}, E.~A., {Lawler}, J.~E., {Sneden}, C., {Cowan}, J.~J., \&
  {Brukhovesky}, A. 2019, \apjs, 243, 33, \dodoi{10.3847/1538-4365/ab322e}

\bibitem[{{Den Hartog} {et~al.}(2011){Den Hartog}, {Lawler}, {Sobeck},
  {Sneden}, \& {Cowan}}]{DenHartog11Mn}
{Den Hartog}, E.~A., {Lawler}, J.~E., {Sobeck}, J.~S., {Sneden}, C., \&
  {Cowan}, J.~J. 2011, \apjs, 194, 35, \dodoi{10.1088/0067-0049/194/2/35}

\bibitem[{{Den Hartog} {et~al.}(2014){Den Hartog}, {Ruffoni}, {Lawler},
  {Pickering}, {Lind}, \& {Brewer}}]{DenHartog14Fe}
{Den Hartog}, E.~A., {Ruffoni}, M.~P., {Lawler}, J.~E., {et~al.} 2014, \apjs,
  215, 23, \dodoi{10.1088/0067-0049/215/2/23}

\bibitem[{{DES Collaboration} {et~al.}(2018){DES Collaboration}, {Abbott},
  {Abdalla}, {Allam}, {Amara}, {Annis}, {Asorey}, {Avila}, {Ballester},
  {Banerji}, {Barkhouse}, {Baruah}, {Baumer}, {Bechtol}, {Becker},
  {Benoit-L{\'e}vy}, {Bernstein}, {Bertin}, {Blazek}, {Bocquet}, {Brooks},
  {Brout}, {Buckley-Geer}, {Burke}, {Busti}, {Campisano}, {Cardiel-Sas},
  {Carnero Rosell}, {Carrasco Kind}, {Carretero}, {Castander}, {Cawthon},
  {Chang}, {Chen}, {Conselice}, {Costa}, {Crocce}, {Cunha}, {D'Andrea}, {da
  Costa}, {Das}, {Daues}, {Davis}, {Davis}, {De Vicente}, {DePoy}, {DeRose},
  {Desai}, {Diehl}, {Dietrich}, {Dodelson}, {Doel}, {Drlica-Wagner}, {Eifler},
  {Elliott}, {Evrard}, {Farahi}, {Fausti Neto}, {Fernandez}, {Finley},
  {Flaugher}, {Foley}, {Fosalba}, {Friedel}, {Frieman}, {Garc{\'\i}a-Bellido},
  {Gaztanaga}, {Gerdes}, {Giannantonio}, {Gill}, {Glazebrook}, {Goldstein},
  {Gower}, {Gruen}, {Gruendl}, {Gschwend}, {Gupta}, {Gutierrez}, {Hamilton},
  {Hartley}, {Hinton}, {Hislop}, {Hollowood}, {Honscheid}, {Hoyle}, {Huterer},
  {Jain}, {James}, {Jeltema}, {Johnson}, {Johnson}, {Kacprzak}, {Kent},
  {Khullar}, {Klein}, {Kovacs}, {Koziol}, {Krause}, {Kremin}, {Kron}, {Kuehn},
  {Kuhlmann}, {Kuropatkin}, {Lahav}, {Lasker}, {Li}, {Li}, {Liddle}, {Lima},
  {Lin}, {L{\'o}pez-Reyes}, {MacCrann}, {Maia}, {Maloney}, {Manera}, {March},
  {Marriner}, {Marshall}, {Martini}, {McClintock}, {McKay}, {McMahon},
  {Melchior}, {Menanteau}, {Miller}, {Miquel}, {Mohr}, {Morganson}, {Mould},
  {Neilsen}, {Nichol}, {Nogueira}, {Nord}, {Nugent}, {Nunes}, {Ogand o}, {Old},
  {Pace}, {Palmese}, {Paz-Chinch{\'o}n}, {Peiris}, {Percival}, {Petravick},
  {Plazas}, {Poh}, {Pond}, {Porredon}, {Pujol}, {Refregier}, {Reil}, {Ricker},
  {Rollins}, {Romer}, {Roodman}, {Rooney}, {Ross}, {Rykoff}, {Sako}, {Sanchez},
  {Sanchez}, {Santiago}, {Saro}, {Scarpine}, {Scolnic}, {Serrano},
  {Sevilla-Noarbe}, {Sheldon}, {Shipp}, {Silveira}, {Smith}, {Smith}, {Smith},
  {Soares-Santos}, {Sobreira}, {Song}, {Stebbins}, {Suchyta}, {Sullivan},
  {Swanson}, {Tarle}, {Thaler}, {Thomas}, {Thomas}, {Troxel}, {Tucker},
  {Vikram}, {Vivas}, {Walker}, {Wechsler}, {Weller}, {Wester}, {Wolf}, {Wu},
  {Yanny}, {Zenteno}, {Zhang}, {Zuntz}, {DES Collaboration}, {Juneau},
  {Fitzpatrick}, {Nikutta}, {Nidever}, {Olsen}, {Scott}, \& {NOAO Data
  Lab}}]{desdr1}
{DES Collaboration}, {Abbott}, T.~M.~C., {Abdalla}, F.~B., {et~al.} 2018,
  \apjs, 239, 18, \dodoi{10.3847/1538-4365/aae9f0}

\bibitem[{{Dotter} {et~al.}(2008){Dotter}, {Chaboyer}, {Jevremovi{\'c}},
  {Kostov}, {Baron}, \& {Ferguson}}]{Dotter08}
{Dotter}, A., {Chaboyer}, B., {Jevremovi{\'c}}, D., {et~al.} 2008, \apjs, 178,
  89, \dodoi{10.1086/589654}

\bibitem[{{Ezzeddine} {et~al.}(2017){Ezzeddine}, {Frebel}, \&
  {Plez}}]{Ezzeddine17}
{Ezzeddine}, R., {Frebel}, A., \& {Plez}, B. 2017, \apj, 847, 142,
  \dodoi{10.3847/1538-4357/aa8875}

\bibitem[{{Foreman-Mackey} {et~al.}(2013){Foreman-Mackey}, {Hogg}, {Lang}, \&
  {Goodman}}]{emcee}
{Foreman-Mackey}, D., {Hogg}, D.~W., {Lang}, D., \& {Goodman}, J. 2013, \pasp,
  125, 306, \dodoi{10.1086/670067}

\bibitem[{{Frebel} {et~al.}(2013{\natexlab{a}}){Frebel}, {Casey}, {Jacobson},
  \& {Yu}}]{Frebel13}
{Frebel}, A., {Casey}, A.~R., {Jacobson}, H.~R., \& {Yu}, Q.
  2013{\natexlab{a}}, \apj, 769, 57, \dodoi{10.1088/0004-637X/769/1/57}

\bibitem[{{Frebel} {et~al.}(2010){Frebel}, {Kirby}, \& {Simon}}]{Frebel10a}
{Frebel}, A., {Kirby}, E.~N., \& {Simon}, J.~D. 2010, \nat, 464, 72,
  \dodoi{10.1038/nature08772}

\bibitem[{{Frebel} {et~al.}(2013{\natexlab{b}}){Frebel}, {Lunnan}, {Casey},
  {Norris}, {Wyse}, \& {Gilmore}}]{Frebel13_300S}
{Frebel}, A., {Lunnan}, R., {Casey}, A.~R., {et~al.} 2013{\natexlab{b}}, \apj,
  771, 39, \dodoi{10.1088/0004-637X/771/1/39}

\bibitem[{{Frebel} {et~al.}(2016){Frebel}, {Norris}, {Gilmore}, \&
  {Wyse}}]{Frebel16}
{Frebel}, A., {Norris}, J.~E., {Gilmore}, G., \& {Wyse}, R.~F.~G. 2016, \apj,
  826, 110, \dodoi{10.3847/0004-637X/826/2/110}

\bibitem[{{Frebel} {et~al.}(2014){Frebel}, {Simon}, \& {Kirby}}]{Frebel14}
{Frebel}, A., {Simon}, J.~D., \& {Kirby}, E.~N. 2014, \apj, 786, 74,
  \dodoi{10.1088/0004-637X/786/1/74}

\bibitem[{{Freeman} \& {Bland-Hawthorn}(2002)}]{Freeman02}
{Freeman}, K., \& {Bland-Hawthorn}, J. 2002, \araa, 40, 487,
  \dodoi{10.1146/annurev.astro.40.060401.093840}

\bibitem[{{Fu} {et~al.}(2018){Fu}, {Simon}, {Shetrone}, {Bovy}, {Beers},
  {Fern{\'a}ndez-Trincado}, {Placco}, {Zamora}, {Allende Prieto},
  {Garc{\'\i}a-Hern{\'a}ndez}, {Harding}, {Ivans}, {Lane}, {Nitschelm},
  {Roman-Lopes}, \& {Sobeck}}]{Fu18}
{Fu}, S.~W., {Simon}, J.~D., {Shetrone}, M., {et~al.} 2018, \apj, 866, 42,
  \dodoi{10.3847/1538-4357/aad9f9}

\bibitem[{{Fulbright}(2000)}]{Fulbright00}
{Fulbright}, J.~P. 2000, \aj, 120, 1841, \dodoi{10.1086/301548}

\bibitem[{{Fulbright} {et~al.}(2004){Fulbright}, {Rich}, \&
  {Castro}}]{Fulbright04}
{Fulbright}, J.~P., {Rich}, R.~M., \& {Castro}, S. 2004, \apj, 612, 447,
  \dodoi{10.1086/421712}

\bibitem[{{Gaia Collaboration} {et~al.}(2016){Gaia Collaboration}, {Prusti},
  {de Bruijne}, {Brown}, {Vallenari}, {Babusiaux}, {Bailer-Jones}, {Bastian},
  {Biermann}, {Evans}, \& et~al.}]{GaiaSatellite}
{Gaia Collaboration}, {Prusti}, T., {de Bruijne}, J.~H.~J., {et~al.} 2016,
  \aap, 595, A1, \dodoi{10.1051/0004-6361/201629272}

\bibitem[{{Gaia Collaboration} {et~al.}(2018){Gaia Collaboration}, {Brown},
  {Vallenari}, {Prusti}, {de Bruijne}, {Babusiaux}, {Bailer-Jones}, {Biermann},
  {Evans}, {Eyer}, \& et~al.}]{GaiaDR2}
{Gaia Collaboration}, {Brown}, A.~G.~A., {Vallenari}, A., {et~al.} 2018, \aap,
  616, A1, \dodoi{10.1051/0004-6361/201833051}

\bibitem[{{Gallagher} {et~al.}(2015){Gallagher}, {Ludwig}, {Ryan}, \&
  {Aoki}}]{Gallagher15}
{Gallagher}, A.~J., {Ludwig}, H.~G., {Ryan}, S.~G., \& {Aoki}, W. 2015, \aap,
  579, A94, \dodoi{10.1051/0004-6361/201424803}

\bibitem[{{Garc{\'\i}a P{\'e}rez} {et~al.}(2016){Garc{\'\i}a P{\'e}rez},
  {Allende Prieto}, {Holtzman}, {Shetrone}, {M{\'e}sz{\'a}ros}, {Bizyaev},
  {Carrera}, {Cunha}, {Garc{\'\i}a-Hern{\'a}ndez}, {Johnson}, {Majewski},
  {Nidever}, {Schiavon}, {Shane}, {Smith}, {Sobeck}, {Troup}, {Zamora},
  {Weinberg}, {Bovy}, {Eisenstein}, {Feuillet}, {Frinchaboy}, {Hayden},
  {Hearty}, {Nguyen}, {O'Connell}, {Pinsonneault}, {Wilson}, \&
  {Zasowski}}]{GarciaPerez16}
{Garc{\'\i}a P{\'e}rez}, A.~E., {Allende Prieto}, C., {Holtzman}, J.~A.,
  {et~al.} 2016, \aj, 151, 144, \dodoi{10.3847/0004-6256/151/6/144}

\bibitem[{{Geisler} {et~al.}(2005){Geisler}, {Smith}, {Wallerstein},
  {Gonzalez}, \& {Charbonnel}}]{Geisler05}
{Geisler}, D., {Smith}, V.~V., {Wallerstein}, G., {Gonzalez}, G., \&
  {Charbonnel}, C. 2005, \aj, 129, 1428, \dodoi{10.1086/427540}

\bibitem[{{Gilmore} {et~al.}(2013){Gilmore}, {Norris}, {Monaco}, {Yong},
  {Wyse}, \& {Geisler}}]{Gilmore13}
{Gilmore}, G., {Norris}, J.~E., {Monaco}, L., {et~al.} 2013, \apj, 763, 61,
  \dodoi{10.1088/0004-637X/763/1/61}

\bibitem[{{G{\'o}mez} {et~al.}(2013){G{\'o}mez}, {Helmi}, {Cooper}, {Frenk},
  {Navarro}, \& {White}}]{Gomez13}
{G{\'o}mez}, F.~A., {Helmi}, A., {Cooper}, A.~P., {et~al.} 2013, \mnras, 436,
  3602, \dodoi{10.1093/mnras/stt1838}

\bibitem[{{Gratton} {et~al.}(2019){Gratton}, {Bragaglia}, {Carretta},
  {D'Orazi}, {Lucatello}, \& {Sollima}}]{Gratton19}
{Gratton}, R., {Bragaglia}, A., {Carretta}, E., {et~al.} 2019, \aapr, 27, 8,
  \dodoi{10.1007/s00159-019-0119-3}

\bibitem[{{Gratton} {et~al.}(2004){Gratton}, {Sneden}, \&
  {Carretta}}]{Gratton04}
{Gratton}, R., {Sneden}, C., \& {Carretta}, E. 2004, \araa, 42, 385,
  \dodoi{10.1146/annurev.astro.42.053102.133945}

\bibitem[{{Gratton} {et~al.}(2012){Gratton}, {Carretta}, \&
  {Bragaglia}}]{Gratton12}
{Gratton}, R.~G., {Carretta}, E., \& {Bragaglia}, A. 2012, \aapr, 20, 50,
  \dodoi{10.1007/s00159-012-0050-3}

\bibitem[{{Grillmair} \& {Carlberg}(2016)}]{Grillmair16b}
{Grillmair}, C.~J., \& {Carlberg}, R.~G. 2016, \apjl, 820, L27,
  \dodoi{10.3847/2041-8205/820/2/L27}

\bibitem[{{Hannaford} {et~al.}(1982){Hannaford}, {Lowe}, {Grevesse}, {Biemont},
  \& {Whaling}}]{Hannaford82Y}
{Hannaford}, P., {Lowe}, R.~M., {Grevesse}, N., {Biemont}, E., \& {Whaling}, W.
  1982, \apj, 261, 736, \dodoi{10.1086/160384}

\bibitem[{{Hansen} {et~al.}(2018){Hansen}, {El-Souri}, {Monaco}, {Villanova},
  {Bonifacio}, {Caffau}, \& {Sbordone}}]{JHansen18}
{Hansen}, C.~J., {El-Souri}, M., {Monaco}, L., {et~al.} 2018, \apj, 855, 83,
  \dodoi{10.3847/1538-4357/aa978f}

\bibitem[{{Hansen} {et~al.}(2016){Hansen}, {Andersen}, {Nordstr{\"o}m},
  {Beers}, {Placco}, {Yoon}, \& {Buchhave}}]{Hansen16}
{Hansen}, T.~T., {Andersen}, J., {Nordstr{\"o}m}, B., {et~al.} 2016, \aap, 588,
  A3, \dodoi{10.1051/0004-6361/201527409}

\bibitem[{{Hansen} {et~al.}(in prep){Hansen}, {Ji}, \& {S5
  Collaboration}}]{Hansenprep}
{Hansen}, T.~T., {Ji}, A.~P., \& {S5 Collaboration}. in prep

\bibitem[{{Hartwig} {et~al.}(2018){Hartwig}, {Yoshida}, {Magg}, {Frebel},
  {Glover}, {G{\'o}mez}, {Griffen}, {Ishigaki}, {Ji}, {Klessen}, {O'Shea}, \&
  {Tominaga}}]{Hartwig18}
{Hartwig}, T., {Yoshida}, N., {Magg}, M., {et~al.} 2018, \mnras, 478, 1795,
  \dodoi{10.1093/mnras/sty1176}

\bibitem[{{Hasselquist} {et~al.}(2017){Hasselquist}, {Shetrone}, {Smith},
  {Holtzman}, {McWilliam}, {Fern{\'a}ndez-Trincado}, {Beers}, {Majewski},
  {Nidever}, {Tang}, {Tissera}, {Fern{\'a}ndez Alvar}, {Allende Prieto},
  {Almeida}, {Anguiano}, {Battaglia}, {Carigi}, {Delgado Inglada},
  {Frinchaboy}, {Garc{\'\i}a-Hern{\'a}ndez}, {Geisler}, {Minniti}, {Placco},
  {Schultheis}, {Sobeck}, \& {Villanova}}]{Hasselquist17}
{Hasselquist}, S., {Shetrone}, M., {Smith}, V., {et~al.} 2017, \apj, 845, 162,
  \dodoi{10.3847/1538-4357/aa7ddc}

\bibitem[{{Hayes} {et~al.}(2020){Hayes}, {Majewski}, {Hasselquist}, {Anguiano},
  {Shetrone}, {Law}, {Schiavon}, {Cunha}, {Smith}, {Beaton}, {Price-Whelan},
  {Allende Prieto}, {Battaglia}, {Bizyaev}, {Brownstein}, {Cohen},
  {Frinchaboy}, {Garc{\'\i}a-Hern{\'a}ndez}, {Lacerna}, {Lane},
  {M{\'e}sz{\'a}ros}, {Bidin}, {M{\~{u}}noz}, {Nidever}, {Oravetz}, {Oravetz},
  {Pan}, {Roman-Lopes}, {Sobeck}, \& {Stringfellow}}]{Hayes20}
{Hayes}, C.~R., {Majewski}, S.~R., {Hasselquist}, S., {et~al.} 2020, \apj, 889,
  63, \dodoi{10.3847/1538-4357/ab62ad}

\bibitem[{{Helmi}(2020)}]{Helmi20}
{Helmi}, A. 2020, arXiv e-prints, arXiv:2002.04340.
\newblock \doarXiv{2002.04340}

\bibitem[{{Helmi} {et~al.}(2018){Helmi}, {Babusiaux}, {Koppelman}, {Massari},
  {Veljanoski}, \& {Brown}}]{Helmi18}
{Helmi}, A., {Babusiaux}, C., {Koppelman}, H.~H., {et~al.} 2018, \nat, 563, 85,
  \dodoi{10.1038/s41586-018-0625-x}

\bibitem[{{Helmi} {et~al.}(1999){Helmi}, {White}, {de Zeeuw}, \&
  {Zhao}}]{Helmi99}
{Helmi}, A., {White}, S. D.~M., {de Zeeuw}, P.~T., \& {Zhao}, H. 1999, \nat,
  402, 53, \dodoi{10.1038/46980}

\bibitem[{{Hendricks} {et~al.}(2016){Hendricks}, {Boeche}, {Johnson}, {Frank},
  {Koch}, {Mateo}, \& {Bailey}}]{Hendricks16}
{Hendricks}, B., {Boeche}, C., {Johnson}, C.~I., {et~al.} 2016, \aap, 585, A86,
  \dodoi{10.1051/0004-6361/201526996}

\bibitem[{{Hill} {et~al.}(2019){Hill}, {Sk{\'u}lad{\'o}ttir}, {Tolstoy},
  {Venn}, {Shetrone}, {Jablonka}, {Primas}, {Battaglia}, {de Boer},
  {Fran{\c{c}}ois}, {Helmi}, {Kaufer}, {Letarte}, {Starkenburg}, \&
  {Spite}}]{Hill19}
{Hill}, V., {Sk{\'u}lad{\'o}ttir}, {\'A}., {Tolstoy}, E., {et~al.} 2019, \aap,
  626, A15, \dodoi{10.1051/0004-6361/201833950}

\bibitem[{Hunter(2007)}]{matplotlib}
Hunter, J.~D. 2007, Computing in Science \& Engineering, 9, 90,
  \dodoi{http://dx.doi.org/10.1109/MCSE.2007.55}

\bibitem[{{Husser} {et~al.}(2013){Husser}, {Wende-von Berg}, {Dreizler},
  {Homeier}, {Reiners}, {Barman}, \& {Hauschildt}}]{Husser13}
{Husser}, T.~O., {Wende-von Berg}, S., {Dreizler}, S., {et~al.} 2013, \aap,
  553, A6, \dodoi{10.1051/0004-6361/201219058}

\bibitem[{{Ibata} {et~al.}(2019){Ibata}, {Malhan}, \& {Martin}}]{Ibata19}
{Ibata}, R.~A., {Malhan}, K., \& {Martin}, N.~F. 2019, \apj, 872, 152,
  \dodoi{10.3847/1538-4357/ab0080}

\bibitem[{{Ishigaki} {et~al.}(2014){Ishigaki}, {Aoki}, {Arimoto}, \&
  {Okamoto}}]{Ishigaki14}
{Ishigaki}, M.~N., {Aoki}, W., {Arimoto}, N., \& {Okamoto}, S. 2014, \aap, 562,
  A146, \dodoi{10.1051/0004-6361/201322796}

\bibitem[{{Jablonka} {et~al.}(2015){Jablonka}, {North}, {Mashonkina}, {Hill},
  {Revaz}, {Shetrone}, {Starkenburg}, {Irwin}, {Tolstoy}, {Battaglia}, {Venn},
  {Helmi}, {Primas}, \& {Fran{\c c}ois}}]{Jablonka15}
{Jablonka}, P., {North}, P., {Mashonkina}, L., {et~al.} 2015, \aap, 583, A67,
  \dodoi{10.1051/0004-6361/201525661}

\bibitem[{{Jahandar} {et~al.}(2017){Jahandar}, {Venn}, {Shetrone}, {Irwin},
  {Bovy}, {Sakari}, {Kielty}, {Digby}, \& {Frinchaboy}}]{Jahandar17}
{Jahandar}, F., {Venn}, K.~A., {Shetrone}, M.~D., {et~al.} 2017, \mnras, 470,
  4782, \dodoi{10.1093/mnras/stx1592}

\bibitem[{{Ji} {et~al.}(2016){Ji}, {Frebel}, {Simon}, \& {Chiti}}]{Ji16c}
{Ji}, A.~P., {Frebel}, A., {Simon}, J.~D., \& {Chiti}, A. 2016, \apj, 830, 93,
  \dodoi{10.3847/0004-637X/830/2/93}

\bibitem[{{Ji} {et~al.}(2020){Ji}, {Li}, {Simon}, {Marshall}, {Vivas}, {Pace},
  {Bechtol}, {Drlica-Wagner}, {Koposov}, {Hansen}, {Allam}, {Gruendl},
  {Johnson}, {McNanna}, {No{\"e}l}, {Tucker}, \& {Walker}}]{Ji20}
{Ji}, A.~P., {Li}, T.~S., {Simon}, J.~D., {et~al.} 2020, \apj, 889, 27,
  \dodoi{10.3847/1538-4357/ab6213}

\bibitem[{{Johnston} {et~al.}(2008){Johnston}, {Bullock}, {Sharma}, {Font},
  {Robertson}, \& {Leitner}}]{Johnston08}
{Johnston}, K.~V., {Bullock}, J.~S., {Sharma}, S., {et~al.} 2008, \apj, 689,
  936, \dodoi{10.1086/592228}

\bibitem[{Jones {et~al.}(2001)Jones, Oliphant, Peterson, {et~al.}}]{scipy}
Jones, E., Oliphant, T., Peterson, P., {et~al.} 2001, {SciPy}: Open source
  scientific tools for {Python}.
\newblock \url{http://www.scipy.org/}

\bibitem[{{Keller} {et~al.}(2010){Keller}, {Yong}, \& {Da Costa}}]{Keller10}
{Keller}, S.~C., {Yong}, D., \& {Da Costa}, G.~S. 2010, \apj, 720, 940,
  \dodoi{10.1088/0004-637X/720/1/940}

\bibitem[{{Kelson}(2003)}]{Kelson03}
{Kelson}, D.~D. 2003, \pasp, 115, 688, \dodoi{10.1086/375502}

\bibitem[{{Kemp} {et~al.}(2018){Kemp}, {Casey}, {Miles}, {Norfolk},
  {Lattanzio}, {Karakas}, {Schlaufman}, {Ho}, {Tout}, {Ness}, \& {Ji}}]{Kemp18}
{Kemp}, A.~J., {Casey}, A.~R., {Miles}, M.~T., {et~al.} 2018, \mnras, 480,
  1384, \dodoi{10.1093/mnras/sty1915}

\bibitem[{{Kirby} \& {Cohen}(2012)}]{Kirby12}
{Kirby}, E.~N., \& {Cohen}, J.~G. 2012, \aj, 144, 168,
  \dodoi{10.1088/0004-6256/144/6/168}

\bibitem[{{Kirby} {et~al.}(2011){Kirby}, {Cohen}, {Smith}, {Majewski}, {Sohn},
  \& {Guhathakurta}}]{Kirby11}
{Kirby}, E.~N., {Cohen}, J.~G., {Smith}, G.~H., {et~al.} 2011, \apj, 727, 79,
  \dodoi{10.1088/0004-637X/727/2/79}

\bibitem[{{Kirby} {et~al.}(2009){Kirby}, {Guhathakurta}, {Bolte}, {Sneden}, \&
  {Geha}}]{Kirby09}
{Kirby}, E.~N., {Guhathakurta}, P., {Bolte}, M., {Sneden}, C., \& {Geha}, M.~C.
  2009, \apj, 705, 328, \dodoi{10.1088/0004-637X/705/1/328}

\bibitem[{{Kirby} {et~al.}(2019){Kirby}, {Xie}, {Guo}, {de los Reyes},
  {Bergemann}, {Kovalev}, {Shen}, {Piro}, \& {McWilliam}}]{Kirby19}
{Kirby}, E.~N., {Xie}, J.~L., {Guo}, R., {et~al.} 2019, \apj, 881, 45,
  \dodoi{10.3847/1538-4357/ab2c02}

\bibitem[{{Koposov}(2019)}]{rvspecfit}
{Koposov}, S.~E. 2019, {RVSpecFit: Radial velocity and stellar atmospheric
  parameter fitting}.
\newblock \doeprint{1907.013}

\bibitem[{{Koposov} {et~al.}(2014){Koposov}, {Irwin}, {Belokurov},
  {Gonzalez-Solares}, {Yoldas}, {Lewis}, {Metcalfe}, \& {Shanks}}]{Koposov14}
{Koposov}, S.~E., {Irwin}, M., {Belokurov}, V., {et~al.} 2014, \mnras, 442,
  L85, \dodoi{10.1093/mnrasl/slu060}

\bibitem[{{Koposov} {et~al.}(2011){Koposov}, {Gilmore}, {Walker}, {Belokurov},
  {Evans}, {Fellhauer}, {Gieren}, {Geisler}, {Monaco}, {Norris}, {Okamoto},
  {Pe{\~n}arrubia}, {Wilkinson}, {Wyse}, \& {Zucker}}]{Koposov11}
{Koposov}, S.~E., {Gilmore}, G., {Walker}, M.~G., {et~al.} 2011, \apj, 736,
  146, \dodoi{10.1088/0004-637X/736/2/146}

\bibitem[{{Koposov} {et~al.}(2019){Koposov}, {Belokurov}, {Li}, {Mateu},
  {Erkal}, {Grillmair}, {Hendel}, {Price-Whelan}, {Laporte}, {Hawkins}, {Sohn},
  {del Pino}, {Evans}, {Slater}, {Kallivayalil}, {Navarro}, \& {Orphan Aspen
  Treasury Collaboration}}]{Koposov19}
{Koposov}, S.~E., {Belokurov}, V., {Li}, T.~S., {et~al.} 2019, \mnras, 485,
  4726, \dodoi{10.1093/mnras/stz457}

\bibitem[{{Koposov} {et~al.}(2020){Koposov}, {Boubert}, {Li}, {Erkal}, {Da
  Costa}, {Zucker}, {Ji}, {Kuehn}, {Lewis}, {Mackey}, {Simpson}, {Shipp},
  {Wan}, {Belokurov}, {Bland-Hawthorn}, {Martell}, {Nordlander}, {Pace}, {De
  Silva}, {Wang}, \& {S5 collaboration}}]{Koposov20}
{Koposov}, S.~E., {Boubert}, D., {Li}, T.~S., {et~al.} 2020, \mnras, 491, 2465,
  \dodoi{10.1093/mnras/stz3081}

\bibitem[{{Kos} {et~al.}(2018){Kos}, {Bland-Hawthorn}, {Freeman}, {Buder},
  {Traven}, {De Silva}, {Sharma}, {Asplund}, {Duong}, {Lin}, {Lind}, {Martell},
  {Simpson}, {Stello}, {Zucker}, {Zwitter}, {Anguiano}, {Da Costa}, {D'Orazi},
  {Horner}, {Kafle}, {Lewis}, {Munari}, {Nataf}, {Ness}, {Reid}, {Schlesinger},
  {Ting}, \& {Wyse}}]{Kos18}
{Kos}, J., {Bland-Hawthorn}, J., {Freeman}, K., {et~al.} 2018, \mnras, 473,
  4612, \dodoi{10.1093/mnras/stx2637}

\bibitem[{{Kramida} {et~al.}(2019){Kramida}, {Ralchenko}, {Reader}, \& {NIST
  ASD Team}}]{NIST}
{Kramida}, A., {Ralchenko}, Y., {Reader}, J., \& {NIST ASD Team}. 2019, {NIST
  Atomic Spectra Database (version 5.7.1) (Gaithersburg, MD: National Institute
  of Standards and Technology)}, \dodoi{10.18434/T4W30F}

\bibitem[{{Kruijssen}(2019)}]{Kruijssen19}
{Kruijssen}, J.~M.~D. 2019, \mnras, 486, L20, \dodoi{10.1093/mnrasl/slz052}

\bibitem[{{Kruijssen} {et~al.}(2019){Kruijssen}, {Pfeffer}, {Reina-Campos},
  {Crain}, \& {Bastian}}]{Kruijssen19b}
{Kruijssen}, J.~M.~D., {Pfeffer}, J.~L., {Reina-Campos}, M., {Crain}, R.~A., \&
  {Bastian}, N. 2019, \mnras, 486, 3180, \dodoi{10.1093/mnras/sty1609}

\bibitem[{{Krumholz} {et~al.}(2019){Krumholz}, {McKee}, \& {Bland
  -Hawthorn}}]{Krumholz19}
{Krumholz}, M.~R., {McKee}, C.~F., \& {Bland -Hawthorn}, J. 2019, \araa, 57,
  227, \dodoi{10.1146/annurev-astro-091918-104430}

\bibitem[{{Kurucz} \& {Bell}(1995)}]{Kurucz95}
{Kurucz}, R., \& {Bell}, B. 1995, Atomic Line Data (R.L.~Kurucz and B.~Bell)
  Kurucz CD-ROM No.~23.~Cambridge, Mass.: Smithsonian Astrophysical
  Observatory, 1995., 23

\bibitem[{{Larsen} {et~al.}(2014){Larsen}, {Brodie}, {Grundahl}, \&
  {Strader}}]{Larsen14}
{Larsen}, S.~S., {Brodie}, J.~P., {Grundahl}, F., \& {Strader}, J. 2014, \apj,
  797, 15, \dodoi{10.1088/0004-637X/797/1/15}

\bibitem[{{Lawler} {et~al.}(2001{\natexlab{a}}){Lawler}, {Bonvallet}, \&
  {Sneden}}]{Lawler01La}
{Lawler}, J.~E., {Bonvallet}, G., \& {Sneden}, C. 2001{\natexlab{a}}, \apj,
  556, 452, \dodoi{10.1086/321549}

\bibitem[{Lawler \& Dakin(1989)}]{Lawler89}
Lawler, J.~E., \& Dakin, J.~T. 1989, J. Opt. Soc. Am. B, 6, 1457,
  \dodoi{10.1364/JOSAB.6.001457}

\bibitem[{{Lawler} {et~al.}(2006){Lawler}, {Den Hartog}, {Sneden}, \&
  {Cowan}}]{Lawler06Sm}
{Lawler}, J.~E., {Den Hartog}, E.~A., {Sneden}, C., \& {Cowan}, J.~J. 2006,
  \apjs, 162, 227, \dodoi{10.1086/498213}

\bibitem[{{Lawler} {et~al.}(2013){Lawler}, {Guzman}, {Wood}, {Sneden}, \&
  {Cowan}}]{Lawler13Ti}
{Lawler}, J.~E., {Guzman}, A., {Wood}, M.~P., {Sneden}, C., \& {Cowan}, J.~J.
  2013, \apjs, 205, 11, \dodoi{10.1088/0067-0049/205/2/11}

\bibitem[{{Lawler} {et~al.}(2019){Lawler}, {Hala}, {Sneden}, {Nave}, {Wood}, \&
  {Cowan}}]{Lawler19Sc}
{Lawler}, J.~E., {Hala}, {Sneden}, C., {et~al.} 2019, \apjs, 241, 21,
  \dodoi{10.3847/1538-4365/ab08ef}

\bibitem[{{Lawler} {et~al.}(2015){Lawler}, {Sneden}, \& {Cowan}}]{Lawler15Co}
{Lawler}, J.~E., {Sneden}, C., \& {Cowan}, J.~J. 2015, \apjs, 220, 13,
  \dodoi{10.1088/0067-0049/220/1/13}

\bibitem[{{Lawler} {et~al.}(2009){Lawler}, {Sneden}, {Cowan}, {Ivans}, \& {Den
  Hartog}}]{Lawler09Ce}
{Lawler}, J.~E., {Sneden}, C., {Cowan}, J.~J., {Ivans}, I.~I., \& {Den Hartog},
  E.~A. 2009, \apjs, 182, 51, \dodoi{10.1088/0067-0049/182/1/51}

\bibitem[{{Lawler} {et~al.}(2017){Lawler}, {Sneden}, {Nave}, {Den Hartog},
  {Emraho{\u{g}}lu}, \& {Cowan}}]{Lawler17Cr}
{Lawler}, J.~E., {Sneden}, C., {Nave}, G., {et~al.} 2017, \apjs, 228, 10,
  \dodoi{10.3847/1538-4365/228/1/10}

\bibitem[{{Lawler} {et~al.}(2001{\natexlab{b}}){Lawler}, {Wickliffe}, {den
  Hartog}, \& {Sneden}}]{Lawler01Eu}
{Lawler}, J.~E., {Wickliffe}, M.~E., {den Hartog}, E.~A., \& {Sneden}, C.
  2001{\natexlab{b}}, \apj, 563, 1075, \dodoi{10.1086/323407}

\bibitem[{{Lawler} {et~al.}(2014){Lawler}, {Wood}, {Den Hartog}, {Feigenson},
  {Sneden}, \& {Cowan}}]{Lawler14V}
{Lawler}, J.~E., {Wood}, M.~P., {Den Hartog}, E.~A., {et~al.} 2014, \apjs, 215,
  20, \dodoi{10.1088/0067-0049/215/2/20}

\bibitem[{{Leaman}(2012)}]{Leaman12}
{Leaman}, R. 2012, \aj, 144, 183, \dodoi{10.1088/0004-6256/144/6/183}

\bibitem[{{Lemasle} {et~al.}(2012){Lemasle}, {Hill}, {Tolstoy}, {Venn},
  {Shetrone}, {Irwin}, {de Boer}, {Starkenburg}, \& {Salvadori}}]{Lemasle12}
{Lemasle}, B., {Hill}, V., {Tolstoy}, E., {et~al.} 2012, \aap, 538, A100,
  \dodoi{10.1051/0004-6361/201118132}

\bibitem[{{Letarte} {et~al.}(2010){Letarte}, {Hill}, {Tolstoy}, {Jablonka},
  {Shetrone}, {Venn}, {Spite}, {Irwin}, {Battaglia}, {Helmi}, {Primas},
  {Fran{\c{c}}ois}, {Kaufer}, {Szeifert}, {Arimoto}, \& {Sadakane}}]{Letarte10}
{Letarte}, B., {Hill}, V., {Tolstoy}, E., {et~al.} 2010, \aap, 523, A17,
  \dodoi{10.1051/0004-6361/200913413}

\bibitem[{{Lewis} {et~al.}(2002){Lewis}, {Cannon}, {Taylor}, {Glazebrook},
  {Bailey}, {Baldry}, {Barton}, {Bridges}, {Dalton}, {Farrell}, {Gray},
  {Lankshear}, {McCowage}, {Parry}, {Sharples}, {Shortridge}, {Smith},
  {Stevenson}, {Straede}, {Waller}, {Whittard}, {Wilcox}, \&
  {Willis}}]{Lewis02}
{Lewis}, I.~J., {Cannon}, R.~D., {Taylor}, K., {et~al.} 2002, \mnras, 333, 279,
  \dodoi{10.1046/j.1365-8711.2002.05333.x}

\bibitem[{{Li} {et~al.}(2019){Li}, {Koposov}, {Zucker}, {Lewis}, {Kuehn},
  {Simpson}, {Ji}, {Shipp}, {Mao}, {Geha}, {Pace}, {Mackey}, {Allam}, {Tucker},
  {Da Costa}, {Erkal}, {Simon}, {Mould}, {Martell}, {Wan}, {De Silva},
  {Bechtol}, {Balbinot}, {Belokurov}, {Bland-Hawthorn}, {Casey}, {Cullinane},
  {Drlica-Wagner}, {Sharma}, {Vivas}, {Wechsler}, {Yanny}, \& {S5
  Collaboration}}]{Li19S5}
{Li}, T.~S., {Koposov}, S.~E., {Zucker}, D.~B., {et~al.} 2019, \mnras, 490,
  3508, \dodoi{10.1093/mnras/stz2731}

\bibitem[{{Li} {et~al.}(2020){Li}, {Koposov}, {Erkal}, {Ji}, {Shipp}, {Pace},
  {Hilmi}, {Kuehn}, {Lewis}, {Mackey}, {Simpson}, {Wan}, {Zucker}, {Bland
  -Hawthorn}, {Cullinane}, {Da Costa}, {Drlica-Wagner}, {Hattori}, {Martell},
  \& {Sharma}}]{Liprep}
{Li}, T.~S., {Koposov}, S.~E., {Erkal}, D., {et~al.} 2020, arXiv e-prints,
  arXiv:2006.10763.
\newblock \doarXiv{2006.10763}

\bibitem[{{Lind} {et~al.}(2011){Lind}, {Asplund}, {Barklem}, \&
  {Belyaev}}]{Lind11}
{Lind}, K., {Asplund}, M., {Barklem}, P.~S., \& {Belyaev}, A.~K. 2011, \aap,
  528, A103, \dodoi{10.1051/0004-6361/201016095}

\bibitem[{{Ljung} {et~al.}(2006){Ljung}, {Nilsson}, {Asplund}, \&
  {Johansson}}]{Ljung06Zr}
{Ljung}, G., {Nilsson}, H., {Asplund}, M., \& {Johansson}, S. 2006, \aap, 456,
  1181, \dodoi{10.1051/0004-6361:20065212}

\bibitem[{{Mackereth} \& {Bovy}(2020)}]{Mackereth20}
{Mackereth}, J.~T., \& {Bovy}, J. 2020, \mnras, 492, 3631,
  \dodoi{10.1093/mnras/staa047}

\bibitem[{{Majewski} {et~al.}(2003){Majewski}, {Skrutskie}, {Weinberg}, \&
  {Ostheimer}}]{Majewski03}
{Majewski}, S.~R., {Skrutskie}, M.~F., {Weinberg}, M.~D., \& {Ostheimer}, J.~C.
  2003, \apj, 599, 1082, \dodoi{10.1086/379504}

\bibitem[{{Majewski} {et~al.}(2017){Majewski}, {Schiavon}, {Frinchaboy},
  {Allende Prieto}, {Barkhouser}, {Bizyaev}, {Blank}, {Brunner}, {Burton},
  {Carrera}, {Chojnowski}, {Cunha}, {Epstein}, {Fitzgerald}, {Garc{\'\i}a
  P{\'e}rez}, {Hearty}, {Henderson}, {Holtzman}, {Johnson}, {Lam}, {Lawler},
  {Maseman}, {M{\'e}sz{\'a}ros}, {Nelson}, {Nguyen}, {Nidever}, {Pinsonneault},
  {Shetrone}, {Smee}, {Smith}, {Stolberg}, {Skrutskie}, {Walker}, {Wilson},
  {Zasowski}, {Anders}, {Basu}, {Beland}, {Blanton}, {Bovy}, {Brownstein},
  {Carlberg}, {Chaplin}, {Chiappini}, {Eisenstein}, {Elsworth}, {Feuillet},
  {Fleming}, {Galbraith-Frew}, {Garc{\'\i}a}, {Garc{\'\i}a-Hern{\'a}ndez},
  {Gillespie}, {Girardi}, {Gunn}, {Hasselquist}, {Hayden}, {Hekker}, {Ivans},
  {Kinemuchi}, {Klaene}, {Mahadevan}, {Mathur}, {Mosser}, {Muna}, {Munn},
  {Nichol}, {O'Connell}, {Parejko}, {Robin}, {Rocha-Pinto}, {Schultheis},
  {Serenelli}, {Shane}, {Silva Aguirre}, {Sobeck}, {Thompson}, {Troup},
  {Weinberg}, \& {Zamora}}]{Majewski17}
{Majewski}, S.~R., {Schiavon}, R.~P., {Frinchaboy}, P.~M., {et~al.} 2017, \aj,
  154, 94, \dodoi{10.3847/1538-3881/aa784d}

\bibitem[{{Marino} {et~al.}(2008){Marino}, {Villanova}, {Piotto}, {Milone},
  {Momany}, {Bedin}, \& {Medling}}]{Marino08}
{Marino}, A.~F., {Villanova}, S., {Piotto}, G., {et~al.} 2008, \aap, 490, 625,
  \dodoi{10.1051/0004-6361:200810389}

\bibitem[{{Marshall} {et~al.}(2019){Marshall}, {Hansen}, {Simon}, {Li},
  {Bernstein}, {Kuehn}, {Pace}, {DePoy}, {Palmese}, {Pieres}, {Strigari},
  {Drlica-Wagner}, {Bechtol}, {Lidman}, {Nagasawa}, {Bertin}, {Brooks},
  {Buckley-Geer}, {Burke}, {Carnero Rosell}, {Carrasco Kind}, {Carretero},
  {Cunha}, {D'Andrea}, {da Costa}, {De Vicente}, {Desai}, {Doel}, {Eifler},
  {Flaugher}, {Fosalba}, {Frieman}, {Garc{\'\i}a-Bellido}, {Gaztanaga},
  {Gerdes}, {Gruendl}, {Gschwend}, {Gutierrez}, {Hartley}, {Hollowood},
  {Honscheid}, {Hoyle}, {James}, {Kuropatkin}, {Maia}, {Menanteau}, {Miller},
  {Miquel}, {Plazas}, {Sanchez}, {Santiago}, {Scarpine}, {Schubnell},
  {Serrano}, {Sevilla-Noarbe}, {Smith}, {Soares-Santos}, {Suchyta}, {Swanson},
  {Tarle}, {Wester}, \& {DES Collaboration}}]{Marshall19}
{Marshall}, J.~L., {Hansen}, T., {Simon}, J.~D., {et~al.} 2019, \apj, 882, 177,
  \dodoi{10.3847/1538-4357/ab3653}

\bibitem[{{Mashonkina} \& {Belyaev}(2019)}]{Mashonkina19}
{Mashonkina}, L.~I., \& {Belyaev}, A.~K. 2019, Astronomy Letters, 45, 341,
  \dodoi{10.1134/S1063773719060033}

\bibitem[{{Mashonkina} {et~al.}(2016){Mashonkina}, {Sitnova}, \&
  {Pakhomov}}]{Mashonkina16}
{Mashonkina}, L.~I., {Sitnova}, T.~N., \& {Pakhomov}, Y.~V. 2016, Astronomy
  Letters, 42, 606, \dodoi{10.1134/S1063773716080028}

\bibitem[{{Masseron} {et~al.}(2014){Masseron}, {Plez}, {Van Eck}, {Colin},
  {Daoutidis}, {Godefroid}, {Coheur}, {Bernath}, {Jorissen}, \&
  {Christlieb}}]{Masseron14}
{Masseron}, T., {Plez}, B., {Van Eck}, S., {et~al.} 2014, \aap, 571, A47,
  \dodoi{10.1051/0004-6361/201423956}

\bibitem[{{Mateu} {et~al.}(2018){Mateu}, {Read}, \& {Kawata}}]{Mateu18}
{Mateu}, C., {Read}, J.~I., \& {Kawata}, D. 2018, \mnras, 474, 4112,
  \dodoi{10.1093/mnras/stx2937}

\bibitem[{{Matsuno} {et~al.}(2019){Matsuno}, {Aoki}, \& {Suda}}]{Matsuno19}
{Matsuno}, T., {Aoki}, W., \& {Suda}, T. 2019, \apjl, 874, L35,
  \dodoi{10.3847/2041-8213/ab0ec0}

\bibitem[{{Matteucci} \& {Brocato}(1990)}]{Matteucci90}
{Matteucci}, F., \& {Brocato}, E. 1990, \apj, 365, 539, \dodoi{10.1086/169508}

\bibitem[{McKinney(2010)}]{pandas}
McKinney, W. 2010, in Python in {{Science Conference}}, {Austin, Texas},
  56--61, \dodoi{10.25080/Majora-92bf1922-00a}

\bibitem[{{McWilliam}(1998)}]{McWilliam98}
{McWilliam}, A. 1998, \aj, 115, 1640, \dodoi{10.1086/300289}

\bibitem[{{McWilliam} {et~al.}(2018){McWilliam}, {Piro}, {Badenes}, \&
  {Bravo}}]{McWilliam18}
{McWilliam}, A., {Piro}, A.~L., {Badenes}, C., \& {Bravo}, E. 2018, \apj, 857,
  97, \dodoi{10.3847/1538-4357/aab772}

\bibitem[{{McWilliam} {et~al.}(1995){McWilliam}, {Preston}, {Sneden}, \&
  {Searle}}]{McWilliam95}
{McWilliam}, A., {Preston}, G.~W., {Sneden}, C., \& {Searle}, L. 1995, \aj,
  109, 2757, \dodoi{10.1086/117486}

\bibitem[{{McWilliam} {et~al.}(2013){McWilliam}, {Wallerstein}, \&
  {Mottini}}]{McWilliam13}
{McWilliam}, A., {Wallerstein}, G., \& {Mottini}, M. 2013, \apj, 778, 149,
  \dodoi{10.1088/0004-637X/778/2/149}

\bibitem[{{Mel{\'e}ndez} \& {Barbuy}(2009)}]{Melendez09Fe2}
{Mel{\'e}ndez}, J., \& {Barbuy}, B. 2009, \aap, 497, 611,
  \dodoi{10.1051/0004-6361/200811508}

\bibitem[{{Monaco} {et~al.}(2007){Monaco}, {Bellazzini}, {Bonifacio},
  {Buzzoni}, {Ferraro}, {Marconi}, {Sbordone}, \& {Zaggia}}]{Monaco07}
{Monaco}, L., {Bellazzini}, M., {Bonifacio}, P., {et~al.} 2007, \aap, 464, 201,
  \dodoi{10.1051/0004-6361:20066228}

\bibitem[{{Mu{\~n}oz} {et~al.}(2018){Mu{\~n}oz}, {C{\^o}t{\'e}}, {Santana},
  {Geha}, {Simon}, {Oyarz{\'u}n}, {Stetson}, \& {Djorgovski}}]{Munoz18}
{Mu{\~n}oz}, R.~R., {C{\^o}t{\'e}}, P., {Santana}, F.~A., {et~al.} 2018, \apj,
  860, 66, \dodoi{10.3847/1538-4357/aac16b}

\bibitem[{{Mucciarelli} {et~al.}(2012){Mucciarelli}, {Bellazzini}, {Ibata},
  {Merle}, {Chapman}, {Dalessandro}, \& {Sollima}}]{Mucciarelli12}
{Mucciarelli}, A., {Bellazzini}, M., {Ibata}, R., {et~al.} 2012, \mnras, 426,
  2889, \dodoi{10.1111/j.1365-2966.2012.21847.x}

\bibitem[{{Myeong} {et~al.}(2019){Myeong}, {Vasiliev}, {Iorio}, {Evans}, \&
  {Belokurov}}]{Myeong19}
{Myeong}, G.~C., {Vasiliev}, E., {Iorio}, G., {Evans}, N.~W., \& {Belokurov},
  V. 2019, \mnras, 488, 1235, \dodoi{10.1093/mnras/stz1770}

\bibitem[{{Naidu} {et~al.}(2020){Naidu}, {Conroy}, {Bonaca}, {Johnson}, {Ting},
  {Caldwell}, {Zaritsky}, \& {Cargile}}]{Naidu20}
{Naidu}, R.~P., {Conroy}, C., {Bonaca}, A., {et~al.} 2020, arXiv e-prints,
  arXiv:2006.08625.
\newblock \doarXiv{2006.08625}

\bibitem[{{Nidever} {et~al.}(2015){Nidever}, {Holtzman}, {Allende Prieto},
  {Beland}, {Bender}, {Bizyaev}, {Burton}, {Desphande}, {Fleming}, {Garc{\'\i}a
  P{\'e}rez}, {Hearty}, {Majewski}, {M{\'e}sz{\'a}ros}, {Muna}, {Nguyen},
  {Schiavon}, {Shetrone}, {Skrutskie}, {Sobeck}, \& {Wilson}}]{Nidever15}
{Nidever}, D.~L., {Holtzman}, J.~A., {Allende Prieto}, C., {et~al.} 2015, \aj,
  150, 173, \dodoi{10.1088/0004-6256/150/6/173}

\bibitem[{{Nordlander} \& {Lind}(2017)}]{Nordlander17}
{Nordlander}, T., \& {Lind}, K. 2017, \aap, 607, A75,
  \dodoi{10.1051/0004-6361/201730427}

\bibitem[{{Norris} {et~al.}(2010{\natexlab{a}}){Norris}, {Wyse}, {Gilmore},
  {Yong}, {Frebel}, {Wilkinson}, {Belokurov}, \& {Zucker}}]{Norris10a}
{Norris}, J.~E., {Wyse}, R.~F.~G., {Gilmore}, G., {et~al.} 2010{\natexlab{a}},
  \apj, 723, 1632, \dodoi{10.1088/0004-637X/723/2/1632}

\bibitem[{{Norris} {et~al.}(2010{\natexlab{b}}){Norris}, {Yong}, {Gilmore}, \&
  {Wyse}}]{Norris10b}
{Norris}, J.~E., {Yong}, D., {Gilmore}, G., \& {Wyse}, R.~F.~G.
  2010{\natexlab{b}}, \apj, 711, 350, \dodoi{10.1088/0004-637X/711/1/350}

\bibitem[{{Norris} {et~al.}(2017){Norris}, {Yong}, {Venn}, {Gilmore},
  {Casagrande}, \& {Dotter}}]{Norris17b}
{Norris}, J.~E., {Yong}, D., {Venn}, K.~A., {et~al.} 2017, \apjs, 230, 28,
  \dodoi{10.3847/1538-4365/aa755e}

\bibitem[{{O'Brian} {et~al.}(1991){O'Brian}, {Wickliffe}, {Lawler}, {Whaling},
  \& {Brault}}]{Obrian91Fe}
{O'Brian}, T.~R., {Wickliffe}, M.~E., {Lawler}, J.~E., {Whaling}, W., \&
  {Brault}, J.~W. 1991, Journal of the Optical Society of America B Optical
  Physics, 8, 1185, \dodoi{10.1364/JOSAB.8.001185}

\bibitem[{{Pace} {et~al.}(in prep){Pace}, {Li}, \& {S5
  Collaboration}}]{Paceprep}
{Pace}, A., {Li}, T., \& {S5 Collaboration}. in prep

\bibitem[{{Placco} {et~al.}(2014){Placco}, {Frebel}, {Beers}, \&
  {Stancliffe}}]{Placco14}
{Placco}, V.~M., {Frebel}, A., {Beers}, T.~C., \& {Stancliffe}, R.~J. 2014,
  \apj, 797, 21, \dodoi{10.1088/0004-637X/797/1/21}

\bibitem[{{Price-Whelan} {et~al.}(2018){Price-Whelan}, {Sip{\H{o}}cz},
  {G{\"u}nther}, {Lim}, {Crawford}, {Conseil}, {Shupe}, {Craig}, {Dencheva},
  {Ginsburg}, {VanderPlas}, {Bradley}, {P{\'e}rez-Su{\'a}rez}, {de Val-Borro},
  {Paper Contributors}, {Aldcroft}, {Cruz}, {Robitaille}, {Tollerud},
  {Coordination Committee}, {Ardelean}, {Babej}, {Bach}, {Bachetti}, {Bakanov},
  {Bamford}, {Barentsen}, {Barmby}, {Baumbach}, {Berry}, {Biscani}, {Boquien},
  {Bostroem}, {Bouma}, {Brammer}, {Bray}, {Breytenbach}, {Buddelmeijer},
  {Burke}, {Calderone}, {Cano Rodr{\'\i}guez}, {Cara}, {Cardoso}, {Cheedella},
  {Copin}, {Corrales}, {Crichton}, {D{\textquoteright}Avella}, {Deil},
  {Depagne}, {Dietrich}, {Donath}, {Droettboom}, {Earl}, {Erben}, {Fabbro},
  {Ferreira}, {Finethy}, {Fox}, {Garrison}, {Gibbons}, {Goldstein}, {Gommers},
  {Greco}, {Greenfield}, {Groener}, {Grollier}, {Hagen}, {Hirst}, {Homeier},
  {Horton}, {Hosseinzadeh}, {Hu}, {Hunkeler}, {Ivezi{\'c}}, {Jain}, {Jenness},
  {Kanarek}, {Kendrew}, {Kern}, {Kerzendorf}, {Khvalko}, {King}, {Kirkby},
  {Kulkarni}, {Kumar}, {Lee}, {Lenz}, {Littlefair}, {Ma}, {Macleod},
  {Mastropietro}, {McCully}, {Montagnac}, {Morris}, {Mueller}, {Mumford},
  {Muna}, {Murphy}, {Nelson}, {Nguyen}, {Ninan}, {N{\"o}the}, {Ogaz}, {Oh},
  {Parejko}, {Parley}, {Pascual}, {Patil}, {Patil}, {Plunkett}, {Prochaska},
  {Rastogi}, {Reddy Janga}, {Sabater}, {Sakurikar}, {Seifert}, {Sherbert},
  {Sherwood-Taylor}, {Shih}, {Sick}, {Silbiger}, {Singanamalla}, {Singer},
  {Sladen}, {Sooley}, {Sornarajah}, {Streicher}, {Teuben}, {Thomas},
  {Tremblay}, {Turner}, {Terr{\'o}n}, {van Kerkwijk}, {de la Vega}, {Watkins},
  {Weaver}, {Whitmore}, {Woillez}, {Zabalza}, \& {Contributors}}]{astropy:2018}
{Price-Whelan}, A.~M., {Sip{\H{o}}cz}, B.~M., {G{\"u}nther}, H.~M., {et~al.}
  2018, \aj, 156, 123, \dodoi{10.3847/1538-3881/aabc4f}

\bibitem[{Pritzl {et~al.}(2005)Pritzl, Venn, \& Irwin}]{Pritzl05}
Pritzl, B.~J., Venn, K.~A., \& Irwin, M. 2005, \aj, 130, 2140,
  \dodoi{10.1086/432911}

\bibitem[{{Reggiani} {et~al.}(2019){Reggiani}, {Amarsi}, {Lind}, {Barklem},
  {Zatsarinny}, {Bartschat}, {Fursa}, {Bray}, {Spina}, \&
  {Mel{\'e}ndez}}]{Reggiani19}
{Reggiani}, H., {Amarsi}, A.~M., {Lind}, K., {et~al.} 2019, \aap, 627, A177,
  \dodoi{10.1051/0004-6361/201935156}

\bibitem[{{Rocha} {et~al.}(2012){Rocha}, {Peter}, \& {Bullock}}]{Rocha12}
{Rocha}, M., {Peter}, A. H.~G., \& {Bullock}, J. 2012, \mnras, 425, 231,
  \dodoi{10.1111/j.1365-2966.2012.21432.x}

\bibitem[{{Roederer} \& {Gnedin}(2019)}]{Roederer19}
{Roederer}, I.~U., \& {Gnedin}, O.~Y. 2019, \apj, 883, 84,
  \dodoi{10.3847/1538-4357/ab365c}

\bibitem[{{Roederer} \& {Lawler}(2012)}]{Roederer12Zn}
{Roederer}, I.~U., \& {Lawler}, J.~E. 2012, \apj, 750, 76,
  \dodoi{10.1088/0004-637X/750/1/76}

\bibitem[{{Roederer} {et~al.}(2014){Roederer}, {Preston}, {Thompson},
  {Shectman}, {Sneden}, {Burley}, \& {Kelson}}]{Roederer14c}
{Roederer}, I.~U., {Preston}, G.~W., {Thompson}, I.~B., {et~al.} 2014, \aj,
  147, 136, \dodoi{10.1088/0004-6256/147/6/136}

\bibitem[{{Roederer} {et~al.}(2010){Roederer}, {Sneden}, {Thompson}, {Preston},
  \& {Shectman}}]{Roederer10}
{Roederer}, I.~U., {Sneden}, C., {Thompson}, I.~B., {Preston}, G.~W., \&
  {Shectman}, S.~A. 2010, \apj, 711, 573, \dodoi{10.1088/0004-637X/711/2/573}

\bibitem[{{Roederer} {et~al.}(2016){Roederer}, {Mateo}, {Bailey}, {Song},
  {Bell}, {Crane}, {Loebman}, {Nidever}, {Olszewski}, {Shectman}, {Thompson},
  {Valluri}, \& {Walker}}]{Roederer16b}
{Roederer}, I.~U., {Mateo}, M., {Bailey}, III, J.~I., {et~al.} 2016, \aj, 151,
  82, \dodoi{10.3847/0004-6256/151/3/82}

\bibitem[{{Roediger} {et~al.}(2014){Roediger}, {Courteau}, {Graves}, \&
  {Schiavon}}]{Roediger14}
{Roediger}, J.~C., {Courteau}, S., {Graves}, G., \& {Schiavon}, R.~P. 2014,
  \apjs, 210, 10, \dodoi{10.1088/0067-0049/210/1/10}

\bibitem[{{Ruffoni} {et~al.}(2014){Ruffoni}, {Den Hartog}, {Lawler}, {Brewer},
  {Lind}, {Nave}, \& {Pickering}}]{Ruffoni14Fe}
{Ruffoni}, M.~P., {Den Hartog}, E.~A., {Lawler}, J.~E., {et~al.} 2014, \mnras,
  441, 3127, \dodoi{10.1093/mnras/stu780}

\bibitem[{{Ryabchikova} {et~al.}(2015){Ryabchikova}, {Piskunov}, {Kurucz},
  {Stempels}, {Heiter}, {Pakhomov}, \& {Barklem}}]{VALD}
{Ryabchikova}, T., {Piskunov}, N., {Kurucz}, R.~L., {et~al.} 2015, \physscr,
  90, 054005, \dodoi{10.1088/0031-8949/90/5/054005}

\bibitem[{{Sakari} {et~al.}(2018){Sakari}, {Placco}, {Hansen}, {Holmbeck},
  {Beers}, {Frebel}, {Roederer}, {Venn}, {Wallerstein}, {Davis}, {Farrell}, \&
  {Yong}}]{Sakari18}
{Sakari}, C.~M., {Placco}, V.~M., {Hansen}, T., {et~al.} 2018, \apjl, 854, L20,
  \dodoi{10.3847/2041-8213/aaa9b4}

\bibitem[{{Schlegel} {et~al.}(1998){Schlegel}, {Finkbeiner}, \&
  {Davis}}]{SFD98}
{Schlegel}, D.~J., {Finkbeiner}, D.~P., \& {Davis}, M. 1998, \apj, 500, 525,
  \dodoi{10.1086/305772}

\bibitem[{Sharp {et~al.}(2006)Sharp, Saunders, Smith, Churilov, Correll,
  Dawson, Farrel, Frost, Haynes, Heald, Lankshear, Mayfield, Waller, \&
  Whittard}]{Sharp06}
Sharp, R., Saunders, W., Smith, G., {et~al.} 2006, in {{SPIE Astronomical
  Telescopes}} + {{Instrumentation}}, ed. I.~S. McLean \& M.~Iye, {Orlando,
  Florida , USA}, 62690G, \dodoi{10.1117/12.671022}

\bibitem[{{Shetrone} {et~al.}(2003){Shetrone}, {Venn}, {Tolstoy}, {Primas},
  {Hill}, \& {Kaufer}}]{Shetrone03}
{Shetrone}, M., {Venn}, K.~A., {Tolstoy}, E., {et~al.} 2003, \aj, 125, 684,
  \dodoi{10.1086/345966}

\bibitem[{{Shetrone} {et~al.}(2015){Shetrone}, {Bizyaev}, {Lawler}, {Allende
  Prieto}, {Johnson}, {Smith}, {Cunha}, {Holtzman}, {Garc{\'\i}a P{\'e}rez},
  {M{\'e}sz{\'a}ros}, {Sobeck}, {Zamora}, {Garc{\'\i}a-Hern{\'a}ndez}, {Souto},
  {Chojnowski}, {Koesterke}, {Majewski}, \& {Zasowski}}]{Shetrone15}
{Shetrone}, M., {Bizyaev}, D., {Lawler}, J.~E., {et~al.} 2015, \apjs, 221, 24,
  \dodoi{10.1088/0067-0049/221/2/24}

\bibitem[{{Shetrone} {et~al.}(2001){Shetrone}, {C{\^o}t{\'e}}, \&
  {Sargent}}]{Shetrone01}
{Shetrone}, M.~D., {C{\^o}t{\'e}}, P., \& {Sargent}, W.~L.~W. 2001, \apj, 548,
  592, \dodoi{10.1086/319022}

\bibitem[{{Shipp} {et~al.}(2018){Shipp}, {Drlica-Wagner}, {Balbinot},
  {Ferguson}, {Erkal}, {Li}, {Bechtol}, {Belokurov}, {Buncher}, {Carollo},
  {Carrasco Kind}, {Kuehn}, {Marshall}, {Pace}, {Rykoff}, {Sevilla-Noarbe},
  {Sheldon}, {Strigari}, {Vivas}, {Yanny}, {Zenteno}, {Abbott}, {Abdalla},
  {Allam}, {Avila}, {Bertin}, {Brooks}, {Burke}, {Carretero}, {Castander},
  {Cawthon}, {Crocce}, {Cunha}, {D'Andrea}, {da Costa}, {Davis}, {De Vicente},
  {Desai}, {Diehl}, {Doel}, {Evrard}, {Flaugher}, {Fosalba}, {Frieman},
  {Garc{\'{\i}}a-Bellido}, {Gaztanaga}, {Gerdes}, {Gruen}, {Gruendl},
  {Gschwend}, {Gutierrez}, {Hoyle}, {James}, {Johnson}, {Krause}, {Kuropatkin},
  {Lahav}, {Lin}, {Maia}, {March}, {Martini}, {Menanteau}, {Miller}, {Miquel},
  {Nichol}, {Plazas}, {Romer}, {Sako}, {Sanchez}, {Scarpine}, {Schindler},
  {Schubnell}, {Smith}, {Smith}, {Sobreira}, {Suchyta}, {Swanson}, {Tarle},
  {Thomas}, {Tucker}, {Walker}, {Wechsler}, \& {the DES
  Collaboration}}]{Shipp18}
{Shipp}, N., {Drlica-Wagner}, A., {Balbinot}, E., {et~al.} 2018, ArXiv
  e-prints.
\newblock \doarXiv{1801.03097}

\bibitem[{{Shipp} {et~al.}(2019){Shipp}, {Li}, {Pace}, {Erkal},
  {Drlica-Wagner}, {Yanny}, {Belokurov}, {Wester}, {Koposov}, {Kuehn}, {Lewis},
  {Simpson}, {Wan}, {Zucker}, {Martell}, {Wang}, \& {S<SUP>5</SUP>
  Collaboration}}]{Shipp19}
{Shipp}, N., {Li}, T.~S., {Pace}, A.~B., {et~al.} 2019, \apj, 885, 3,
  \dodoi{10.3847/1538-4357/ab44bf}

\bibitem[{{Simmerer} {et~al.}(2004){Simmerer}, {Sneden}, {Cowan}, {Collier},
  {Woolf}, \& {Lawler}}]{Simmerer04}
{Simmerer}, J., {Sneden}, C., {Cowan}, J.~J., {et~al.} 2004, \apj, 617, 1091,
  \dodoi{10.1086/424504}

\bibitem[{{Simon}(2019)}]{Simon19}
{Simon}, J.~D. 2019, \araa, 57, 375,
  \dodoi{10.1146/annurev-astro-091918-104453}

\bibitem[{{Simon} {et~al.}(2015){Simon}, {Jacobson}, {Frebel}, {Thompson},
  {Adams}, \& {Shectman}}]{Simon15Scl}
{Simon}, J.~D., {Jacobson}, H.~R., {Frebel}, A., {et~al.} 2015, \apj, 802, 93,
  \dodoi{10.1088/0004-637X/802/2/93}

\bibitem[{{Simpson} {et~al.}(2020){Simpson}, {Martell}, {Da Costa}, {Horner},
  {Wyse}, {Ting}, {Asplund}, {Bland-Hawthorn}, {Buder}, {De Silva}, {Freeman},
  {Kos}, {Lewis}, {Lind}, {Sharma}, {Zucker}, {Zwitter}, {{\v{C}}otar},
  {Cottrell}, \& {Nordlander}}]{Simpson20}
{Simpson}, J.~D., {Martell}, S.~L., {Da Costa}, G., {et~al.} 2020, \mnras, 491,
  3374, \dodoi{10.1093/mnras/stz3105}

\bibitem[{{Sk{\'u}lad{\'o}ttir} {et~al.}(2015){Sk{\'u}lad{\'o}ttir}, {Tolstoy},
  {Salvadori}, {Hill}, {Pettini}, {Shetrone}, \& {Starkenburg}}]{Skuladottir15}
{Sk{\'u}lad{\'o}ttir}, {\'A}., {Tolstoy}, E., {Salvadori}, S., {et~al.} 2015,
  \aap, 574, A129, \dodoi{10.1051/0004-6361/201424782}

\bibitem[{{Sneden} {et~al.}(2008){Sneden}, {Cowan}, \& {Gallino}}]{Sneden08}
{Sneden}, C., {Cowan}, J.~J., \& {Gallino}, R. 2008, \araa, 46, 241,
  \dodoi{10.1146/annurev.astro.46.060407.145207}

\bibitem[{{Sneden} {et~al.}(2016){Sneden}, {Cowan}, {Kobayashi}, {Pignatari},
  {Lawler}, {Den Hartog}, \& {Wood}}]{Sneden16}
{Sneden}, C., {Cowan}, J.~J., {Kobayashi}, C., {et~al.} 2016, \apj, 817, 53,
  \dodoi{10.3847/0004-637X/817/1/53}

\bibitem[{{Sneden} {et~al.}(2009){Sneden}, {Lawler}, {Cowan}, {Ivans}, \& {Den
  Hartog}}]{Sneden09}
{Sneden}, C., {Lawler}, J.~E., {Cowan}, J.~J., {Ivans}, I.~I., \& {Den Hartog},
  E.~A. 2009, \apjs, 182, 80, \dodoi{10.1088/0067-0049/182/1/80}

\bibitem[{{Sneden} {et~al.}(2014){Sneden}, {Lucatello}, {Ram}, {Brooke}, \&
  {Bernath}}]{Sneden14}
{Sneden}, C., {Lucatello}, S., {Ram}, R.~S., {Brooke}, J.~S.~A., \& {Bernath},
  P. 2014, \apjs, 214, 26, \dodoi{10.1088/0067-0049/214/2/26}

\bibitem[{{Sneden}(1973)}]{Sneden73}
{Sneden}, C.~A. 1973, PhD thesis, The University of Texas at Austin.

\bibitem[{{Sobeck} {et~al.}(2007){Sobeck}, {Lawler}, \& {Sneden}}]{Sobeck07Cr}
{Sobeck}, J.~S., {Lawler}, J.~E., \& {Sneden}, C. 2007, \apj, 667, 1267,
  \dodoi{10.1086/519987}

\bibitem[{{Sobeck} {et~al.}(2011){Sobeck}, {Kraft}, {Sneden}, {Preston},
  {Cowan}, {Smith}, {Thompson}, {Shectman}, \& {Burley}}]{Sobeck11}
{Sobeck}, J.~S., {Kraft}, R.~P., {Sneden}, C., {et~al.} 2011, \aj, 141, 175,
  \dodoi{10.1088/0004-6256/141/6/175}

\bibitem[{{Stan Development Team}(2018)}]{pystan}
{Stan Development Team}. 2018, {PyStan: the Python interface to Stan, Version
  2.17.1.0}.
\newblock \url{http://mc-stan.org}

\bibitem[{{Stoughton} {et~al.}(2002){Stoughton}, {Lupton}, {Bernardi},
  {Blanton}, {Burles}, {Castand er}, {Connolly}, {Eisenstein}, {Frieman},
  {Hennessy}, {Hindsley}, {Ivezi{\'c}}, {Kent}, {Kunszt}, {Lee}, {Meiksin},
  {Munn}, {Newberg}, {Nichol}, {Nicinski}, {Pier}, {Richards}, {Richmond},
  {Schlegel}, {Smith}, {Strauss}, {SubbaRao}, {Szalay}, {Thakar}, {Tucker},
  {Vand en Berk}, {Yanny}, {Adelman}, {Anderson}, {Anderson}, {Annis},
  {Bahcall}, {Bakken}, {Bartelmann}, {Bastian}, {Bauer}, {Berman},
  {B{\"o}hringer}, {Boroski}, {Bracker}, {Briegel}, {Briggs}, {Brinkmann},
  {Brunner}, {Carey}, {Carr}, {Chen}, {Christian}, {Colestock}, {Crocker},
  {Csabai}, {Czarapata}, {Dalcanton}, {Davidsen}, {Davis}, {Dehnen},
  {Dodelson}, {Doi}, {Dombeck}, {Donahue}, {Ellman}, {Elms}, {Evans}, {Eyer},
  {Fan}, {Federwitz}, {Friedman}, {Fukugita}, {Gal}, {Gillespie}, {Glazebrook},
  {Gray}, {Grebel}, {Greenawalt}, {Greene}, {Gunn}, {de Haas}, {Haiman},
  {Haldeman}, {Hall}, {Hamabe}, {Hansen}, {Harris}, {Harris}, {Harvanek},
  {Hawley}, {Hayes}, {Heckman}, {Helmi}, {Henden}, {Hogan}, {Hogg}, {Holmgren},
  {Holtzman}, {Huang}, {Hull}, {Ichikawa}, {Ichikawa}, {Johnston}, {Kauffmann},
  {Kim}, {Kimball}, {Kinney}, {Klaene}, {Kleinman}, {Klypin}, {Knapp},
  {Korienek}, {Krolik}, {Kron}, {Krzesi{\'n}ski}, {Lamb}, {Leger},
  {Limmongkol}, {Lindenmeyer}, {Long}, {Loomis}, {Loveday}, {MacKinnon},
  {Mannery}, {Mantsch}, {Margon}, {McGehee}, {McKay}, {McLean}, {Menou},
  {Merelli}, {Mo}, {Monet}, {Nakamura}, {Narayanan}, {Nash}, {Neilsen},
  {Newman}, {Nitta}, {Odenkirchen}, {Okada}, {Okamura}, {Ostriker}, {Owen},
  {Pauls}, {Peoples}, {Peterson}, {Petravick}, {Pope}, {Pordes}, {Postman},
  {Prosapio}, {Quinn}, {Rechenmacher}, {Rivetta}, {Rix}, {Rockosi}, {Rosner},
  {Ruthmansdorfer}, {Sandford}, {Schneider}, {Scranton}, {Sekiguchi}, {Sergey},
  {Sheth}, {Shimasaku}, {Smee}, {Snedden}, {Stebbins}, {Stubbs}, {Szapudi},
  {Szkody}, {Szokoly}, {Tabachnik}, {Tsvetanov}, {Uomoto}, {Vogeley}, {Voges},
  {Waddell}, {Walterbos}, {Wang}, {Watanabe}, {Weinberg}, {White}, {White},
  {Wilhite}, {Wolfe}, {Yasuda}, {York}, {Zehavi}, \& {Zheng}}]{StoughtonSDSS}
{Stoughton}, C., {Lupton}, R.~H., {Bernardi}, M., {et~al.} 2002, \aj, 123, 485,
  \dodoi{10.1086/324741}

\bibitem[{{Tafelmeyer} {et~al.}(2010){Tafelmeyer}, {Jablonka}, {Hill},
  {Shetrone}, {Tolstoy}, {Irwin}, {Battaglia}, {Helmi}, {Starkenburg}, {Venn},
  {Abel}, {Francois}, {Kaufer}, {North}, {Primas}, \&
  {Szeifert}}]{Tafelmeyer10}
{Tafelmeyer}, M., {Jablonka}, P., {Hill}, V., {et~al.} 2010, \aap, 524, A58,
  \dodoi{10.1051/0004-6361/201014733}

\bibitem[{{Tinsley}(1980)}]{Tinsley80}
{Tinsley}, B.~M. 1980, \fcp, 5, 287

\bibitem[{{Tolstoy} {et~al.}(2009){Tolstoy}, {Hill}, \& {Tosi}}]{Tolstoy09}
{Tolstoy}, E., {Hill}, V., \& {Tosi}, M. 2009, \araa, 47, 371,
  \dodoi{10.1146/annurev-astro-082708-101650}

\bibitem[{{Tsujimoto} {et~al.}(2015){Tsujimoto}, {Ishigaki}, {Shigeyama}, \&
  {Aoki}}]{Tsujimoto15a}
{Tsujimoto}, T., {Ishigaki}, M.~N., {Shigeyama}, T., \& {Aoki}, W. 2015, \pasj,
  67, L3, \dodoi{10.1093/pasj/psv035}

\bibitem[{{Tsujimoto} {et~al.}(2017){Tsujimoto}, {Matsuno}, {Aoki}, {Ishigaki},
  \& {Shigeyama}}]{Tsujimoto17}
{Tsujimoto}, T., {Matsuno}, T., {Aoki}, W., {Ishigaki}, M.~N., \& {Shigeyama},
  T. 2017, \apjl, 850, L12, \dodoi{10.3847/2041-8213/aa9886}

\bibitem[{{Ural} {et~al.}(2015){Ural}, {Cescutti}, {Koch}, {Kleyna},
  {Feltzing}, \& {Wilkinson}}]{Ural15}
{Ural}, U., {Cescutti}, G., {Koch}, A., {et~al.} 2015, \mnras, 449, 761,
  \dodoi{10.1093/mnras/stv294}

\bibitem[{{van~der~Walt} {et~al.}(2011){van~der~Walt}, Colbert, \&
  Varoquaux}]{numpy}
{van~der~Walt}, S., Colbert, S.~C., \& Varoquaux, G. 2011, Computing in Science
  \& Engineering, 13, 22, \dodoi{http://dx.doi.org/10.1109/MCSE.2011.37}

\bibitem[{{Venn} {et~al.}(2004){Venn}, {Irwin}, {Shetrone}, {Tout}, {Hill}, \&
  {Tolstoy}}]{Venn04}
{Venn}, K.~A., {Irwin}, M., {Shetrone}, M.~D., {et~al.} 2004, \aj, 128, 1177,
  \dodoi{10.1086/422734}

\bibitem[{{Venn} {et~al.}(2017){Venn}, {Starkenburg}, {Malo}, {Martin}, \&
  {Laevens}}]{Venn17}
{Venn}, K.~A., {Starkenburg}, E., {Malo}, L., {Martin}, N., \& {Laevens},
  B.~P.~M. 2017, \mnras, 466, 3741, \dodoi{10.1093/mnras/stw3198}

\bibitem[{{Venn} {et~al.}(2012){Venn}, {Shetrone}, {Irwin}, {Hill}, {Jablonka},
  {Tolstoy}, {Lemasle}, {Divell}, {Starkenburg}, {Letarte}, {Baldner},
  {Battaglia}, {Helmi}, {Kaufer}, \& {Primas}}]{Venn12}
{Venn}, K.~A., {Shetrone}, M.~D., {Irwin}, M.~J., {et~al.} 2012, \apj, 751,
  102, \dodoi{10.1088/0004-637X/751/2/102}

\bibitem[{Wan {et~al.}(2020)Wan, Lewis, Li, Simpson, Martell, Zucker, Mould,
  Erkal, Pace, Mackey, Ji, Koposov, Kuehn, Shipp, Balbinot, Bland-Hawthorn,
  Casey, Da~Costa, Kafle, Sharma, \& De~Silva}]{Wan20}
Wan, Z., Lewis, G.~F., Li, T.~S., {et~al.} 2020, Nature, 583, 768

\bibitem[{Waskom {et~al.}(2016)Waskom, Botvinnik, O'Kane, Hobson, Halchenko,
  Lukauskas, Warmenhoven, Cole, Hoyer, Vanderplas, gkunter, Villalba, Quintero,
  Martin, Miles, Meyer, Augspurger, Yarkoni, Bachant, Evans, Fitzgerald, Nagy,
  Ziegler, Megies, Wehner, St-Jean, Coelho, Hitz, Lee, \& Rocher}]{seaborn}
Waskom, M., Botvinnik, O., O'Kane, D., {et~al.} 2016, seaborn: v0.7.0 (January
  2016), \dodoi{10.5281/zenodo.45133}

\bibitem[{{Wenger} {et~al.}(2000){Wenger}, {Ochsenbein}, {Egret}, {Dubois},
  {Bonnarel}, {Borde}, {Genova}, {Jasniewicz}, {Lalo{\"e}}, {Lesteven}, \&
  {Monier}}]{Simbad}
{Wenger}, M., {Ochsenbein}, F., {Egret}, D., {et~al.} 2000, \aaps, 143, 9,
  \dodoi{10.1051/aas:2000332}

\bibitem[{{Willman} \& {Strader}(2012)}]{Willman12}
{Willman}, B., \& {Strader}, J. 2012, \aj, 144, 76,
  \dodoi{10.1088/0004-6256/144/3/76}

\bibitem[{{Wilson} {et~al.}(2019){Wilson}, {Hearty}, {Skrutskie}, {Majewski},
  {Holtzman}, {Eisenstein}, {Gunn}, {Blank}, {Henderson}, {Smee}, {Nelson},
  {Nidever}, {Arns}, {Barkhouser}, {Barr}, {Beland}, {Bershady}, {Blanton},
  {Brunner}, {Burton}, {Carey}, {Carr}, {Colque}, {Crane}, {Damke}, {Davidson},
  {Dean}, {Di Mille}, {Don}, {Ebelke}, {Evans}, {Fitzgerald}, {Gillespie},
  {Hall}, {Harding}, {Harding}, {Hammond}, {Hancock}, {Harrison}, {Hope},
  {Horne}, {Karakla}, {Lam}, {Leger}, {MacDonald}, {Maseman}, {Matsunari},
  {Melton}, {Mitcheltree}, {O'Brien}, {O'Connell}, {Patten}, {Richardson},
  {Rieke}, {Rieke}, {Roman-Lopes}, {Schiavon}, {Sobeck}, {Stolberg}, {Stoll},
  {Tembe}, {Trujillo}, {Uomoto}, {Vernieri}, {Walker}, {Weinberg}, {Young},
  {Anthony-Brumfield}, {Bizyaev}, {Breslauer}, {De Lee}, {Downey}, {Halverson},
  {Huehnerhoff}, {Klaene}, {Leon}, {Long}, {Mahadevan}, {Malanushenko},
  {Nguyen}, {Owen}, {S{\'a}nchez-Gallego}, {Sayres}, {Shane}, {Shectman},
  {Shetrone}, {Skinner}, {Stauffer}, \& {Zhao}}]{Wilson19}
{Wilson}, J.~C., {Hearty}, F.~R., {Skrutskie}, M.~F., {et~al.} 2019, \pasp,
  131, 055001, \dodoi{10.1088/1538-3873/ab0075}

\bibitem[{{Wood} {et~al.}(2014{\natexlab{a}}){Wood}, {Lawler}, {Den Hartog},
  {Sneden}, \& {Cowan}}]{Wood14V}
{Wood}, M.~P., {Lawler}, J.~E., {Den Hartog}, E.~A., {Sneden}, C., \& {Cowan},
  J.~J. 2014{\natexlab{a}}, \apjs, 214, 18, \dodoi{10.1088/0067-0049/214/2/18}

\bibitem[{{Wood} {et~al.}(2013){Wood}, {Lawler}, {Sneden}, \&
  {Cowan}}]{Wood13Ti}
{Wood}, M.~P., {Lawler}, J.~E., {Sneden}, C., \& {Cowan}, J.~J. 2013, \apjs,
  208, 27, \dodoi{10.1088/0067-0049/208/2/27}

\bibitem[{{Wood} {et~al.}(2014{\natexlab{b}}){Wood}, {Lawler}, {Sneden}, \&
  {Cowan}}]{Wood14Ni}
---. 2014{\natexlab{b}}, \apjs, 211, 20, \dodoi{10.1088/0067-0049/211/2/20}

\bibitem[{{Yong} {et~al.}(2013){Yong}, {Mel{\'e}ndez}, {Grundahl}, {Roederer},
  {Norris}, {Milone}, {Marino}, {Coelho}, {McArthur}, {Lind}, {Collet}, \&
  {Asplund}}]{Yong13b}
{Yong}, D., {Mel{\'e}ndez}, J., {Grundahl}, F., {et~al.} 2013, \mnras, 434,
  3542, \dodoi{10.1093/mnras/stt1276}

\bibitem[{{York} {et~al.}(2000){York}, {Adelman}, {Anderson}, {Anderson},
  {Annis}, {Bahcall}, {Bakken}, {Barkhouser}, {Bastian}, {Berman}, {Boroski},
  {Bracker}, {Briegel}, {Briggs}, {Brinkmann}, {Brunner}, {Burles}, {Carey},
  {Carr}, {Castander}, {Chen}, {Colestock}, {Connolly}, {Crocker}, {Csabai},
  {Czarapata}, {Davis}, {Doi}, {Dombeck}, {Eisenstein}, {Ellman}, {Elms},
  {Evans}, {Fan}, {Federwitz}, {Fiscelli}, {Friedman}, {Frieman}, {Fukugita},
  {Gillespie}, {Gunn}, {Gurbani}, {de Haas}, {Haldeman}, {Harris}, {Hayes},
  {Heckman}, {Hennessy}, {Hindsley}, {Holm}, {Holmgren}, {Huang}, {Hull},
  {Husby}, {Ichikawa}, {Ichikawa}, {Ivezi{\'c}}, {Kent}, {Kim}, {Kinney},
  {Klaene}, {Kleinman}, {Kleinman}, {Knapp}, {Korienek}, {Kron}, {Kunszt},
  {Lamb}, {Lee}, {Leger}, {Limmongkol}, {Lindenmeyer}, {Long}, {Loomis},
  {Loveday}, {Lucinio}, {Lupton}, {MacKinnon}, {Mannery}, {Mantsch}, {Margon},
  {McGehee}, {McKay}, {Meiksin}, {Merelli}, {Monet}, {Munn}, {Narayanan},
  {Nash}, {Neilsen}, {Neswold}, {Newberg}, {Nichol}, {Nicinski}, {Nonino},
  {Okada}, {Okamura}, {Ostriker}, {Owen}, {Pauls}, {Peoples}, {Peterson},
  {Petravick}, {Pier}, {Pope}, {Pordes}, {Prosapio}, {Rechenmacher}, {Quinn},
  {Richards}, {Richmond}, {Rivetta}, {Rockosi}, {Ruthmansdorfer}, {Sand ford},
  {Schlegel}, {Schneider}, {Sekiguchi}, {Sergey}, {Shimasaku}, {Siegmund},
  {Smee}, {Smith}, {Snedden}, {Stone}, {Stoughton}, {Strauss}, {Stubbs},
  {SubbaRao}, {Szalay}, {Szapudi}, {Szokoly}, {Thakar}, {Tremonti}, {Tucker},
  {Uomoto}, {Vanden Berk}, {Vogeley}, {Waddell}, {Wang}, {Watanabe},
  {Weinberg}, {Yanny}, {Yasuda}, \& {SDSS Collaboration}}]{YorkSDSS}
{York}, D.~G., {Adelman}, J., {Anderson}, John~E., J., {et~al.} 2000, \aj, 120,
  1579, \dodoi{10.1086/301513}

\bibitem[{{Yuan} {et~al.}(2020){Yuan}, {Myeong}, {Beers}, {Evans}, {Lee},
  {Banerjee}, {Gudin}, {Hattori}, {Li}, {Matsuno}, {Placco}, {Smith},
  {Whitten}, \& {Zhao}}]{Yuan20}
{Yuan}, Z., {Myeong}, G.~C., {Beers}, T.~C., {et~al.} 2020, \apj, 891, 39,
  \dodoi{10.3847/1538-4357/ab6ef7}

\bibitem[{{Zasowski} {et~al.}(2017){Zasowski}, {Cohen}, {Chojnowski},
  {Santana}, {Oelkers}, {Andrews}, {Beaton}, {Bender}, {Bird}, {Bovy},
  {Carlberg}, {Covey}, {Cunha}, {Dell'Agli}, {Fleming}, {Frinchaboy},
  {Garc{\'\i}a-Hern{\'a}ndez}, {Harding}, {Holtzman}, {Johnson}, {Kollmeier},
  {Majewski}, {M{\'e}sz{\'a}ros}, {Munn}, {Mu{\~n}oz}, {Ness}, {Nidever},
  {Poleski}, {Rom{\'a}n-Z{\'u}{\~n}iga}, {Shetrone}, {Simon}, {Smith},
  {Sobeck}, {Stringfellow}, {Szigeti{\'a}ros}, {Tayar}, \&
  {Troup}}]{Zakowski17}
{Zasowski}, G., {Cohen}, R.~E., {Chojnowski}, S.~D., {et~al.} 2017, \aj, 154,
  198, \dodoi{10.3847/1538-3881/aa8df9}

\end{thebibliography}



\appendix

\section{Stellar Parameter Comparisons}\label{app:params_comparison}

\subsection{Comparison to LTE spectroscopic-only parameters}

\begin{figure}[h!]
    \centering
    \includegraphics[width=0.7\linewidth]{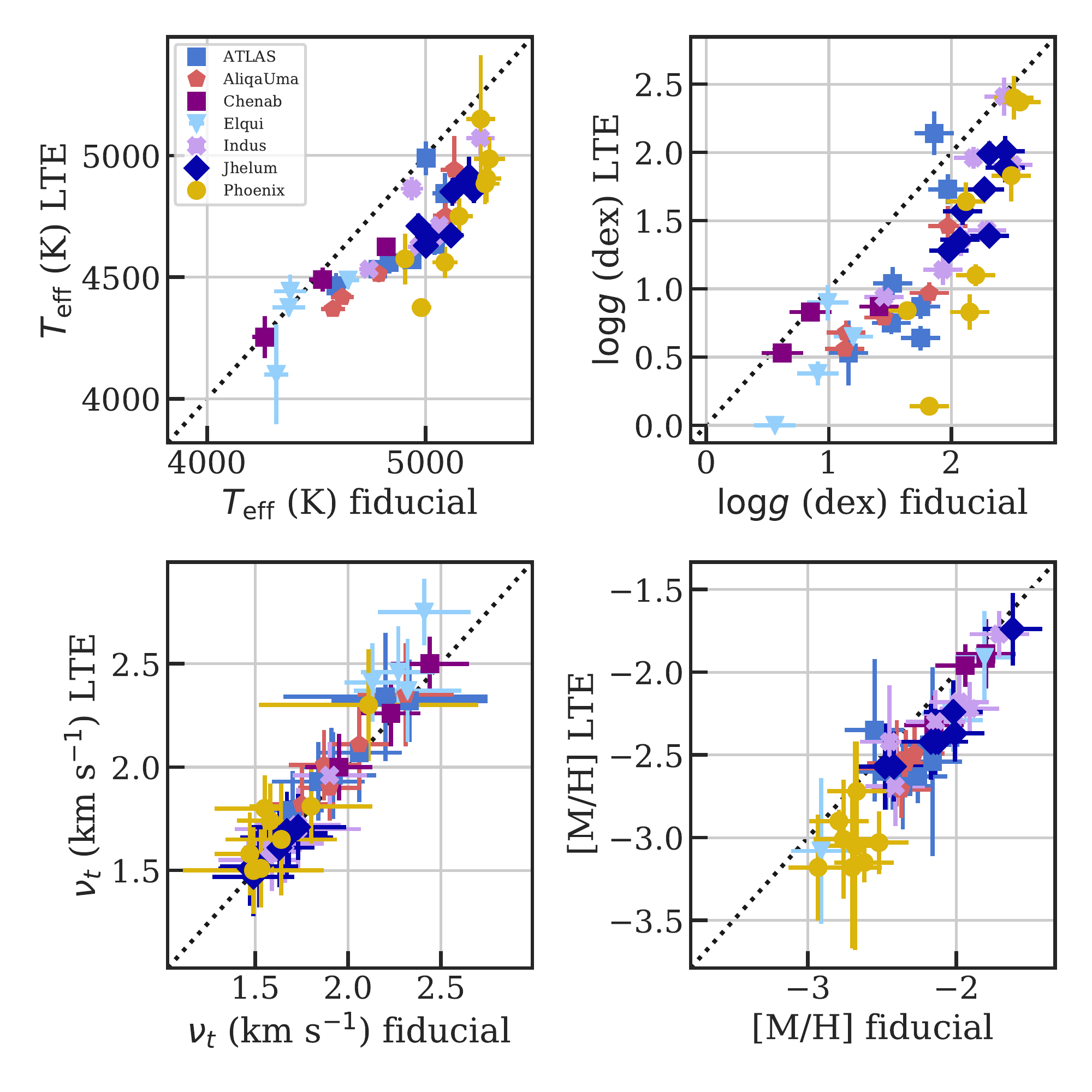}
    \caption{Comparison of the adopted fiducial stellar parameters to parameters from a standard 1D-LTE analysis. See text for details.}
    \label{appfig:lte_params}
\end{figure}

A standard LTE stellar parameter analysis is done for comparison and verification.
We determine $\Teff$ by balancing Fe\,I abundance vs. excitation potential, $\logg$ by balancing Fe\,I and Fe\,II abundances, $\vt$ by balancing Fe\,II abundance vs. reduced equivalent width, and set [M/H] to the Fe\,II abundance.
The LTE-only stellar parameters are compared to the fiducial parameters in Figure~\ref{appfig:lte_params}.
Because of NLTE effects of Fe\,I, such LTE analysis in cool, metal-poor stars like ours tends to produce cooler temperatures and lower $\logg$ compared to photometric temperatures and theoretical isochrones \citep[e.g.,][]{Ezzeddine17}.
A pure LTE analysis thus also would shift $\vt$ and [M/H] to higher and lower values, respectively.
In our sample, the median offset and half-of-68\% scatter is $\Delta \Teff = -272 \pm 129$ K, $\Delta \logg = -0.55 \pm 0.32$ dex, $\Delta \vt = 0.04 \pm 0.08 \kms$, and $\Delta \mbox{[M/H]} = -0.22 \pm 0.13$, where the sign of $\Delta$ is $\text{LTE} - \text{fiducial}$.

The LTE stellar parameters rely only on spectroscopy and show all stars to be red giants. This verifies that our stars are not foreground dwarf interlopers and justifies the use of photometric stellar parameters.

Note that the photometric and isochrone-based parameters suggest that a linear correction to an LTE-only $\Teff$ \citep[e.g.,][]{Frebel13} is insufficient to describe the transformation to a photometric $\Teff$.

\subsection{Comparison to \code{rvspecfit} stellar parameters}
\begin{figure}[h]
    \centering
    \includegraphics[width=0.7\linewidth]{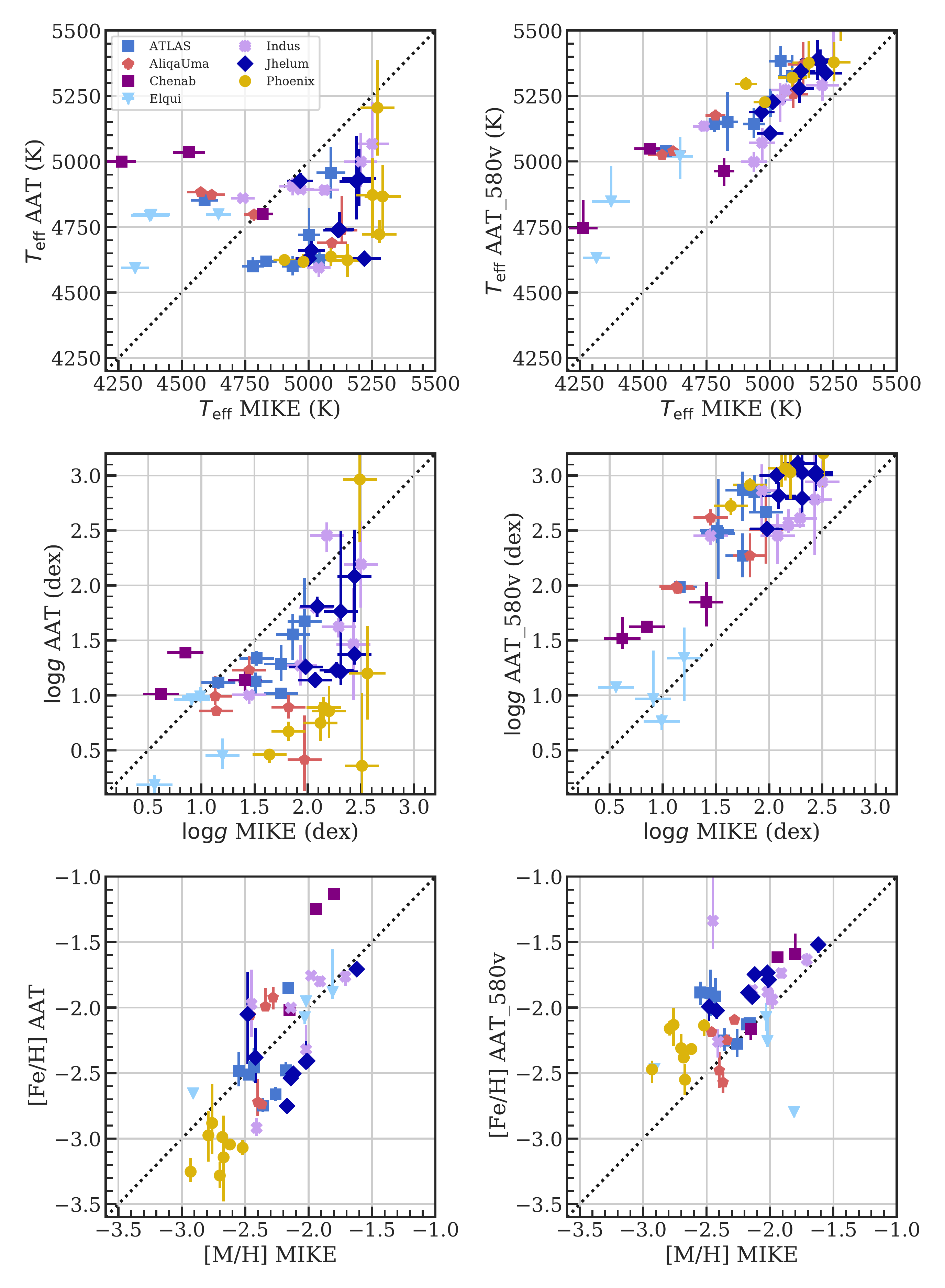}
    \caption{Comparison of MIKE to AAT stellar parameters. The left column is the fit to the Ca triplet (1700D grating, $R \sim 10,000$, 8420--8820\AA), and the right column is to the blue arm (580V grating, $R \sim 1,300$, 3800--5800\AA).}
    \label{fig:aatsp}
\end{figure}

The AAT spectra used to identify these stream targets \citep{Li19S5} had stellar parameters and metallicities determined by \code{rvspecfit} \citep{Koposov11,rvspecfit}.
This is a full-spectrum fit using the \code{PHOENIX-2.0} spectral grid \citep{Husser13}.
The comparison is shown in Figure~\ref{fig:aatsp}.
The left column shows the comparison to values determined from the red 1700D grating ($R \sim 10000$, 8420--8820\AA) while the right shows values determined from the blue 580V grating ($R \sim 1300$, 3800-5800\AA).

In general, there are clear differences in the AAT stellar parameters compared to the MIKE stellar parameters. On the red side, the differences could be attributed to the fact that the AAT is effectively doing an LTE spectroscopic parameter determination.
Comparing the left column of Figure~\ref{fig:aatsp} to Figure~\ref{appfig:lte_params}, the $\Teff$ and $\logg$ trends are similar for the bulk of stars, failing mostly on the coolest stars.
On the blue side, \code{rvspecfit} prefers higher $\Teff$, $\logg$, and [M/H] compared to the derived MIKE values. The origin of this difference is less clear, but could be due to difficulties modeling Balmer line shapes biasing temperatures to be high.
However in both cases, the metallicities are reasonably consistent, especially the relative metallicities.

\section{A New Framework for Abundance Means and Uncertainties}\label{app:estimator}

Here we describe and derive how to combine individual line measurements into final abundances and uncertainties.
We consider individual line errors and responses to stellar parameters, self-consistently estimate and include systematic uncertainties, and fully propagate all stellar parameter correlations to \emph{both} the mean and error of the final abundance.
The propagation of stellar parameters to abundance errors is similar to that in the literature \citep[e.g.,][]{McWilliam95}. However, previous treatments did not consider the effect of stellar parameter correlations on the abundance mean.

Consider a star with measured stellar parameters $\theta_k$ for $k=1$ to $4$ (i.e. $\theta_1 = \Teff$, $\theta_2 = \logg$, $\theta_3 = \vt$, $\theta_4 = \MH$). We assume $\theta$ is drawn from a multivariate normal distribution $\theta \sim \mathcal{N}(\Theta, \Sigma_\theta)$ where $\Theta$ is the true stellar parameters and $\Sigma_\theta$ is the covariance matrix taken from combining the individual stellar parameter uncertainties $\sigma_{\theta,}$ and the correlation matrix $\rho_{kl}$.
Define $\delta\theta = \theta - \Theta$ to be the stellar parameter error, which has the distribution $\delta\theta \sim \mathcal{N}(0, \Sigma_\theta)$.

Consider a species $X$ in this star that is measured by $N$ lines indexed by $i=1, \ldots N$.
Each line has a measured abundance $x_i$, a statistical error $e_i$, and gradients with respect to each stellar parameter $G_{i,k} = \delta_{i,k}/\sigma_{\theta,k}$ where $\delta_{i,k}$ is defined as in Table~\ref{tab:lines}.
Our model for $x_i$ is
\begin{equation}\label{eq:ximodel}
\begin{split}
    x_i &= \xtrue + \epsilon_i + \sum_k G_{i,k} \delta\theta_k \\
        &= \xtrue + \epsilon_i + G_i^T \delta\theta
\end{split}
\end{equation}
where 
$\xtrue$ is the true abundance of species $X$, 
$\epsilon_i \sim \mathcal{N}(0, e_i^2)$ is the random offset from the true value, and
$\delta\theta \sim \mathcal{N}(0,\Sigma_\theta)$ as above.
In other words, we assume $x_i$ has a linear dependence on the stellar parameters.

Our aim is to derive the best estimator for the mean and variance of $\xtrue$, i.e. $\hat x$ and $\Var(\hat x)$. As all distributions are multivariate Gaussians, it is convenient to rewrite Equation~\ref{eq:ximodel} in vector/matrix form as 
\begin{equation}\label{eq:ximatrixmodel}
    x = \xtrue I + M \psi
\end{equation}
where $x$ is the vector of $x_i$; $I$ is defined as the vector of $N$ 1's; the vector $\psi$ is a vector of all the random offsets with size $N + 4$,
\begin{equation}
    \psi = \begin{pmatrix} \epsilon_1 \\ \vdots \\ \epsilon_N \\ \delta\theta_1 \\ \vdots \\ \delta\theta_4 \end{pmatrix},\ \ \psi \sim \mathcal{N}(0, \Sigma_\psi)
\end{equation}
where the covariance matrix $\Sigma_\psi$ has $e_i^2$ on the diagonal augmented by the stellar parameter covariances, i.e.,
\begin{equation}
    \Sigma_\psi = 
    \begin{pmatrix}
    \begin{matrix}
    e_1^2 & 0 & \hdots & 0 \\
    0 & e_2^2 &  & 0 \\
    \vdots & & \ddots & \vdots \\
    0 & \hdots & 0 & e_N^2
    \end{matrix} & \multicolumn{2}{c}{0} \\
    \multirow{2}{*}{0} & \multicolumn{2}{c}{\multirow{2}{*}{$\Sigma_\theta$}}\\
    \end{pmatrix}
\end{equation}
and the matrix $M$ projects from $N+4$ to $N$ dimensions:
\begin{equation}
    M = 
    \begin{pmatrix}
    1 & 0 & \hdots & 0 & G_{11} & G_{12} & G_{13} & G_{14} \\
    0 & 1 & \hdots & \vdots & \vdots & \vdots & \vdots & \vdots \\ 
    \vdots &  & \ddots & \vdots & \vdots & \vdots & \vdots & \vdots \\
    0 & \hdots & 0 & 1 & G_{N1} & G_{N2} & G_{N3} & G_{N4}
    \end{pmatrix}
\end{equation}

Since $M$ is a constant matrix and $\psi$ is a multivariate Gaussian random vector, the distribution of $M\psi$ is
\begin{align}
    M\psi \sim \mathcal{N}(0, \Sigmatilde) \\
    \Sigmatilde = M \Sigma_\psi M^T
\end{align}
and thus our observed vector $x$ is distributed $x \sim \mathcal{N}(\xtrue I, \Sigmatilde)$.
The best estimator for the mean and variance of $\xtrue$ is then
\begin{align}
    \hat{x} &= \frac{I^T \Sigmatilde^{-1} x}{I^T \Sigmatilde^{-1} I} \label{eq:appendixxhat} \\
    \Var(\hat x) &= \frac{1}{I^T \Sigmatilde^{-1} I} \label{eq:appendixvarxhat}
\end{align}

Rather than construct and project down the augmented matrix,
some tedious but straightforward algebra shows
\begin{equation}\label{eq:appendixsigmatilde}
    \Sigmatilde = 
    % \begin{pmatrix}
    % e_1^2 & 0 & \hdots & 0 \\
    % 0 & e_2^2 &  & \vdots \\
    % \vdots & & \ddots & 0 \\
    % 0 & \hdots & 0 & e_N^2
    % \end{pmatrix}
    \text{diag}(e_i^2)
    +
    \delta \rho \delta^T
\end{equation}
where $\delta$ is the matrix of $\delta_{i,k} = G_{i,k} \sigma_k$ and $\rho$ is the correlation matrix.
Computationally, we use this form to calculate $\Sigmatilde$ rather than creating the augmented $\Sigma_\psi$ and $M$ matrices.

To get some intuition on this result, note that we can rewrite Equations \ref{eq:appendixxhat} and \ref{eq:appendixvarxhat} in terms of a weighted sum. If we define
\begin{equation}
    \wtilde = I^T \Sigmatilde^{-1}
\end{equation}
or $\wtilde_i = \sum_j \Sigmatilde_{ij}^{-1}$,
then we see that
\begin{align}
    \hat x &= \frac{\sum_i \wtilde_i x_i}{\sum_i \wtilde_i} \\
    \Var(\hat x) &= \frac{1}{\sum_i \wtilde_i}
\end{align}
which looks like the usual inverse-variance weighted sum but using a different covariance matrix to determine the weights. Note that unlike an inverse variance, these weights can be negative, and they depend on the whole set of lines used to estimate the mean.
The weights $\wtilde_i$ are provided in Table~\ref{tab:lines}.

The above calculations assume that each line provides an unbiased estimate of the total error. In reality, several additional systematic issues (e.g., atomic data uncertainties, 1D model atmospheres, and the LTE assumption) can cause substantial biases that are not averaged away.
This is especially important when many lines are measured for a species (e.g., Fe I), as the systematic floor is well above the naive precision.
To account for this, we use the observed line-to-line scatter to add a systematic floor to the per-line errors.
We modify the model for $\epsilon_i$ to be $\epsilon_i \sim \mathcal{N}(0, e_i^2 + s_X^2)$, where we have added a systematic uncertainty floor for each line of $s_X \geq 0$.
We can solve for $s_X$ in terms of $x_i$, $e_i$, and the optimal estimator $\hat x$ by maximizing the log likelihood:
\begin{equation}
    \log \mathcal{L} = -0.5\sum_i \frac{(x_i - \hat x)^2}{e_i^2+s_X^2} - 0.5 \sum_i \log({e_i^2+s_X^2}) + \text{constants}
\end{equation}
or, after taking the derivative with respect to $s_X$ and setting to zero, solving
\begin{equation}
    \sum_i \frac{(x_i - \hat x)^2}{(e_i^2+s_X^2)^2} = \sum_i \frac{1}{e_i^2+s_X^2}
\end{equation}
for $s_X$, which has to be done numerically.
Since $\hat x$ depends on $s_X$, we iterate between calculating $\hat x$ and numerically solving for $s_X$ until reaching a precision $<0.001$ dex on $s_X$. Then in Equation~\ref{eq:appendixsigmatilde} we simply replace $\text{diag}(e_i^2)$ with $\text{diag}(e_i^2 + s_X^2)$.

This model for the systematic errors is purely empirical, but in principle a more physically motivated approach could be applied under this framework. As the simplest example, $s_X$ can explicitly be set to include the $\log gf$ uncertainties reported in atomic data measurements \citep[e.g.,][]{McWilliam13}. As a more complicated example, NLTE corrections could propagate uncertainties in collisional or radiative rates to line-by-line corrections, which can both modify the mean and systematic uncertainty of a particular line.

Now that we have the optimal estimator for any species $X$, we now consider the covariance between two species $X$ and $Y$. Let their optimal estimators be defined by $\hat x = U_X^T x$ where $U_{X,i} = \wtilde_i/\sum_j \wtilde_j$; and similarly for $Y$. Also let the gradient/difference matrices be $G_X$ and $\delta_X = G_X \text{diag}(\sigma_\theta)$, respectively, which are each $N \times 4$ matrices. Then
\begin{equation}\label{eq:appendixcovar}
\begin{split}
    \Cov(\hat x, \hat y) &= U_X^T G_X \Sigma_\theta G_Y^T U_Y \\
                         &= U_X^T \delta_X \rho \delta_Y^T U_Y \\
                         &\equiv \Delta_X \rho \Delta_Y^T
\end{split}
\end{equation}
where we have defined $\Delta_X = U_X^T \delta_X$.
Table~\ref{tab:abunds} tabulates $\Delta_X$ for all $X$, which we call $\Delta_T$, $\Delta_g$, $\Delta_v$, and $\Delta_M$ in that table. These are morally equivalent to the table of stellar parameter uncertainty given in most high-resolution spectroscopy papers but include proper line weighting.
Note that if calculating $\Cov(\hat x, \hat x)$ make sure to use Equation~\ref{eq:appendixvarxhat}, which includes an extra statistical error term.

Finally to wrap it all up, the error on $\log\epsilon(X)$ (and thus also [X/H], since we assume the solar normalization is error-free) is simply $\sqrt{\Var(\hat x)}$ from Equation~\ref{eq:appendixvarxhat}, which automatically includes all stellar parameter uncertainties and correlations.
To find the error on the ratio of two species [X/Y], we use the fact that [X/Y] = [X/H] - [Y/H], so
\begin{equation}
    \Var([X/Y]) = \Var(\hat x - \hat y) = \Var(\hat x) + \Var(\hat y) - 2 \Cov(\hat x, \hat y)
\end{equation}
and can be evaluated using Equations~\ref{eq:appendixvarxhat} and \ref{eq:appendixcovar}.
Similarly, we can take the covariance between any set of element ratios, e.g. for elements $A,B,C,D$ with estimators $\hat a, \hat b, \hat c,\hat d$
\begin{equation}
    \Cov([A/B],[C/D]) = \Cov(\hat a, \hat c) + \Cov(\hat b, \hat d) - \Cov(\hat a, \hat d) - \Cov(\hat b, \hat c)
\end{equation}

For pedagogical purposes, let us compare to two alternate estimators for $\hat x$ and $\Var(\hat x)$ used in the literature.
Most high-resolution studies do not calculate line-by-line uncertainties, instead taking a straight mean of all measured lines, i.e. $\hat x = \sum_i x_i / N = \sum_i x_i / \sum_i 1$.
The error on the mean is usually found as the standard error, imposing a systematic floor (e.g., 0.1 dex) that is supposed to account for uncertainties in model atmospheres, atomic data, or other model uncertainties.
This standard procedure weights every line equally, which is justifiable in the limit of carefully selected line measurements in very high-S/N data where systematic uncertainties (other than uncertain stellar parameters) dominate.
However, in our red giants, where the blue flux is much lower than the red flux, lines are clearly measured in regions of different S/N.
Furthermore, in low S/N data, this procedure neglects the fact that the error on an individual line measurement is often much larger than the empirical deviation, especially if there are few lines for an element.
The estimator provided here accounts for these issues, at the considerable cost of having to compute uncertainties for individual lines.

To account for some of the issues described above, \citet{McWilliam95} computed individual line uncertainties and combined them with a weighted mean.
% Each line was assigned an error $\sigma_i^2 = \sigma_{i,\text{stat}}^2 + \sum_k \delta_{i,k}^2$, i.e., the quadrature sum of random uncertainties and stellar parameter uncertainties.
Each line was assigned an error $\sigma_i^2 = \sigma_{i,\text{stat}}^2 + \sum_{k,l} \delta_{i,k}\delta_{i,l}\rho_{kl}$, i.e., the quadrature sum of random uncertainties and stellar parameter uncertainties including all cross terms.
Then using weights $w_i = 1/\sigma_i^2$, the mean was found with $\hat x = \sum_i (w_i x_i) / \sum_i w_i$ with uncertainty $\Var (\hat x) = 1/\sum_i w_i$.
However, this procedure ignores correlations between line abundances due to the fact that the same stellar parameters are used for all lines. In other words, it neglects the off-diagonal terms of $\Sigmatilde$, which usually results in moderately underestimated uncertainties.
\citet{Ji20} used the above procedure but added a systematic error that was estimated with the weighted standard error of the lines and added in quadrature to the statistical error. Compared to the analysis here, their overall error is a slight overestimate of the total uncertainty because it double-counts the random error.

\section{Equivalent width and abundance verification}\label{app:equivalentwidths}

To verify the equivalent width and corresponding abundance measurements, we performed an independent check of equivalent width and abundance measurements.
Equivalent widths for 2/3rds of our program stars were independently analyzed using \code{IRAF} and \code{MOOG} by T.T.H., including normalization, equivalent widths, model atmosphere interpolation, and abundance measurements. Equivalent widths were measured by fitting Gaussian profiles to the absorption lines in the continuum-normalized spectra using the \code{splot} task in \code{IRAF}.

\begin{figure}
    \centering
    \includegraphics[width=0.8\linewidth]{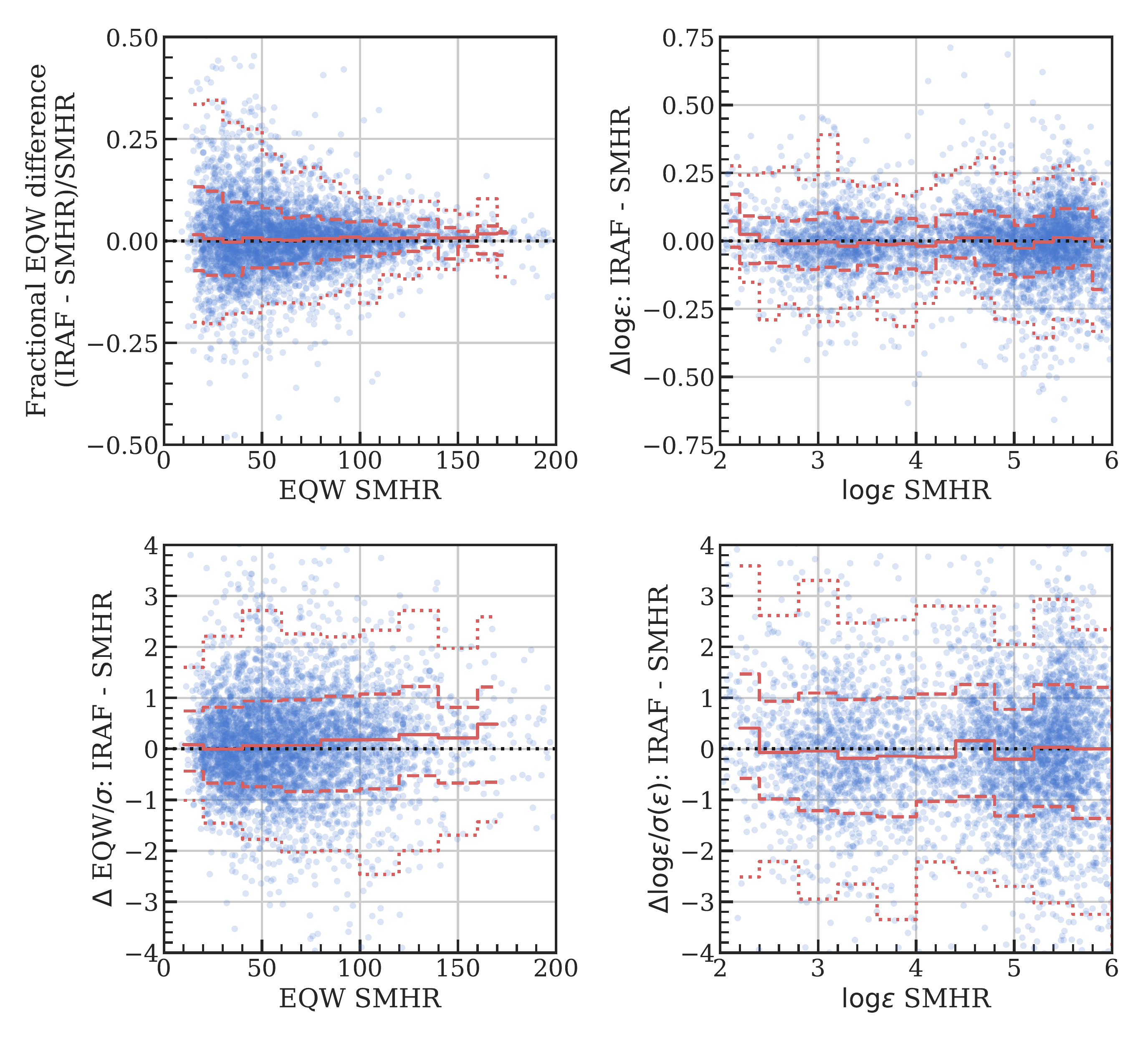}
    \caption{Verification of equivalent widths and abundances. Two of the authors (A.P.J. using \code{SMHR} and T.T.H. using \code{IRAF}) measured equivalent widths, interpolated model atmospheres, and measured abundances with independent methods; but using the same stellar parameters, model atmosphere grid, and radiative transfer code.
    In all panels, blue points show differences between individual matched lines while red dashed (dotted) lines show binned 68\% (95\%) scatter in bins of the x-axis.
    The top-left panel shows the fractional difference in equivalent width of the measurements, showing the expected increase in scatter towards lower equivalent widths.
    The bottom-left panel shows that after normalizing by the equivalent width uncertainties in Table~\ref{tab:lines}, the differences are well-described by Gaussian uncertainties.
    The top-right panel shows the typical scatter between equivalent width abundance measurements is about 0.1 dex, while the bottom-right shows the statistical abundance uncertainties are a good description of the differences.}
    \label{fig:tthapj_eqw}
\end{figure}

Figure~\ref{fig:tthapj_eqw} shows the resulting equivalent width and abundance differences.
The left column plots the difference between TTH's equivalent widths and APJ's equivalent widths.
The right column plots the difference between TTH's abundances and APJ's abundances.
The red solid, dashed, and dotted lines show the median, 68\% scatter, and 95\% scatter in the difference, computed in bins of the x-axis.
The top-left panel shows the fractional equivalent width difference between the two measurements.
The 1$\sigma$ scatter is about 10-15\% at the lowest equivalent widths, decreasing to 5-10\% at higher equivalent widths.
The top-right panel shows the typical $1\sigma$ scatter between individual line abundances is about 0.1 dex.
There is no significant bias in the mean.
The bottom two panels show the difference between equivalent width and abundance, normalized by the uncertainties in Table~\ref{tab:lines}.
As in the top panels, the median, 68\% scatter, and 95\% scatter in bins of the x-axis are plotted as red solid, dashed, and dotted lines respectively.
If the uncertainties are approximately Gaussian with no bias, then the dashed red lines should line up with $\pm1$ units on the y-axis, and the dotted red lines should line up at $\pm2$ units.
The equivalent width uncertainties do indeed line up quite well with these values.
The 68\% scatter in abundance uncertainties also lines up well, but the tails are a little heavier as the dashed red lines in the bottom-right panel are around 2.5-3.0 instead of 2.0.

To verify the synthetic spectrum abundances, we selected seven stars covering the stellar parameter and S/N range of our stars: Chenab\_12, Elqui\_1, ATLAS\_12, Indus\_15, AliqaUma\_7, Phoenix\_2, and Phoenix\_8.
For these seven stars, abundances were independently derived using spectral synthesis via MOOG by T.T.H.  The spectrum normalization, stellar parameters, and model atmospheres were independently determined. 

\begin{figure}
    \centering
    \includegraphics[width=0.5\linewidth]{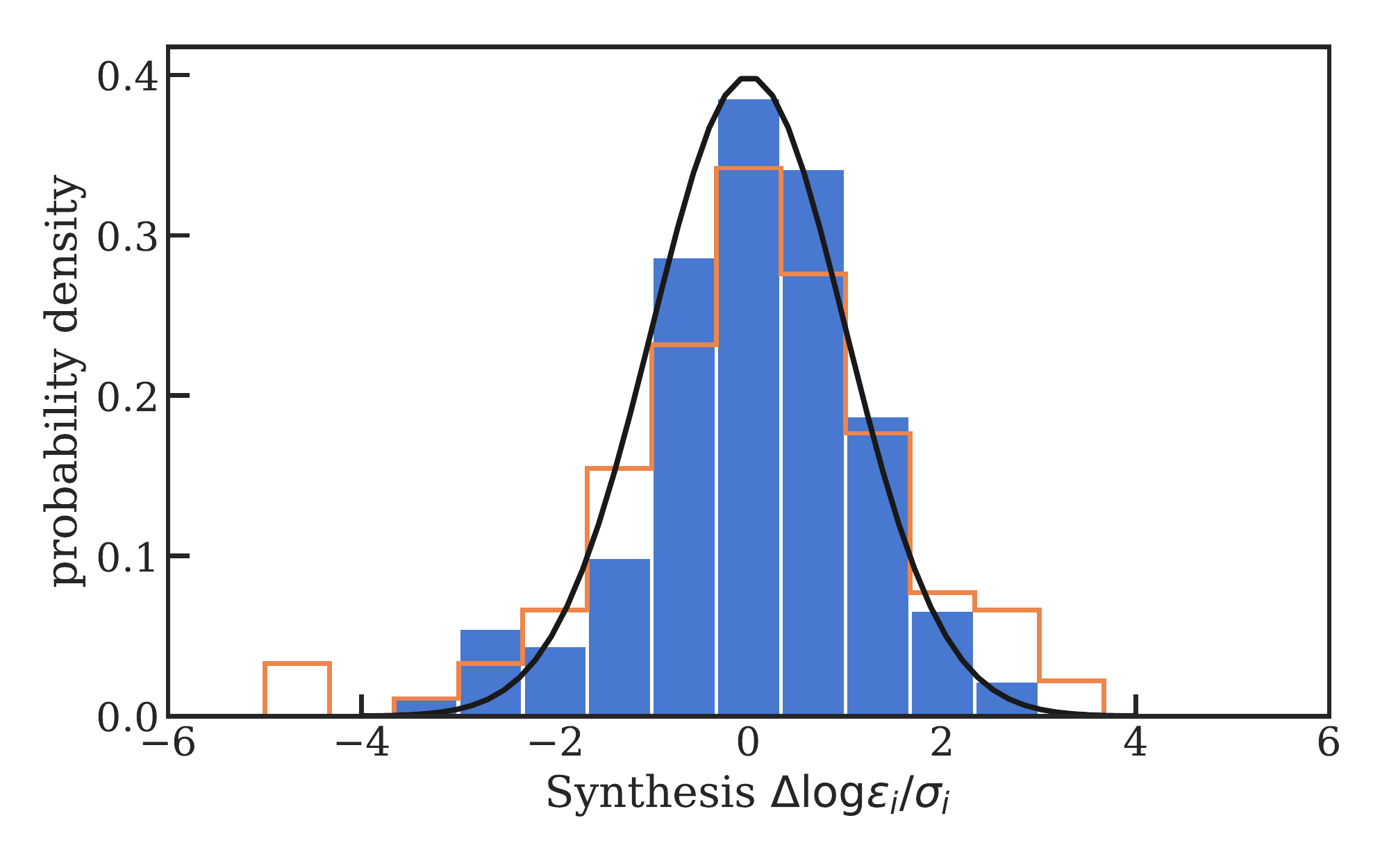}
    \caption{Difference between abundances in Table~\ref{tab:lines} and independent verification. The solid blue histogram is normalized by the total error $\sigma_i$ in Table~\ref{tab:lines}, while the slightly wider orange open histogram is normalized by the pure statistical error $e_i$.
    The black line indicates the unit normal distribution.}
    \label{fig:synchi2}
\end{figure}

The difference between the independent syntheses of individual features is shown in Figure~\ref{fig:synchi2}.
We show the differences normalized by two different abundance uncertainties of Table~\ref{tab:lines}, the pure statistical error reported by SMHR $e_i$ and the adjusted systematic error $\sigma_i$, plotted as orange and blue histograms respectively.
The pure statistical uncertainties (orange) somewhat underestimate the observed dispersion.
Line-by-line investigation shows the differences are primarily due to statistical errors that are too small for some Al, Sc, Mn, and Ba lines.
For Al, the differences are mostly driven by systematics in continuum fitting, especially for the 3961\Ang line that is in the wing of a H line. An extra 0.3 dex systematic error is added to account for this.
The Sc and Mn lines have significant hyperfine splitting, and their abundance is more affected by the smoothing kernel applied to the synthetic spectrum. Reasonable changes in the smoothing kernel affect the abundances by up to 0.1 dex, so 0.1 dex systematic uncertainty is added to Sc and Mn.
For Ba, we use strong lines with hyperfine splitting, and the resulting abundances are also sensitive to the smoothing kernel so an extra 0.1 dex systematic uncertainty is added.
Other lines with hyperfine structure are V, Co, La, and Eu. The existing statistical and systematic errors for these elements appear adequate, so we did not include any extra uncertainty for them.
Including these systematic uncertainties, the normalized abundance differences (blue histogram in Figure~\ref{fig:synchi2}) are close to normally distributed.

\section{Abundance Correlations with Stellar Parameters}\label{app:params_errors}

\begin{figure*}
    \centering
    \includegraphics[width=0.99\linewidth]{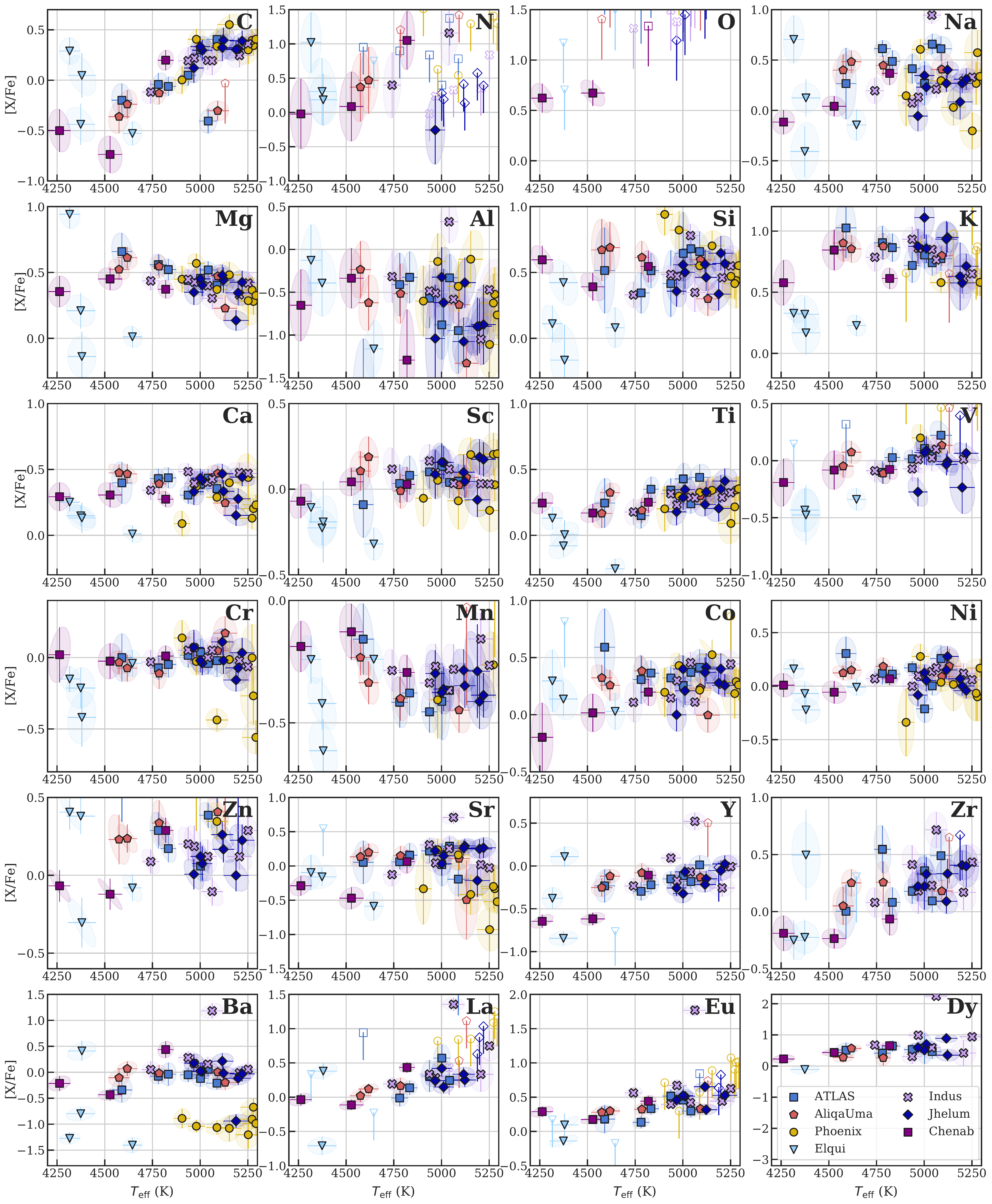}
    \caption{[X/Fe] vs $\Teff$.}
    \label{fig:Teff_correlations}
\end{figure*}

\begin{figure*}
    \centering
    \includegraphics[width=0.99\linewidth]{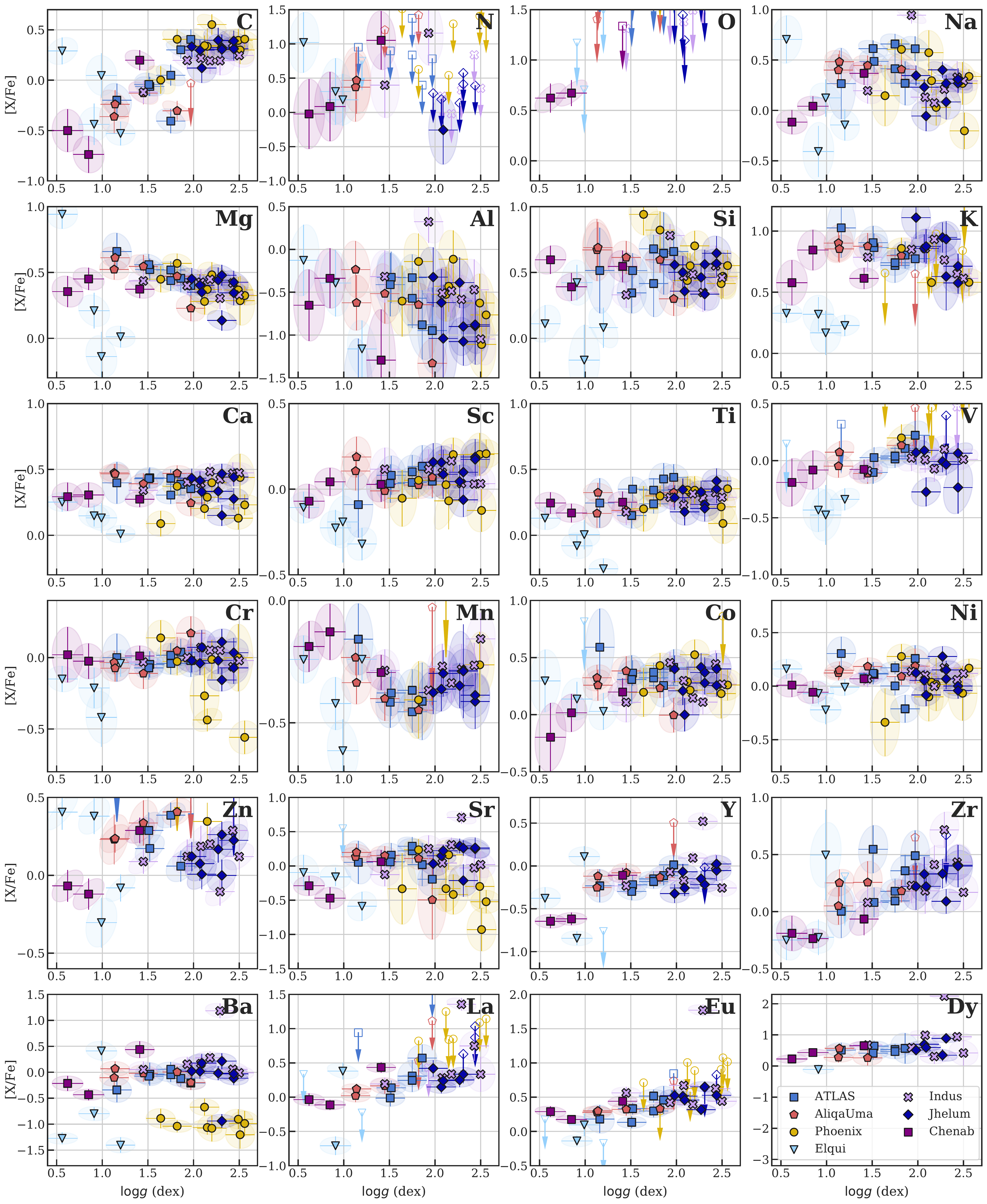}
    \caption{[X/Fe] vs $\logg$.}
    \label{fig:logg_correlations}
\end{figure*}

\begin{figure*}
    \centering
    \includegraphics[width=0.99\linewidth]{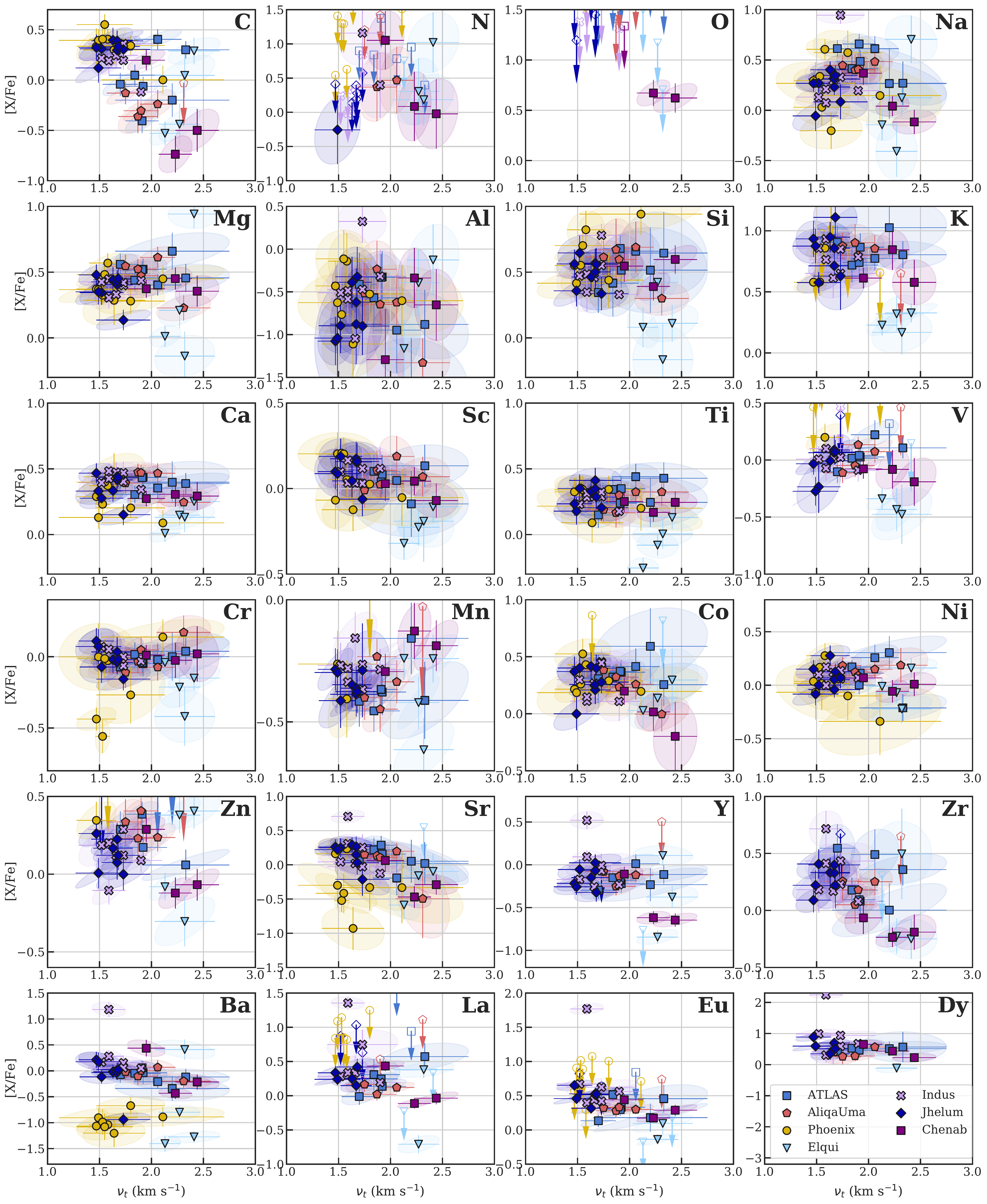}
    \caption{[X/Fe] vs $\vt$.}
    \label{fig:vt_correlations}
\end{figure*}

Figures~\ref{fig:Teff_correlations}, \ref{fig:logg_correlations}, and \ref{fig:vt_correlations} show the abundance trends and correlations with respect to $\Teff$, $\logg$, and $\vt$.
The $1\sigma$ error ovals include correlations between the [X/Fe] abundance and a given stellar parameter.
These are provided primarily as a way for users of the abundances to check for any systematic effects or estimate correlation effects due to stellar parameter uncertainties.

There are some important intrinsic correlations to mention.
First, $\Teff$, $\logg$, and $\vt$ are all highly correlated (Table~\ref{tab:spcorr}, Figure~\ref{fig:sp}).
Thus apparent correlations are not necessarily causal, and should be checked against the typical orientation of the error ellipses.
Second, warmer giants both tend to have weaker lines and are intrinsically less luminous.
Third, due to intrinsic distance differences between the streams, stars in a given stream do not all occupy the same stellar parameters.
The coolest stars in our sample (and thus lowest $\logg$ and highest $\vt$ stars) are in Chenab and Elqui;
the warmest stars in our sample are in Phoenix, Jhelum, and Indus; and ATLAS and Aliqa Uma are in between.
The differing intrinsic abundance trends in these streams thus clearly imprint on the correlations with stellar parameters.

\end{document}